\documentclass[review]{elsarticle}

\usepackage[colorlinks,citecolor=blue,linktoc=all,linkcolor=cyan]{hyperref}
\usepackage{graphicx}

\usepackage[T1]{fontenc}
\usepackage{dsfont}       % use mathds instead of mathbb for outline fonts
\usepackage{mathrsfs}     % provides mathscr without overwriting mathcal
\usepackage{slashed}      % For Dirac slash notation.
\usepackage{amsmath}
\usepackage{amssymb}
\usepackage{amsbsy}
\usepackage{amsfonts}

\numberwithin{equation}{section}
\numberwithin{table}{section}
\numberwithin{figure}{section}

\journal{Progress in Particle and Nuclear Physics}

\usepackage{titlesec}
\usepackage{sectsty}
\titleformat{\section}{\normalfont\Large\bfseries}{\thesection}{1em}{}
\titleformat{\subsection}{\normalfont\large\bfseries}{\thesubsection}{1em}{}
\titleformat{\subsubsection}{\normalfont\normalsize\bfseries}{\thesubsubsection}{1em}{}
\graphicspath{{figs/}}
\DeclareMathOperator{\SU}{\mathrm{SU}}
\DeclareMathOperator{\cep}{\mathrm{cep}}
\DeclareMathOperator{\Tr}{\mathrm{Tr}}

\newcommand{\Tcr}{T_{\mbox{\tiny{cr}}}}
\newcommand{\mucr}{\mu_{\mbox{\tiny{cr}}}}
\newcommand{\LambdaTC}{\Lambda_{\mbox{\tiny{TC}}}}
\newcommand{\LambdaETC}{\Lambda_{\mbox{\tiny{ETC}}}}

\usepackage[stable]{footmisc}

\usepackage{multibib}
\newcites{Talk}{Workshop ``\emph{Phase Transitions in Particle Physics}'' Talks}

\usepackage[font=small, labelfont=bf, textfont={small,it}]{caption}

\usepackage{bm}

\begin{document}
	
\begin{frontmatter}
		
\title{\textbf{\textsc{Phase Transitions in Particle Physics}}\\\textsc{Results and Perspectives from Lattice Quantum Chromo-Dynamics}}

\author[1,2]{\textbf{Gert~Aarts}}
\author[3]{\textbf{Joerg~Aichelin}}
\author[1]{\textbf{Chris~Allton}}
\author[4,5]{\textbf{Andreas~Athenodorou}}
\author[6]{\textbf{Dimitrios~Bachtis}}
\author[7,7a]{\textbf{Claudio~Bonanno}}
%\author[8]{\textbf{Vitaly~Bornyakov}}
\author[9]{\textbf{Nora~Brambilla}}
\author[10,11,12]{\textbf{Elena~Bratkovskaya}}
\author[13,14]{\textbf{Mattia~Bruno}}
\author[15]{\textbf{Michele~Caselle}}
\author[16]{\textbf{Costanza~Conti}}
\author[17]{\textbf{Roberto~Contino}}
\author[18]{\textbf{Leonardo~Cosmai}}
\author[11]{\textbf{Francesca~Cuteri}}
\author[19]{\textbf{Luigi~Del~Debbio}}
\author[4]{\textbf{Massimo~D'Elia}}
\author[20]{\textbf{Petros~Dimopoulos}}
\author[20]{\textbf{Francesco~Di~Renzo}}
\author[10,21]{\textbf{Tetyana~Galatyuk}}
\author[22]{\textbf{Jana~N.~Guenther}}
\author[23]{\textbf{Rachel~Houtz}}
\author[24]{\textbf{Frithjof~Karsch}}
\author[25]{\textbf{Andrey~Yu.~Kotov}}
% CORRESPONDING AUTHOR
\author[7]{\textbf{Maria~Paola~Lombardo}\corref{corrauth}}
\cortext[corrauth]{Corresponding author}
\ead{mariapaola.lombardo@unifi.it}
\author[6]{\textbf{Biagio~Lucini}}
\author[4]{\textbf{Lorenzo~Maio}}
\author[15]{\textbf{Marco~Panero}}
\author[27]{\textbf{Jan~M.~Pawlowski}}
\author[17]{\textbf{Andrea~Pelissetto}}
\author[11]{\textbf{Owe~Philipsen}}
\author[28,29]{\textbf{Antonio Rago}}
\author[30]{\textbf{Claudia~Ratti}}
%\author[8]{\textbf{Roman~Rogalyov}}
\author[31]{\textbf{Sin\'ead~M.~Ryan}}
\author[32]{\textbf{Francesco~Sannino}}
\author[33,33a]{\textbf{Chihiro~Sasaki}}
\author[11,34]{\textbf{Philipp~Schicho}}
\author[24]{\textbf{Christian~Schmidt}}
\author[24]{\textbf{Sipaz~Sharma}}
\author[11,12]{\textbf{Olga~Soloveva}}
\author[35]{\textbf{Marianna~Sorba}}
\author[36]{\textbf{Uwe-Jens~Wiese}}

\address[1]{\scriptsize Department of Physics, Swansea University, Swansea, SA2 8PP, United Kingdom}
\address[2]{European Centre for Theoretical Studies in Nuclear Physics and Related Areas (ECT$^\star$) \& Fondazione Bruno Kessler, 38123 Villazzano (TN), Italy}
\address[3]{SUBATECH, Universit\'e de Nantes, IMT Atlantique, IN2P3/CNRS, 4 rue Alfred Kastler, 44307 Nantes cedex 3, France}
\address[4]{Universit\`a di Pisa and INFN Sezione di Pisa, Largo B.~Pontecorvo 3, I-56127 Pisa, Italy}
\address[5]{Computation-based Science and Technology Research Center, The Cyprus Institute, 20 Kavafi Str., Nicosia 2121, Cyprus}
\address[6]{Department of Mathematics, Swansea University, Bay Campus, SA1 8EN, Swansea, UK}
\address[7]{INFN Sezione di Firenze, Via G.~Sansone 1, I-50019 Sesto Fiorentino, Firenze, Italy}
\address[7a]{Instituto de Física Teórica UAM-CSIC, c/ Nicolás Cabrera 13-15, Universidad Autónoma de Madrid, Cantoblanco, E-28049 Madrid, Spain}
%\address[8]{Institute for High Energy Physics of the NRC ``Kurchatov Institute'', 142281 Protvino, Russia}
\address[9]{Physik Department, Technische Universität München,
	James-Franck-Strasse 1, 85748 Garching, Germany}
\address[10]{GSI Helmholtzzentrum für Schwerionenforschung GmbH, Planckstr. 1, 64291 Darmstadt, Germany}
\address[11]{Institut für Theoretische Physik, Johann Wolfgang Goethe-Universität, Max-von-Laue-Str. 1, 60438 Frankfurt am Main, Germany}
\address[12]{Helmholtz Research Academy Hessen for FAIR (HFHF),GSI Helmholtz Center for Heavy Ion Physics. Campus Frankfurt, 60438 Frankfurt, Germany}
\address[13]{Dipartimento di Fisica, Universit\`a di Milano-Bicocca, Piazza della Scienza 3, I-20126 Milano, Italy}
\address[14]{INFN, Sezione di Milano-Bicocca, Piazza della Scienza 3, I-20126 Milano, Italy}
\address[15]{Department of Physics, University of Turin \& INFN Turin, Via Pietro Giuria 1, I-10125 Turin, Italy}
\address[16]{Dipartimento di Fisica, Universit\`a di Firenze, via G. Sansone 1, 50019 Sesto Fiorentino, Italy}
\address[17]{Dipartimento di Fisica dell’Universit\`a di Roma Sapienza and INFN Sezione di Roma I, I-00185 Roma, Italy}
\address[18]{INFN Sezione di Bari, I-70126 Bari, Italy}
\address[19]{Higgs Centre for Theoretical Physics, School of Physics \& Astronomy, The University of Edinburgh, Peter Guthrie Tait Road, Edinburgh EH9 3FD, United Kingdom}
\address[20]{Dipartimento di Scienze Matematiche, Fisiche e Informatiche,
	Universit\`a di Parma and INFN, Gruppo Collegato di Parma I-43100 Parma, Italy}
\address[21]{Technische Universität Darmstadt, 64289 Darmstadt, Germany}
\address[22]{Department of Physics, Wuppertal University, Gaussstr. 20, D-42119, Wuppertal, Germany}
\address[23]{Institute for Particle Physics Phenomenology, Durham University, Durham DH1 3LE, UK}
\address[24]{Fakultät für Physik, Universität Bielefeld, D-33615 Bielefeld, Germany}
\address[25]{Jülich Supercomputing Centre, Forschungszentrum Jülich, D-52428 Jülich, Germany}
%\address[26]{Bogoliubov Laboratory of Theoretical Physics, Joint Institute for Nuclear Research, Dubna, 141980, Russia}
\address[27]{Institut für Theoretische Physik, Universität Heidelberg, Philosophenweg 16, 69120 Heidelberg, Germany}
\address[28]{Centre for Mathematical Sciences, Plymouth University, Plymouth, PL4 8AA, United Kingdom}
\address[29]{Theory Department, CERN, Esplanade des Particules 1, 1201 Geneva, Switzerland}
\address[30]{Department of Physics, University of Houston, Houston, TX 77204, USA}
\address[31]{School of Mathematics, Trinity College, Dublin, Ireland}
\address[32]{Dipartimento di Fisica ``Ettore Pancini'', Universit\`a degli studi di Napoli ``Federico II'' \& INFN Napoli, Complesso Univ. Monte S. Angelo, I-80126 Napoli, Italy}
\address[33]{Institute of Theoretical Physics, University of Wroclaw, plac Maksa Borna 9, 50-204 Wroclaw, Poland}
\address[33a]{International Institute for Sustainability with Knotted Chiral Meta Matter (SKCM$^2$), Hiroshima University, Higashi-Hiroshima, Hiroshima 739-8511, Japan}
\address[34]{Department of Physics and Helsinki Institute of Physics, P.O. Box 64, FI-00014 University of Helsinki, Finland}
\address[35]{SISSA and INFN Trieste, Via Bonomea 265, 34136 Trieste, Italy}
\address[36]{Albert Einstein Center for Fundamental Physics, Institute for Theoretical Physics, Bern University Sidlerstrasse 5, CH-3012 Bern, Switzerland}

\begin{abstract}
Phase transitions in a  non-perturbative regime can be studied by \emph{ab initio} Lattice Field Theory methods. The  status and future research directions for LFT investigations of Quantum Chromo-Dynamics under extreme conditions are reviewed, including properties of hadrons and of the hypothesized QCD axion as inferred from QCD topology in different phases. We discuss phase transitions in strong interactions in an extended  parameter space, and the possibility of model building for Dark Matter and Electro-Weak Symmetry Breaking. Methodological challenges are addressed as well, including new developments in Artificial Intelligence geared towards the identification of different phases and transitions.

\par\noindent\rule{\textwidth}{0.4pt}
\end{abstract}

		
\end{frontmatter}

\thispagestyle{empty}
\tableofcontents
	
%to begin the line numbers: 
%\linenumbers

\section{Introduction}
\label{sec:introduction}

Gauge theories exist in a variety of different phases. The main focus of this manuscript is Quantum Chromo-Dynamics, QCD, the gauge theory describing strong interactions in elementary particle physics. We will concentrate on an \emph{ab initio} approach, Lattice Field Theory (LFT), and also report on progress within first principles Functional Approaches to QCD (FAs), as well as Effective Field theories (EFTs). We will describe the results that have been obtained, the current challenges, and the future prospects. This overview of the theoretical state of the art is accompanied by reports on the experimental efforts. 

We will consider QCD at finite temperature and/or density, as well as in external magnetic fields. In the space spanned by these parameters, symmetries may be realised in different ways, and the change of symmetry corresponds to phase transitions.

At zero temperature the QCD chiral symmetry is spontaneously broken for massless quarks, with the appearance of composite Goldstone bosons. Further, experimental searches for free quarks have been unsuccessful so far, and the accepted wisdom is that in this regime QCD is confining. The interplay of chiral symmetry and confinement is still poorly understood and is an important subject of current research. When the lightest quarks have non-zero masses, the pseudo-Goldstone becomes massive, with a definite prediction for their dependence on the quark masses.

Temperature induces the restoration of chiral symmetry, with an accompanying
liberation of light degrees of freedom, a dramatic phenomenon probed in
heavy-ion collision experiments. The analysis of the transitions and their characteristics is at the heart of this paper and is described in section~\ref{sec:thermal_phase_transitions_and_critical_points}.

Equally important is the nature of the exotic phase(s) at the high temperatures probed in experiments: a difficult important task is the connection between lattice results, obtained at equilibrium, and experimental observations from heavy ion collisions with their non-equilibrium dynamics. The role of magnetic fields has been investigated as well. These aspects are discussed in section~\ref{sec:nature_and_phenomenology_of_the_quark-gluon_plasma}.
Dense matter poses specific problems: pairing phenomena have been
investigated in a variety of approaches, considering different unbalances, and making also natural a connection with condensed matter. The cold and dense matter is not yet directly accessible with
LFT simulations. Here our knowledge comes mostly from functional approaches to QCD and low energy effective theories, with a wealth of interesting and important phenomena. Since in this manuscript, we focus on topics amenable to LFT studies,  we will not further pursue these important issues. 

The aspects of Strong Interactions outlined so far have experimental and phenomenological relevance. High temperature matter, up to temperatures of about 500 MeV, is created and explored in heavy ion collision experiments. Pushing the temperature at higher values, one reaches regions of cosmological relevance, traversed during the evolution of the primordial Universe. Hypothetically, in this region, the freeze-out of axions occurs: axions are dark matter candidates motivated by a natural extension of QCD, originated by the breaking of an anomalous symmetry. The physics of gravitational waves is an important close-by field. The physics of extremely high matter, beyond experimental capabilities, but still far below the Electroweak Transition, in which the topology of QCD plays a major role, is described in section~\ref{sec:cosmology_topology_axions}.

From a theoretical point of view, QCD is just one among infinitely many non-Abelian gauge theories with chiral symmetries. By simply changing the parameters of the Lagrangian of Strong Interactions, such as the gauge group, i.e.~the number of color charges $N$, the matter field content (including the fermion representation and the number of quark flavors $N_f$), the spacetime dimension $D$, \dots it is possible to investigate different theories and the rich phenomenology they exhibit. Such studies enrich our knowledge and provide helpful inspiration and guidance for devising viable theories beyond the Standard Model. In this manuscript, we will primarily discuss the physics of theories with large $N_f$: the increase of the number of flavors triggers the restoration of chiral symmetry, and the chirally symmetric phase at large $N_f$ is conformally invariant in the infrared. Composite-Higgs models can be built in a specific region of the phase space, i.e., the one close to the conformal window. In the pre-conformal phase the thermal transition may well be stronger, making these theories potentially interesting also
for cosmology. These subjects are discussed in section~\ref{sec:conformal_phase}. 

Lattice methods require the positivity of the Action for the importance sampling involved. This is achieved by rotating 
the time to the imaginary axes, thus making the metric Euclidean. Even in this case, the positivity
of the Action is violated if a chemical potential introduces an imbalance between baryon and anti-baryons, or if a CP violating $\theta$ term is introduced. All these issues are generically known as \emph{sign problem}, i.e. the failure of importance sampling due to a complex statistical weight.  We will highlight the major challenges and some promising avenues in section~\ref{sec:methodological_challenges_spectral_functions_and_sign_problem}. 
Next, we will discuss the application of modern artificial intelligence (AI) techniques to the analysis of phase transitions in section~\ref{sec:machine_learning}. This concerns both the recognition of phase transitions from data samples as well as supporting the importance sampling with machine learning. 

Finally, in section~\ref{sec:statistical_field_theory} we will discuss methods from statistical field theory, which are an essential tool for the analysis of phase transitions. Historically, the main approach to studying the critical and near-critical behavior of a theory has been based on the magnetic equation of state: the starting point is the identification of the order parameter and of the symmetry-breaking pattern at the transition. The key concept is universality and the theoretical framework is that of the renormalization group. Recently, the standard approach has been critically reconsidered, with a deeper analysis of the role of gauge symmetries. In recent years, conformal theories have taken center stage: studies of two-point correlation functions may supplement the analysis of the order parameter, and the conformal bootstrap has led to exciting new developments. We will focus on the very small subset of studies and recent developments that are potentially relevant in the analysis of lattice results on phase transitions,  without any pretense to cover all the vast subjects of statistical field theory. 

In short summary, in this manuscript, we discuss how the properties of the strong interactions depend on the temperature, on different chemical potentials, on the magnetic field, on the quark masses, and on the number of flavors. The material is organized in several  Sections, however, our aim is to
see and present it as different angles of the same phase diagram. Hopefully, the
knowledge of the physical theory -- Quantum Chromo-Dynamics with three families of quarks -- , which remains the main focus of these studies, 
will benefit from this broad view. 

We dispense with introductory material (see e.g.~\cite{Gattringer:2010zz} for a pedagogical introduction to LFTs and~\cite{Philipsen:2021qji,Guenther:2020jwe} for recent LFT reviews,~\cite{Fischer:2018sdj, Dupuis:2020fhh, Fu:2022gou} for recent reviews on functional approaches to QCD), and we concentrate on advanced, state-of-the-art methods 
and results, as well as on promising novel research paths (without any claim to be exhaustive); occasionally the same studies are mentioned in different sections, when they may be looked at from different points of view.

This paper grew out of the workshop ``Phase Transitions in Particle Physics'' 
organized at the GGI in Firenze in Spring 2022. The talks presented there are enlisted and referenced in a dedicated bibliography at the end.

\section{Thermal Phase Transitions and Critical Points\footnote{Editor: Sipaz Sharma}}
\label{sec:thermal_phase_transitions_and_critical_points}
\subsection{QCD Phase Diagram: Expectations}

\begin{figure}[!htb]
\centering
\includegraphics[width=0.8\textwidth]{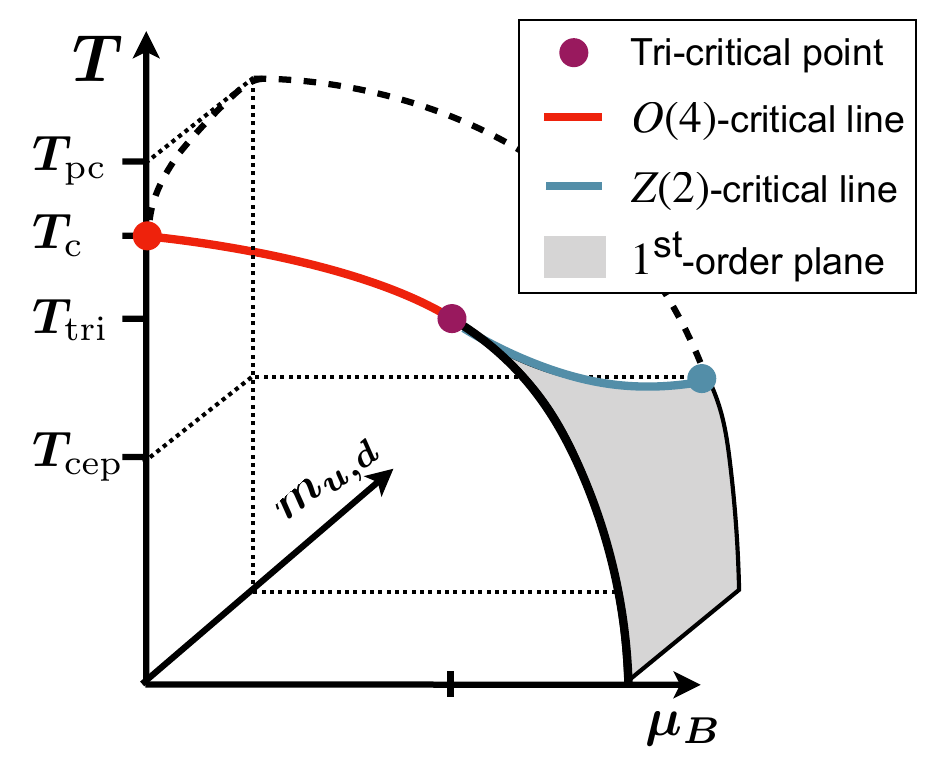}
\caption{Hypothesized phase diagram of QCD assuming a O(4) universality class of the thermal transition in the massless up and down quarks limit, and a physical strange mass.  The axes denote the temperature $T$, the baryon-number chemical potential $\mu_B$, and the light quark masses $m_{u,d}$. In the front, the situation at zero light quark masses is shown, whereas in the back the phase diagram for physical light quark masses is depicted. A hierarchy of important transition temperatures is indicated as $T_{\mbox{\tiny{pc}}}>T_{\mbox{\tiny{c}}}>T_{\mbox{\tiny{tri}}}>T_{\mbox{\tiny{cep}}}$, with the pseudo-critical transition temperature at physical masses $T_{\mbox{\tiny{pc}}}$, the chiral phase-transition temperature $T_{\mbox{\tiny{c}}}$, the temperature of the tri-critical point $T_{\mbox{\tiny{tri}}}$ and the phase-transition temperature of the critical end-point at physical quark masses $T_{\mbox{\tiny{cep}}}$. Source \cite{Karsch:2019mbv}}
\label{fig:diagtmu}
\end{figure}

Thermal phase transitions and critical points are pieces of the QCD phase diagram puzzle. Figure \ref{fig:diagtmu} with three axes denoting temperature $T$, baryon chemical potential $\mu_{B}$, and mass $m_{u,d}$ of degenerate light up and down quarks represents the conjectured QCD phase diagram, as discussed in \cite{Karsch:2019mbv} and references therein.

In the chiral plane, where $m_{u,d}$ vanishes, for vanishing $\mu_B$, restoration of the spontaneously broken $SU(2)_L \times SU(2)_R$ chiral symmetry group -- which is isomorphic to $O(4)$ -- as a function of $T$ is expected to be a genuine second order phase transition belonging to $3$-$d$, $O(4)$ universality class occurring at $T_c$, which is represented by the red dot in Figure \ref{fig:diagtmu} \footnote{Alternative predictions, which also consider a possible role of the chiral anomaly, are discussed in Section 2.2.1}. In the region of small baryon chemical potential, phase transition stays second order belonging to $3$-$d$, $O(4)$ universality class; the transition temperature, $T_c(\mu_B)$ decreases with $\mu_B$, which is clearly depicted by the bending of red curve originating from $T_c(0)$ towards $\mu_B$ axis. After $T_c(\mu_B)$ hits the purple tri-critical point at $T_{tri}$ for some value of $\mu_B$, the transition becomes first order in the higher $\mu_B$ region shown by the black solid line.

Upon adding light quark mass direction to $T$-$\mu_B$ plane in the higher $\mu_B$ region, for a fixed $m_{u,d}$ value, transition stays first order with decreasing $\mu_B$  -- this transition would be a line in the grey first order plane starting from the zero $T$ plane -- until it reaches a certain combination of $T$, $\mu_B$ such that it hits a point on the blue $Z(2)$ critical line and becomes second order belonging to $3$-$d$, $Z(2)$  universality class.

%For small but non-vanishing $\mu_B$,
Due to the explicit breaking of the chiral symmetry, the transition is no longer a genuine phase transition for non-zero $m_{u,d}$ but a crossover depicted by the black dashed line. Notice that at the physical value of the light quark masses -- the backward plane of the shown QCD phase diagram -- and vanishing baryon chemical potential, the crossover transition occurs at a pseudo-critical temperature, $T_{pc}$. This pseudo-critical temperature for the physical value of $m_{u,d}$ decreases as a function of $\mu_B$, and the transition remains a crossover until it meets the blue $Z(2)$ critical line at the temperature $T_{\cep}$ and chemical potential $\mu_{\cep}$, depicted with the blue dot. The existence of this critical endpoint and its location, $(T_{\cep}$, $\mu_{\cep})$, is the modern-day Holy Grail of the experimental as well as the theoretical physics community working on QCD phase diagram and is further discussed in Section~\ref{sec:conformal_phase}.

\subsection{Degree of Understanding}
\subsubsection{Lower Density Region}
One of the ways to understand the phase diagram of QCD is by employing numerical simulations in the framework of Lattice QCD. In recent years, different lattice studies exploiting chiral observables and their universal scaling features have converged on the value of crossover temperature, $T_{\mathrm{pc}}$ at around $\mathrm{156.5}$ MeV at vanishing $\mu_B$ \cite{HotQCD:2018pds, Borsanyi:2020fev}. The value of $T_c$ has been found to be equal to $\mathrm{132^{+3}_{-6}}$ MeV, and the same study argues that the phase transition indeed belongs to $3$-$d$, $O(4)$ universality class~\cite{HotQCD:2019xnw}. Results with Wilson fermions find
a compatible value for $T_c$ and explore the limits of the $O(4)$ scaling window \cite{Kotov:2021ujj,Kotov:2022inz}. 

Reference~\cite{HotQCD:2018pds} is a very accurate study of the curvature of the crossover line in terms of $T_{\mathrm{pc}}$ as a function of $\mu_B$ using Taylor expansion around $\mu_B=0$, which further boosts confidence in the expected phase diagram picture in the lower density region. We will
return to the discussion of the curvature of the crossover line in the next Section. 

It is very important to understand the fate of U$_{\mathrm{A}}(1)$ anomaly at the chiral phase transition of $(2+1)$-flavor QCD -- two degenerate light quarks and physical strange quark -- \cite{Pisarski:1983ms, Rajagopal:1992qz, Rajagopal:2000wf}, \citeTalk{Lahiri_talk}. Model studies can reveal the interplay of the dynamics of spontaneous and anomalous chiral symmetry breaking, see e.g.~\cite{Butti:2003nu, Pelissetto:2013hqa, Nakayama:2014sba, Resch:2017vjs}. In the scenario where U$_{\mathrm{A}}(1)$ gets effectively restored near $T_c$, references~\cite{Pisarski:1983ms, Rajagopal:1992qz, Rajagopal:2000wf} based on one-loop calculation within perturbative $\epsilon$ expansion predict a first-order chiral phase transition for $(2+1)$-flavor QCD, whereas Reference~\cite{Pelissetto:2013hqa} upon employing two different $3$-$d$ perturbative schemes: massive zero-momentum (MZM) scheme and the 3D minimal subtraction scheme MS
without $\epsilon$ expansion converges to the possible existence of a stable fixed point. However, the chiral phase transition can only be continuous belonging to O$(4)\times$ O$(2)$ universality class if the considered model lies within the attractive domain of the stable fixed point -- in other words, the possibility of a first-order transition is not excluded. The issue can only be settled within full QCD as the strength of anomalous chiral symmetry breaking and its dynamics is related to QCD topology or rather the topological density. This calls for lattice QCD simulations or investigations in functional approaches to QCD, and more references and discussions will be given in Section \ref{sec:cosmology_topology_axions}. From the viewpoint 
of this section, we note that the lattice calculations in the chiral and continuum limit of $(2+1)$-flavor QCD find that the U$_{\mathrm{A}}(1)$ remains broken at $T_c$, and therefore further support the second-order nature of the $(2+1)$-flavor chiral phase transition belonging to $3$-$d$, O$(4)$ universality class \cite{Kaczmarek:2021ser, Kaczmarek:2023bxb}. 

Hadronic correlators provide an important complement to the analysis based on the chiral order parameter, and pole, as well as screening masses, are actively investigated~\cite{Lowdon:2022yct,Skullerud:2022yjr,DallaBrida:2021ddx}\citeTalk{Harris_talk}. Hadronic correlators in Euclidean time also serve as input to spectral functions - further discussion can be found in Section~\ref{sec:methodological_challenges_spectral_functions_and_sign_problem}.

Interestingly, an approximate SU$(4)$ chiral spin-flavour symmetry was recently observed in multiplet patterns of QCD mesonic correlation functions \cite{Rohrhofer:2019qwq,Philipsen:2022wjj}. This symmetry disappears at a temperature of about $300$ MeV, approximatively matching other fast crossovers \cite{Alexandru:2019gdm,Kotov:2022inz} in the medium which have not yet been completely understood. 

Further interesting aspects concern ``energy-like" observables which include purely gluonic observables like the Polyakov loop, commonly used as an indicator of confinement/deconfinement crossover for dynamical quarks, as well as heavy quark potential~\cite{Bornyakov:2021enf}~\citeTalk{Lahiri_talk,Kudrov_talk}. Analysis of flux tubes plays an important role as well~\cite{Baker:2022cwb} in this context. Recent studies addressed the sensitivity of these purely gluonic observables to the chiral phase transition~\cite{Clarke:2020htu}.

\subsubsection{Scaling Window}

The standard picture of critical behavior entails a crossover between genuine critical behavior and a mean-field region. The extent of the scaling window is in general regulated by the Ginzburg criterion, and is non-universal, hence it needs to be settled by numerical simulations. From a phenomenological viewpoint, the scaling window is the region where there is still a memory of the underlying critical behavior. This issue has been studied with functional approaches to QCD as well as in low energy EFTs (\citeTalk{Pawlowski_talk}, and references therein). EFT studies with O(4)-models and the quark-meson model in \cite{Braun:2009ruy, Braun:2010vd}, for a review see \cite{Klein:2017shl}, suggest a small critical window with O$(4)$-scaling. Typically, these models assume maximal axial $U(1)$-breaking and the approximations used support O$(4)$ scaling. It has been also argued in these works, that the regime of apparent scaling may be far larger, the difference being hard to extract if the statistical error of the results is sizable. The investigations utilized the functional renormalization group (fRG) that allows direct access to critical scaling. In these models, genuine O(4) scaling was only observed very close to the chiral limit, and it is lost for pion masses $m_\pi\gtrsim 1- 10$\,MeV. These findings were corroborated within functional QCD studies in \cite{Braun:2020ada, Gao:2021vsf}, but a conclusive analysis has not been done yet. The role of the light up and down quark masses, and the extent of the scaling window, were also discussed Ref.~\cite{Kotov:2021ujj, Kotov:2022inz} \citeTalk{Kotov_talk}. Lattice data based on twisted mass Wilson fermions for higher pion masses -- (380-140) MeV -- are consistent with O$(4)$ critical scaling and for pion masses down to the physical value $\mathrm{140}$ MeV, signatures of O$(4)$ scaling can be observed in a temperature range from $\mathrm{120}$ to $\mathrm{300}$ MeV \citeTalk{Kotov_talk}.
While a general consensus has emerged on the $O(4)$-3D universality class in the chiral limit, some differences among different approaches still await clarifications, and this is a subject of current research. 

\subsubsection{Many Flavor QCD at zero $\mu_B$}

\begin{figure}[!htb]
\centering
        \begin{minipage}[b]{0.50\textwidth}
\includegraphics[width=\textwidth]{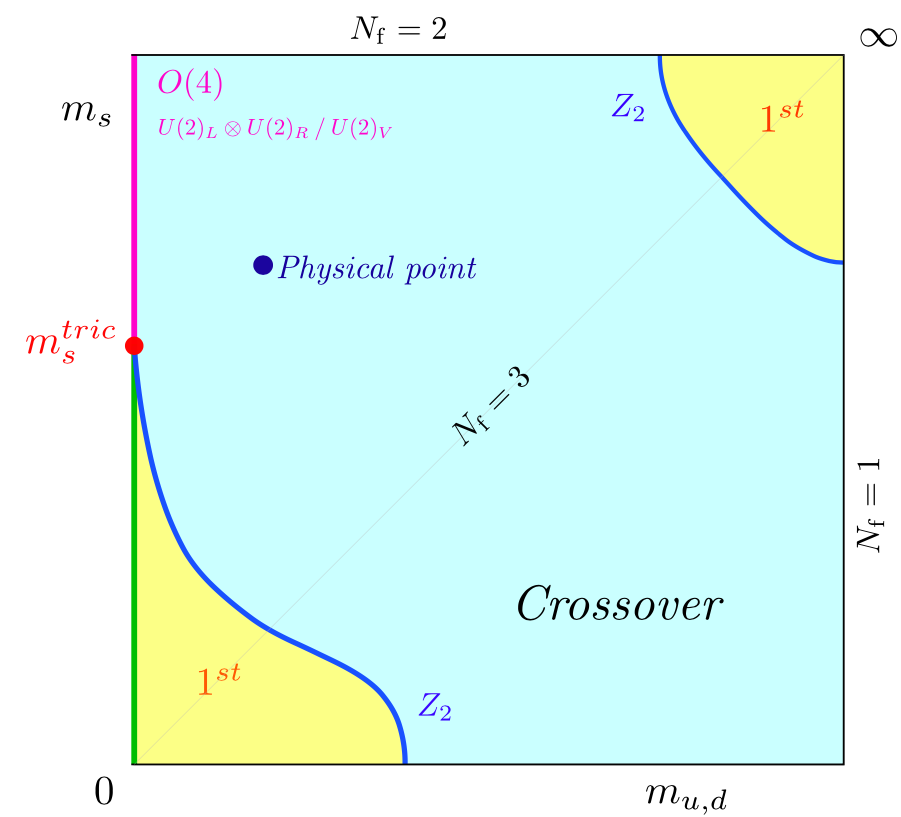}
        \end{minipage}
        \hfill
        \begin{minipage}[b]{0.48\textwidth}
\includegraphics[width=\textwidth]{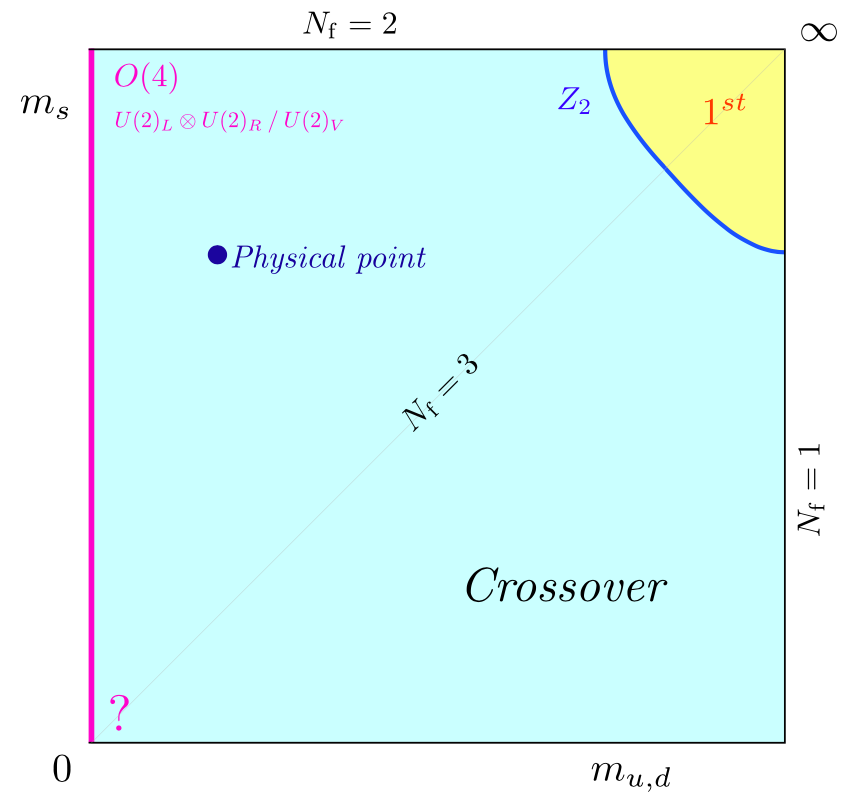}
        \end{minipage}
\caption{Columbia plot [left]. Alternative Columbia plot with a second-order transition in the 3-flavor chiral 
limit [right], as predicted in  \cite{Cuteri:2021ikv}. In this case, nothing is known yet about 
the universality class. Source \cite{Cuteri:2021ikv}.}
\label{fig:columbia}
\end{figure}

The order of the chiral phase transition as a function of the number of massless flavors, $N_f$, 
has been  investigated in \cite{Pisarski:1983ms}, based on the perturbative epsilon expansion
applied to linear sigma models in three dimensions. One popular scenario with up to $N_f=3$, 
see e.g.~\cite{Rajagopal:2000wf}, 
is depicted in the famous Columbia plot \cite{PhysRevLett.65.2491} shown in Figure \ref{fig:columbia} left.
For $N_f\ge3$ massless quark flavors, according to these results, the chiral phase transition is expected to be 
of first-order. The diagonal of the Columbia plot corresponds to the case when all the three quark flavors, u, d, and s are degenerate with masses given by $m_u=m_d=m_s$. When all three quarks have a mass equal to the physical value of the light quark mass $m_{u,d}=m_u=m_d$, the transition is a crossover, but as the quark mass is decreased, 
one expects to hit a $Z(2)$ boundary - the blue line bounding the bottom-left first-order region painted yellow - at some critical mass value $m_c$ \cite{PhysRevLett.65.2491}. 
In this scenario, there is a tri-critical strange quark mass, where the chiral transition changes between first and second order.
Viewing the strange quark mass as a smooth interpolator between $N_f=2$ and $N_f=3$ mass degenerate quarks, 
this corresponds to a situation with $N_f^{tric}<3$.
For a recent review, we refer to \cite{Philipsen:2021qji}.

An interesting and surprising prediction about the second-order 
 nature of the 3-flavor chiral phase transition was made in 
 \cite{Cuteri:2021ikv}\citeTalk{Philipsen_talk}.
This study considers a variable $N_f\in[2,8]$  for various lattice spacings and bare quark masses using unimproved Wilson gauge and staggered fermion actions. 
According to the findings of the previous lattice studies over the 
years, the first-order region shrinks with improved actions as well as 
 with finer lattice spacings. In Reference \cite{Cuteri:2021ikv}, this shrinkage was found to continue to zero, 
leading to the definite existence of a 
tri-critical point. 
In the four-dimensional space of inverse gauge coupling $\beta$, bare quark mass $am$, $N_\tau$ and $N_f$, the bare critical masses $am_c$ form a $Z(2)$ critical surface which separates the first-order region from the crossover. Tri-criticality in the plane of bare critical quark mass $am_c$ and $N_f$ at a fixed lattice spacing $a$ translates to the existence of a $N_{f}^{tri}$ in the chiral limit; in the plane of $am_c$ and $N_\tau^{-1}$ -- where $N_\tau$ is the temporal lattice extent related to temperature $T$ as $N_\tau^{-1}=aT$ -- tri-criticality is encoded in $aT^{tri}$ on the $N_\tau^{-1}$ axis. 
Ref.~\cite{Cuteri:2021ikv} found
$N_{f}^{tri}>6$, implying the disappearance of the bottom-left first-order 
region as shown in Figure \ref{fig:columbia} right. 

Furthermore, Ref.~\cite{Cuteri:2021ikv} pointed out that the $N_f=3$ data generated using $\mathcal{O}(a)$-improved Wilson fermions \cite{Kuramashi:2020meg} is also consistent with tri-critical scaling leading to a finite $aT^{tri}$ in the chiral limit, and hence a second-order transition in the continuum. 
The $N_f=3$ scenario has been recently investigated using Highly Improved Staggered Quark (HISQ) action \citeTalk{Sharma_talk}. The analysis \cite{Dini:2021hug} takes into account the temperature as well as the volume dependence of various chiral observables, such as 3-flavor chiral condensate, chiral susceptibility and observables constructed using some specific combinations of those two. Finally, employment of universal finite-size scaling techniques provides a 3-flavor chiral phase transition temperature for the non-vanishing value of lattice spacing to be $T_c=98^{+3}_{-6}$ MeV  \cite{Dini:2021hug}\citeTalk{Sharma_talk}. Furthermore, no evidence for the first order phase transition is found in the pion mass range explored from 80 MeV up to a physical pion mass value of about 140 MeV, and the results are compatible with $3$-$d$ $O(2)$ universality class, and therefore with a second order phase transition in the 3-flavor chiral limit.
Similarly, no evidence for a first-order transition is seen in the early results of a $N_f=3$ study
using M\"obius domain wall fermions with physical quark masses \cite{Zhang:2022kzb}.  
A recent 5-flavor study \cite{Karsch:2022yka}, based on Machine Learning approach -- extensively discussed in Section \ref{sec:machine_learning} -- finds a non-zero critical endpoint mass marking the boundary of a first-order region in the plane of $\beta$ and $am$, at a fixed temporal lattice extent of $N_\tau=6$. It would be interesting to see how this approach plays out in the four-dimensional space of $(\beta,am,N_\tau,N_f)$. 
Finally, new analytic studies of effective theories along the lines of~\cite{Pisarski:1983ms}, but 
using functional renormalization group~\cite{Fejos:2022mso} or conformal bootstrap~\cite{Kousvos:2022ewl} methods, 
also find
the possibility of a second-order chiral transition for $N_f=3$ under certain conditions. 

In conclusion, according to~\cite{Cuteri:2021ikv}~\citeTalk{Philipsen_talk}, the continuum chiral phase transition is second-order for all $N_f\in[2, 6]$, but no remarks could be made about the universality class of the chiral phase transition. It is also suggested that the phase transition might stay second-order up to the onset of the conformal window at $9\lesssim N_{f}^{*} \lesssim 12$.  These studies connect naturally to the conformal window of strong interactions \cite{Cacciapaglia:2020kgq,Kotov:2021hri,Braun:2010qs}, to be further discussed in Section \ref{sec:conformal_phase}.

In \cite{Pelissetto:2017sfd}, the importance of gauge degrees of freedom in producing a stable fixed point is emphasized, leading to a continuous transition for the antiferromagnetic $\mathrm{CP}^{N-1}$ models when $N\ge4$. A standard Landau-Ginzburg-Wilson (LGW) field-theoretical approach, based on constructing a most general symmetry obeying effective Lagrangian using a gauge-invariant order parameter, predicts a first order transition in such a scenario, whereas numerical results do not sustain this mean field prediction. Furthermore, as pointed out in \cite{Moshe:2003xn}, 
%using different arguments, 
ferromagnetic $\mathrm{CP}^{N-1}$ models in the large $N$ limit behave like an effective Abelian Higgs model for a $N$ component complex scalar field coupled to a $U(1)$ gauge field. This leads to the appearance of a stable fixed point with the possibility of a continuous transition, which again is in contrast to first order prediction of LGW. Possibly, all of the above arguments can be extended to finite temperature QCD for $N_f$ massless flavors, which might settle the disagreements between lattice simulations and theoretical mean-field predictions. We will return to this discussion in Section \ref{sec:statistical_field_theory}. 

\subsubsection{High Density Region}

As discussed above,  the region of high baryon density and lower temperatures is not accessible at the moment to lattice simulations of QCD. In this region we have to rely on functional approaches to QCD, or on low energy EFTs. Alternatively, one may opt to work in QCD-like models such as two-color QCD, or in some (unrealistic) region of the phase space: a dense isospin matter,
with zero baryon density. In the following we discuss some examples of these different situations, to give a flavor of the current research.  

EFTs with different degrees of sophistication are of course an important playground. 
Phenomena such as di-quark condensation and color superconductivity were discovered thanks
to these analyses, see e.g.~Ref.~\cite{Schafer:2000et} for a classic review. A more recent comprehensive report is give in Ref.~\cite{Blaschke:2022lqb}. Topics 
that are close to the discussions on chiral symmetries  are highlighted 
in Refs.~\cite{Sasaki:2019jyh,Marczenko:2022hyt,Sasaki:2022vas}\citeTalk{Sasaki_talk}. A special emphasis is put on the manifestation of (partially) restored chiral symmetry via parity doubling of baryons and mesons in heavy-ion collisions and astrophysical observations.

\begin{figure}
     \centering
%     \vskip -3cm
\hskip -2cm \includegraphics[width=12 cm]{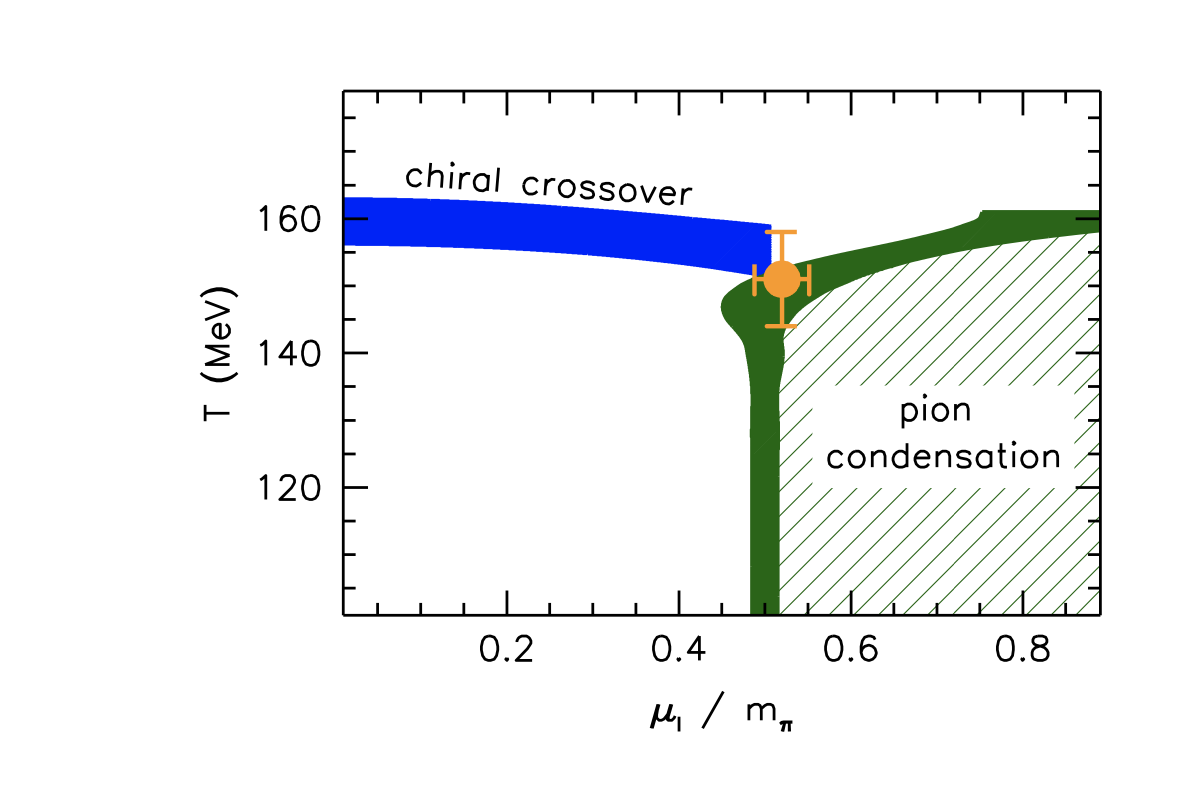}
     \caption{Lattice results for the phase diagram of QCD in the temperature-chemical potential for isospin plane, from Ref\cite{Brandt:2017oyy}.}
     \label{fig:isopd}
 \end{figure}

There are important cases that do not suffer from sign problems on the lattice \cite{Alford:1998sd}: isospin dense matter, and QCD with two colors. Isospin symmetry is a $SU(2)$ rotation in flavor space (QCD interactions are flavor-blind) acting on up and down quarks. In the real world, isospin symmetry is explicitly broken by the (small) mass difference between up and down quarks. In lattice studies, up and down quarks are usually taken as degenerate and an appropriate chemical potential is introduced to create an isospin imbalance \cite{Son:2000xc}. The phase diagram at a finite density of isospin has been studied on the lattice by various authors \cite{Brandt:2017oyy, Brandt:2016zdy, Bornyakov:2021mfj, Braguta:2019noz}. An interesting feature -- see Figure \ref{fig:isopd} -- is that the critical line $T=T(\mu_I)$ has a very small slope -- it is almost horizontal. So, simulations performed at fixed temperature varying $\mu_I$ are  very likely crossing the pion condensation line unless the temperature is really close to $T_c$. Note that  in nuclear matter and in astrophysics isospin imbalance is very important, but smaller than the baryon one. Lattice studies \cite{Brandt:2019hel, Brandt:2017oyy, Brandt:2019ttv, Braguta:2019noz, Detmold:2012wc, Cea:2012ev} which consider  $\mu_I \ne 0, \mu_B = 0$ are thus to some extent artificial, but still interesting: for instance, one can observe 
(1) signatures of the superconducting BCS phase expected on perturbation theory grounds, and (2) the role of pion condensation in the early universe evolution at nonvanishing lepton flavor asymmetries \cite{Vovchenko:2020crk,Brandt:2019hel}\citeTalk{Cuteri_talk}.
 
Two-color QCD is free from the sign problem at nonzero baryon density thanks to its enlarged chiral symmetry:  from the $SU(N_f ) \times SU(N_f ) \times U(1)_B$ to $SU(2N_f)$. Intuitively, baryon and isospin %density
are basically the same symmetry for two colors. For this reason, di-quarks are stable in two color QCD.  Studies of two color matter have been 
reported in \cite{Hands:1999md,Alles:1996nm,Lombardo:2008vc,Hands:2010gd,Hands:2011hd,Hands:2011ye,Astrakhantsev:2018uzd,Iida:2019rah,Astrakhantsev:2020tdl,Begun:2022bxj}.  These studies  have confirmed that baryonic matter forms at an onset $\mu_o = m_\pi/2$,
 whereupon  matter is superfluid. Current studies focus on the understanding of lattice artifacts,
 \cite{Begun:2022bxj}, \citeTalk{Bornyakov_talk}. High quality lattice data allow the study of the interrelation between different pairing patterns, chiral symmetries and gauge dynamics, including signatures of deconfinement. 
 
 Finally, one may consider a chemical potential, $\mu_5 \equiv (\mu_R - \mu_L)/2$ associated with the non-conserved axial current, $\bar \psi_i \gamma_5 \psi$ \cite{Braguta:2015zta}.
 Early lattice studies of chiral density were performed having in mind a toy model for the chiral magnetic effect in heavy ion collisions \cite{Yamamoto_2011}. One first systematic study
 of the phase diagram at equilibrium appeared in Ref.~\cite{Braguta:2015zta}.
Since then the field is developing, also due to the relation with the elusive Chiral Magnetic Effect \cite{Khunjua:2021oxfu}.  Since the axial current is not conserved,  the associated chemical potential, and the related results, need to be 
taken with some care.

\subsection{The Road Ahead}

The analysis of the symmetries, their patterns, and the imprints on the phenomenology of the
related critical points remain an important subject, with several open issues.
In particular,  we have seen that the nature of the phase transition as a function of the number of flavors, and the fate of the axial symmetry are under debate. The nature of the transition
with increasing $N_f$ has also a potential relevance for phenomenology, as models for strong
electroweak breaking often capitalizes on the strong first order transition expected for
large $N_f$. Theoretically, if indeed a second order transition persists till the conformal
window, we will have to understand how a 3D infrared fixed point would morph with  4D conformality.
This latter point -- the fate of the anomaly -- is related
to the topological aspects of QCD, which will be further discussed in Section \ref{sec:conformal_phase}.

Figure 1 shows that, besides the theoretical interest, the chiral behavior in QCD may well constrain the phase diagram, in particular, the location of the critical point
at non-zero density which we will discuss in the Sections \ref{sec:nature_and_phenomenology_of_the_quark-gluon_plasma}, \ref{sec:conformal_phase}. 

There is a growing interest in the approximate $SU(4)$  symmetry observed at high temperatures. 
We may speculate that quarks and gluons are not the
right degrees of freedom for the quark-gluon plasma (QGP) because they are not compatible with this
symmetry. Should this be true,  it would
question all the present transport approaches, which will be further discussed in Section \ref{sec:nature_and_phenomenology_of_the_quark-gluon_plasma}. The crossover from $SU(4)$ symmetry to the $SU(2)XSU(2)$ symmetry of the QGP occurs at a temperature of about  300 MeV, close
to another crossover of an apparent different nature.  An open question is to understand whether
there is a common origin. Several hypotheses have been put forward, none of them completely satisfactory yet. One important aspect of future research is to clarify this point. 

Finally, 
much of the discussions in this section were focused on chiral symmetry. A proper definition
of confinement, and its relation, if any, with chiral symmetry, is  an important theoretical
open problem, going beyond the scope of this review. Here we just note that   steps in this directions require analysis of gauge dynamics, and several studies 
focusing on monopole dynamics, flux tubes and their interrelation with the static potential 
have appeared, see e.g.~\cite{Bornyakov:2019oyq,Bonati:2017anb}\citeTalk{Kudrov_talk}. These analyses may also help in understanding the nature of a threshold in the Quark Gluon Plasma at a temperature of about  300 MeV. 
 
It is of crucial importance for our final understanding of QCD under extreme conditions that all the issues discussed are clarified. Although the results are still not fully conclusive they clearly indicate the research priorities in QCD under extreme conditions in the next future.

\section{Nature and Phenomenology of the Quark-Gluon Plasma\footnote{Editor: Jana N.~Guenther}}\label{sec:nature_and_phenomenology_of_the_quark-gluon_plasma}

Strong interaction matter under extreme conditions can be formed in laboratory: see e.g.~\cite{Busza:2018rrf} for an authoritative overview, as well as the Proceedings of the Quark Matter Conference for updates. A rich and clear discussion with focus on relevant experimental observables for understanding the phase structure of QCD at high $\mu_B$, including a region which is difficult to study on the lattice, can be found in Ref.~\citeTalk{Galatyuk_talk}. 

In this section, we will discuss the region which is still accessible to lattice studies.
In particular, the focus is on the search for the much wanted QCD critical point. 
This has motivated a dedicated collaboration, the  Beam Energy Scan Theory (BEST) collaboration~\citeTalk{Ratti_talk}. The BEST Collaboration "will construct a  theoretical framework for interpreting the results from the ongoing Beam Energy Scan program at the Relativistic Heavy Ion Collider (RHIC). The main goals of this program are to discover, or put constraints on the existence, of a critical point in the QCD phase diagram, and to locate the onset of chiral symmetry restoration by observing correlations related to anomalous hydrodynamic effects in quark gluon plasma.". The WEB page of the BEST Collaboration provides important information, which is reviewed later in this Section. 
 
Central in this discussion is the role of fluctuations: lattice results on fluctuations are reviewed in the next Subsection. Phenomenological applications of lattice studies are discussed next. Let us single out here a  specific point: the calculation of spectral functions. Spectral functions are an important input for
phenomenology; unfortunately their calculation poses specific technical problems, which are discussed in a dedicated Section~\ref{sec:methodological_challenges_spectral_functions_and_sign_problem}. 
Before turning to lattice results,  we would like to mention the cosmological aspects of high temperatures. Temperatures of cosmological relevance may not be
accessible in numerical simulations (see however Sec.~\ref{sec:cosmology_topology_axions}), but
they are amenable to analytic studies or numerical simulations in dimensionally-reduced EFTs. They access phenomena of enormous relevance, including the thermal production of gravitational waves or the existence of electroweak phase transitions beyond the Standard Model. Strictly speaking, this goes beyond the scope of the report, which focuses on strong interactions, however,
the two fields are next to each other and may be bridged by thermal perturbation theory~\cite{Ghiglieri:2021bom,Ghiglieri:2020dpq,Ekstedt:2022bff}\citeTalk{Ghiglieri_talk,Schicho_talk}. 

\subsection{Fluctuations}

Fluctuations are important probes of a phase transition. They are expected to grow large in the
critical region, and the lattice results may be contrasted with predictions from different universality classes \cite{Guenther:2022hmv}\citeTalk{Guenther_talk}. 

Most importantly, they can also offer a starting point to construct various quantities that can be compared to measurements from heavy ion collision experiments.  

Fluctuations are defined as the derivatives of the pressure with respect to various chemical potentials:
\begin{equation}
	\chi^{B,Q,S}_{i,j,k}= \frac{\partial^{i+j+k} (p/T^4)}{
		(\partial \hat\mu_B)^i
		(\partial \hat\mu_Q)^j
		(\partial \hat\mu_S)^k
	}\,, \ \hat\mu=\frac{\mu}{T}
	\label{eq:fluctuations}
\end{equation}

While fluctuations to various order have previously been published on finite lattices for example in Ref.~\cite{Schmidt:2012ka, DElia:2016jqh,Bazavov:2017dus, Borsanyi:2018grb,Bazavov:2020bjn}, now new continuum extrapolated results are available in Ref.~\cite{Bollweg:2021vqf,Bollweg:2022rps}. These results are obtained by the Taylor method and continuum extrapolated from lattices with temporal extend $N_t = 6,8,12$ and 16 with HISQ fermions.  The precision of these results is high enough to allow for a comparison to different models with detailed studies for example on the inclusion or exclusion of various states in a Hadron Resonance Gas (HRG) model. To match the lattice results, for example for $\chi^{BS}_{11}$, it is necessary to add states from quark models to the list of resonances from the PDG \cite{ParticleDataGroup:2020ssz}. On the other hand in Refs~\cite{Bellwied:2021nrt,Bellwied:2021skc}  the coefficients of the fugacity expansion from imaginary chemical potential
\begin{equation}
   \frac{p}{T^4}=\sum_{j=0}^\infty \sum_{k=0}^\infty P^{BS}_{jk} \cosh( j\hat \mu_B  -k\hat \mu_S)
  \end{equation}
are presented. The results are continuum estimates obtained with stout smeared staggered fermions on $N_t=8,10$ and 12 lattices. The analysis is based on a two dimensional fugacity expansion with imaginary $\mu_B$ and $\mu_S$.  The $P^{BS}_{21}$ coefficient includes contributions from $N-\Lambda$ and $N-\Sigma$ scattering where the negative trend indicates the presence of an repulsive interaction that cannot be described with the addition of more resonances. 

Lattice data can also be used as input for parametrizations as done for example in \cite{Noronha-Hostler:2019ayj,Monnai:2019hkn}. The lattice input is especially well suited for the temperature range around the crossover between the hadronic phase that can often be described by the HRG and the QGP-phase.

Moreover, lattice data for fluctuations at low and vanishing density serve as benchmark results for functional QCD computations of fluctuations and that in QCD-assisted EFTs, \cite{Fu:2016tey, Fu:2021oaw, Bernhardt:2022mnx}\citeTalk{Pawlowski_talk}. This allows for an extrapolation of low density lattice results to larger densities, including the regime of the potential critical end point. 

As stated at the beginning of this Section, the ratios of various fluctuations can be used to start a comparison between heavy ion collision experiments and lattice QCD results. The ratios of various fluctuations can be used to express the cumulants of the Baryon number distribution. This offers an observable for comparisons with heavy ion collision measurements of the proton number distribution. At the current precision level this can only be a rough comparison. These cumulants have been published in Refs.~\cite{Borsanyi:2018grb,Bazavov:2020bjn, Bellwied:2021nrt}. If the precision is further increased in the future, other effects should be taken into account, like the continuum limit on the lattice side, or volume fluctuations and non-equilibrium effects on the experimental side (see for example Ref.~\cite{Braun-Munzinger:2016yjz}). However, if the comparisons are done with the necessary care, a deviation between the extrapolated results from the lattice and the experimental measurements can be a hint, that the physics in that area is longer described by an analytic function.

\subsection{Equation of State}
The equation of state is an important quantity both from the purely theoretical point of view as well as input quantity to various models which describe the Quark Gluon plasma. The equation of state at vanishing baryochemical potential $\mu_B$ is known from lattice QCD simulations in the continuum limit (Refs.~\cite{Borsanyi:2010cj,Borsanyi:2013bia,HotQCD:2014kol}) up to high enough temperatures to be matched to perturbative results (Refs.~\cite{Kajantie:2002wa,Andersen:2010wu,Andersen:2011sf}). Its continuation to finite density has posed a significant challenge for several years. When extrapolated to finite $\mu_B$ with a Taylor expansion up to $\mu_B^6$ it shows an increase in the error around the transition temperature, which leaves room for unexpected behavior. This has been observed by different groups and on different data sets (Refs.~\cite{Bazavov:2017dus,Borsanyi:2021sxv}) and with new high precision data (Ref.~\cite{Bollweg:2022rps}) one can observe an increase of the difference between the expansion up to $\mu_B^4$ and $\mu_B^6$.  Some resummation methods (Refs.~\cite{Borsanyi:2021sxv,Borsanyi:2022soo}) hope to mitigate this influence. A comparison in a small volume with direct methods (Ref.~\cite{Borsanyi:2022soo}) shows, that the unexpected behavior does not appear with either the new resummation schemes or higher orders and high precision of the Taylor expansion.

\begin{figure}
	\centering
	\includegraphics[width=0.7\textwidth]{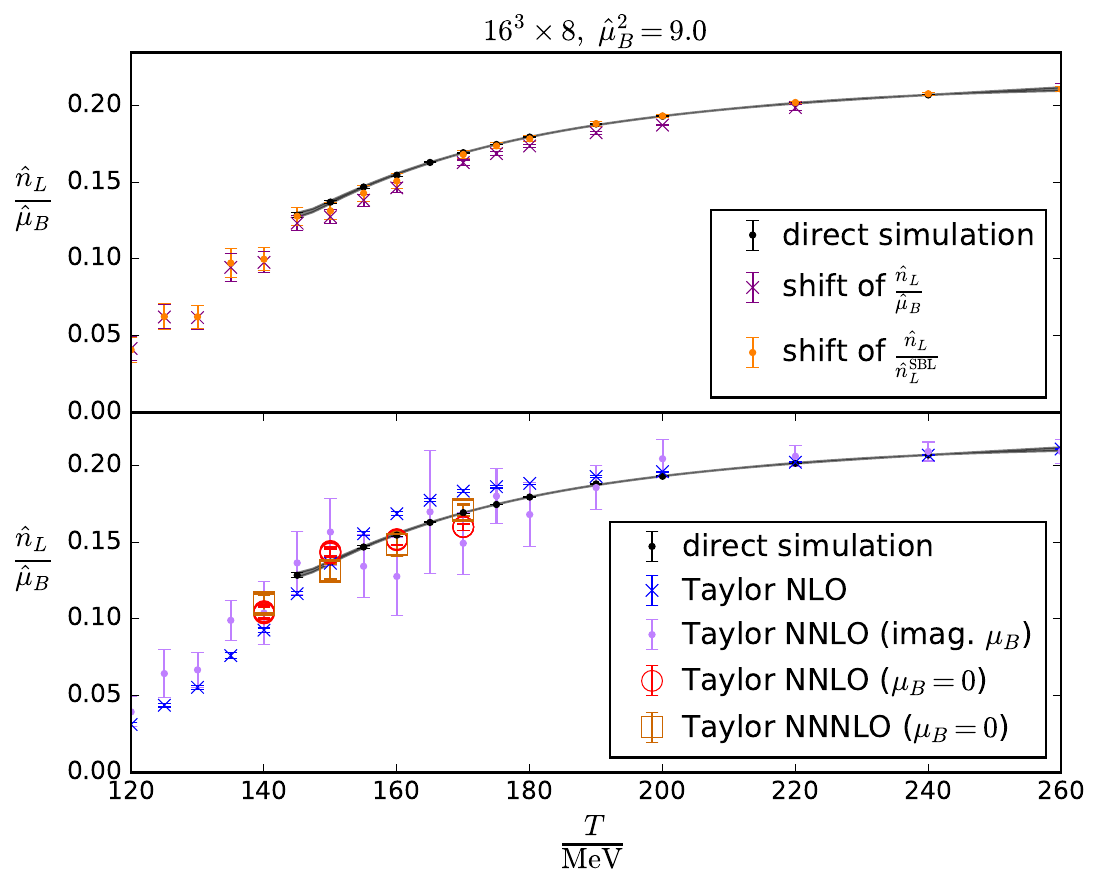}
	\caption{(Ref.~\cite{Borsanyi:2022soo}) Comparison of different extrapolation approaches with direct results on a fixed lattice and in small volume.}
\end{figure}

\subsection{Influence of a Magnetic Field\footnote{Prepared by Lorenzo Maio}}
When trying to match the situation in heavy-ion colliders, an additional important influence on the phase transition is driven by the magnetic field generated in non-central collisions~\cite{Kharzeev:2007jp,Skokov:2009qp,Deng:2012pc}. The simulation of QCD with a magnetic field on the lattice has been a very active field in the last decade (see, e.g. Refs.~\cite{DElia:2010abb,Bali:2011qj,Bali:2012zg,Shovkovy:2012zn,Ilgenfritz:2013ara,Bornyakov:2013eya,Bali:2014kia,Endrodi:2019zrl,Tomiya:2019nym,Ding:2020hxw}). Early results, not yet extrapolated to the continuum limit, showed an increase of the transition temperature as a function of the magnetic field intensity $B$; this agreed well with the expectation resulting from the so-called magnetic catalysis, which describes that at zero temperature chiral symmetry breaking is enhanced by the magnetic field. However, properly continuum extrapolated results revealed a drop of the transition temperature as a function of $B$, an effect which is related to the so-called inverse magnetic catalysis, i.e., the decrease of the chiral condensate in a growing magnetic field for temperatures around and above $T_c$~\cite{Bali:2011qj}. Such a phenomenon induces, furthermore, a strengthening of the crossover, making the gap in the observables between the different phases higher and steeper. This effect was predicted to result, eventually, in the appearance of a real, first order phase transition for magnetic field intensities of the order of $eB\sim10$~GeV$^2$~\cite{Endrodi:2015oba}. Furthermore, studies with various pion masses (Refs.~\cite{DElia:2018xwo,Endrodi:2019zrl}) suggested that the decrease of the pseudocritical temperature with $B$ could be a deconfinement (rather than chirally) driven phenomenon. Indeed, the magnetic field was shown to affect confinement properties, making the string tension anisotropic, in many studies~\cite{Bonati:2014ksa,Bonati:2016kxj,Bonati:2018uwh}.

Very recent lattice results on chiral and confinement properties of $N_f=2+1$ QCD at the physical point in both the vanishing and high temperature cases have been obtained in the presence of unprecedented strong magnetic fields, namely $eB=4$ and $9$~GeV$^2$~\cite{DElia:2021tfb,DElia:2021yvk} \citeTalk{Maio_talk}. Concerning chirality, it was shown that magnetic catalysis maintains its linear behavior in $eB$ in the zero temperature regime, fitting very well to the lowest Landau level prediction. Moreover, the onset of inverse magnetic catalysis is driven to lower and lower temperatures as the magnetic field grows, leading to a drop in the transition temperature larger than expected. Thus, the QGP can be found down to temperatures as low as ${\sim}\,60$~MeV in a $eB=9$~GeV$^2$ magnetic background. Moreover, in the $9$~GeV$^2$ magnetic field simulations, the authors noticed the transition region being extremely narrow. Thus, a deep study on the nature of the transition was performed, through dedicated simulations, providing the first evidence for a first order phase transition of $N_f=2+1$ QCD at the physical point in a magnetic background.

\begin{figure}[!tb]
	\centering
	\includegraphics[width=0.6\textwidth]{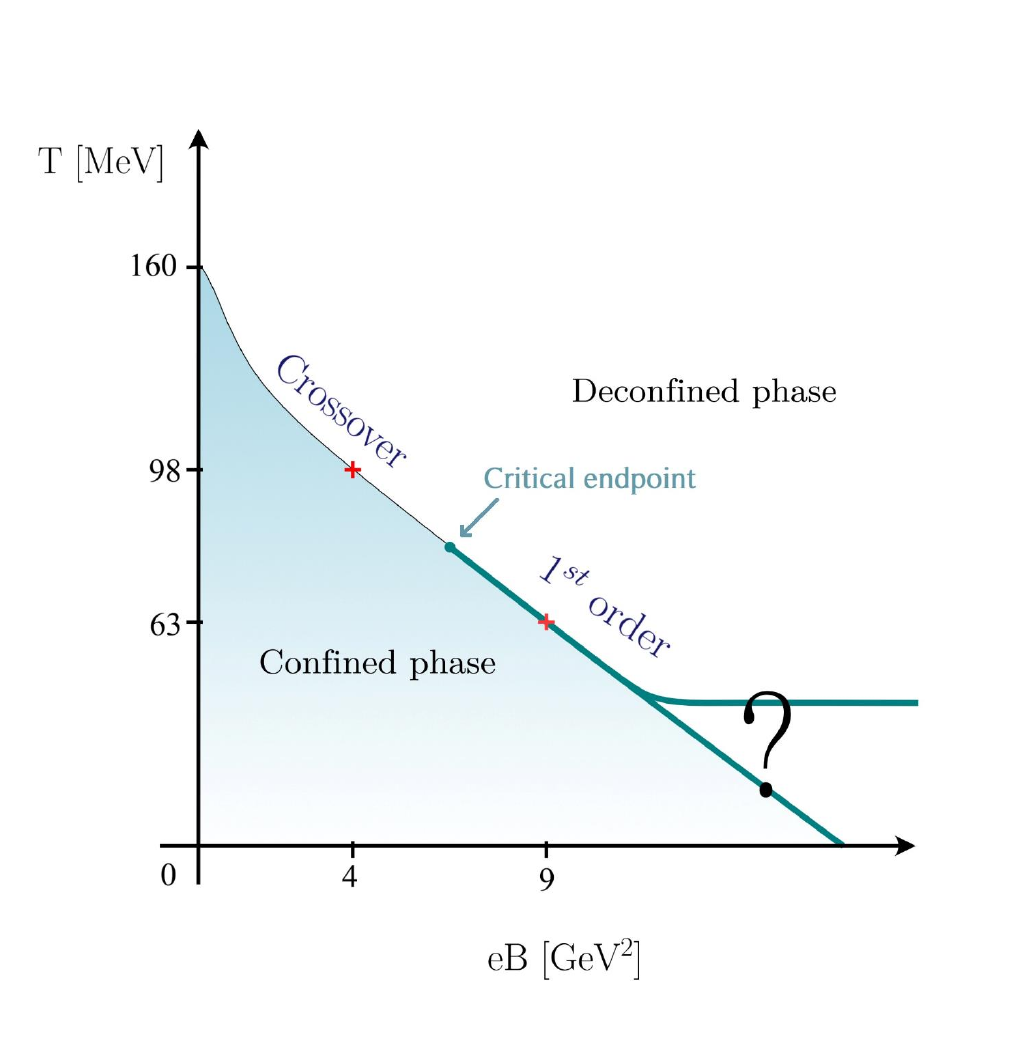}
	\caption{Updated QCD phase diagram in an external magnetic field, based on new facts that emerged in~\cite{DElia:2021tfb,DElia:2021yvk}. The (pseudo)critical temperature continues its steady drop as a function of B, and the transition switches from a crossover to first order at a critical end point located in the range $4$~GeV$^2 < eB_E < 9$~GeV$^2$ (or alternatively $65$~MeV$ < T_E < 95$~MeV). The fate of the critical temperature in the asymptotic magnetic field limit remains an open question.}
	\label{fig:Updated_PD}
\end{figure}

On the confinement side, previous work suggested an anisotropic deconfinement~\cite{Bonati:2016kxj} in the zero temperature regime for magnetic fields ranging up to $eB=4$~GeV$^2$. It was shown that this prediction is not verified and, furthermore, such a partial deconfinement does not happen even for  the largest explored magnetic background, i.e. $eB=9$~GeV$^2$~\cite{DElia:2021tfb}. The authors also studied the confining potential at finite temperature around the phase transition found in~\cite{DElia:2021yvk}. They found, as expected, that the chirally broken phase exhibits confinement in all the directions, while the chirally restored phase appears to be deconfined. To summarize all findings reported above, they proposed an updated version of the $N_f=2+1$ QCD phase diagram at the physical point, as can be seen in Fig.~\ref{fig:Updated_PD}.

The effects of a magnetic field are an active topic \cite{Braguta:2021ucr,Astrakhantsev:2021jpl,TalkWang,Ding:2021cwv}. In addition to studying a magnetic field at zero or finite temperature, also systems where a background magnetic field is considered in combination with a finite density~\cite{Braguta:2019yci,Braguta:2021ucr,Astrakhantsev:2021jpl} or a finite rotation~\cite{Fukushima:2018grm,Chen:2021aiq,Yamamoto:2021oys} are, currently, under investigation. Moreover, recently, also inhomogeneous magnetic backgrounds are taken into consideration because of their phenomenological relevance in the context of heavy-ion scattering experiments~\cite{Brandt:2021vez}.

\subsection{BEST Efforts\footnote{Prepared by Claudia Ratti}}
While a direct comparison between lattice and experiments is challenging, lattice data can also serve as input or benchmark for hydrodynamic evolution models. 
The matter created in heavy-ion collisions can be well-described by relativistic viscous hydrodynamics, which can provide a framework to search for the QCD critical point if modified to take critical phenomena into account.
%These models allow for a search for the critical point at higher chemical potentials as are feasible on the lattice, they can include non-equilibrium effects and compute quantities like freeze-out temperature or resonances. 
The BEST-collaboration combines first-principles lattice QCD calculations and phenomenological approaches, to create a framework for the analysis of experimental data at low collision energies \citeTalk{Ratti_talk},\cite{An:2021wof}. They computed an equation of state that reproduces the lattice QCD one up $\mathcal{O}(\hat{\mu}_B^4)$ and contains a critical point in the 3D Ising model universality class \cite{Parotto:2018pwx} from which they can compute the thermodynamic quantities at various chemical potentials (for example with the strangeness neutrality setting \cite{Karthein:2021nxe}). The equation of state can then be used as an input for hydrodynamical simulations.

To definitively claim or rule out the presence of a QCD critical point or anomalous transport requires a comprehensive framework for modeling the salient features of heavy ion collisions at BES energies, which allows for a quantitative description of the data. 
BEST developed initial conditions, which connect the pre-equilibrium stage of the system to hydrodynamics on a local collision-by-collision basis \cite{Shen:2017bsr,Du:2018mpf,Shen:2020jwv}.
A quantitative understanding of fluctuations near the critical point needs to be developed as well. In fact, the evolution of the long wavelength fluctuations of the order parameter field close to the critical point is not captured by hydrodynamics. Two approaches have been followed within BEST: a stochastic approach with noise \cite{Nahrgang:2018afz}, and a deterministic approach in which correlation functions are treated as additional variables, together with the hydrodynamics ones \cite{Stephanov:2017ghc}. The numerical implementation of the latter are underway \cite{Rajagopal:2019xwg,Du:2020bxp}.

The efforts of the BEST-collaboration also include the particlization after the hydrodynamic phase. The aim is to develop an interface between the hydrodynamic evolution model and the hadronic transport phase, in a way that it preserves fluctuations (see Refs.~\cite{Oliinychenko:2019zfk,Pradeep:2022mkf,Sorensen:2020ygf}).

\subsection{Transport Properties\footnote{Prepared by  Olga Soloveva}}

\begin{figure}
        \centering
	\includegraphics[scale=0.5]{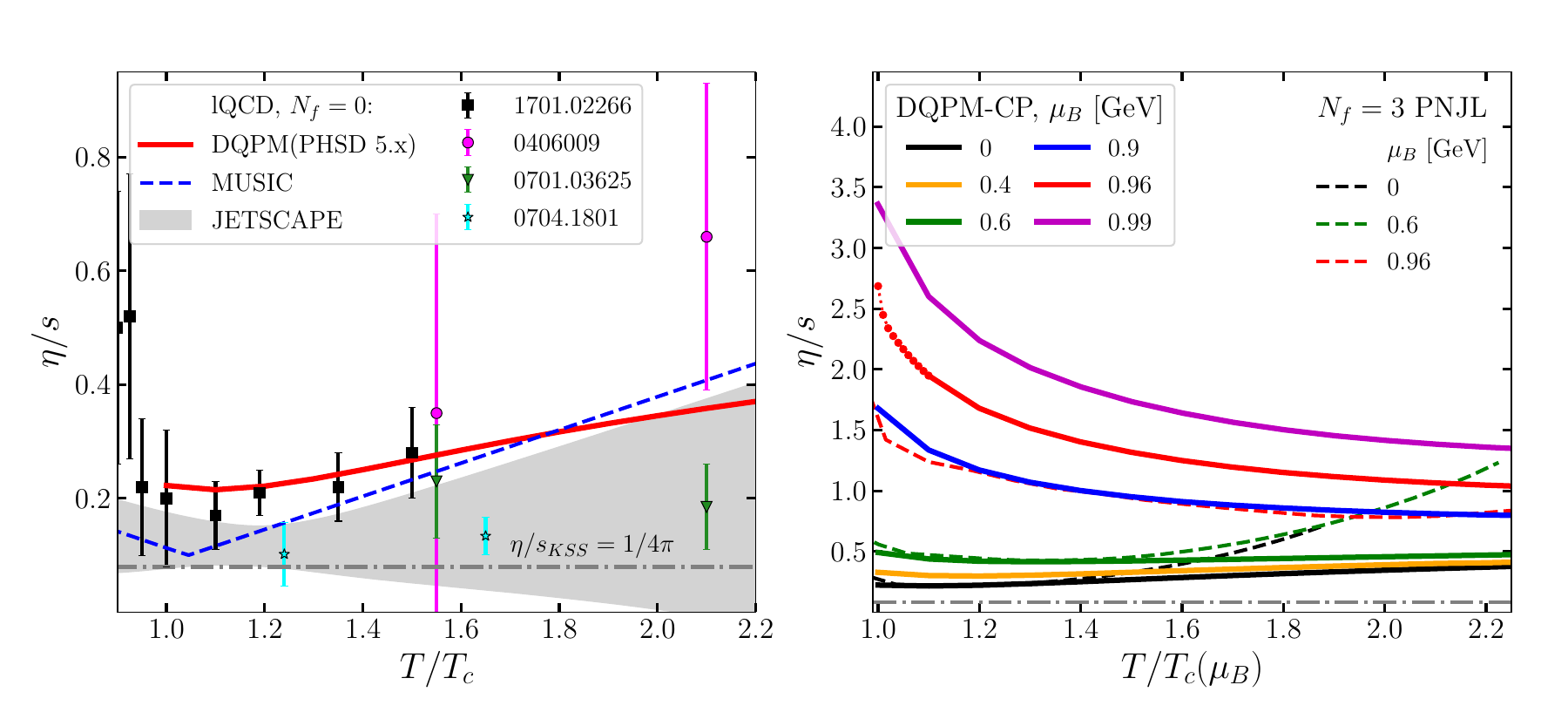}
	\caption{Specific shear viscosity as a function of the scaled temperature $T/T_c$ at $\mu_B=0$ (left) and at finite $\mu_B$ (right). The symbols corresponds to the lQCD results for pure SU(3) gauge theory (black squares) \cite{Astrakhantsev:2017nrs}, (green triangles and magenta circles) \cite{Nakamura:2004sy}, (cyan stars) \cite{Meyer:2007ic}. The dash-dotted gray line demonstrates the Kovtun–Son–Starinets bound $(\eta/s)_{KSS}=  1/(4\pi)$ \cite{Kovtun:2004de}. The grey area represents the model-averaged results from a Bayesian analysis of experimental heavy-ion data \cite{JETSCAPE:2020shq}. The red line corresponds to the DQPM results \cite{Soloveva:2019xph},  while the dashed blue line displays $\eta / s$ parametrization used in hydrodynamic simulations within MUSIC in \cite{Shen:2020jwv}. The model results, obtained by the RTA approach with the interaction rate, for finite $\mu_B$: DQPM-CP results \cite{Soloveva:2021quj}(solid lines) are compared to the  estimates from the $N_f=3$ PNJL model (dashed lines) \cite{Soloveva:2020hpr} as a function of scaled temperature $T/T_c(\mu_B)$.}
	\label{fig:transport_viscosities}
\end{figure}

Experimental and phenomenological aspects of transport are discussed in depth in Ref. \citeTalk{Soloveva_talk}.
The evolution of the QGP phase has been successfully described within hybrid approaches based on relativistic hydrodynamics and transport theory, such as iEBE-VISHNU \cite{Shen:2014vra}, vHLLE + UrQMD/SMASH \cite{Schafer:2021csj, Karpenko:2015xea}
and MUSIC+UrQMD~\cite{Ryu:2015vwa, Shen:2020jwv}. Nevertheless, some advanced transport approaches, such as AMPT~\cite{Sun:2020uoj} and PHSD~\cite{Cassing:2009vt,Moreau:2019vhw} can provide the whole evolution of HIC, including the QGP phase.
In order to perform hydrodynamical simulations of the time evolution of the quark-gluon matter at finite baryon chemical potential, one needs to estimate first the EoS and the transport coefficients of the matter in this region. The transport coefficients depend on the underlying microscopic theory which describes the interaction between quarks and gluons, however, it is notoriously difficult to evaluate microscopic properties of the QGP matter at finite $T$ and $\mu_B$ from first principles. Transport coefficients serve as a bridge between the microscopic transport and hydrodynamics approaches. One can evaluate the transport coefficients by methods of kinetic theory and apply them in the hydrodynamical simulations. 

To examine transport coefficients at finite $\mu_B$ where the phase transition is possibly changing from a crossover to a 1st order one it is necessary to resort to effective models which describe the chiral phase transition. While most of the effective models have similar equations of state (EoS), which match well with available lattice data, the transport coefficients can vary significantly already at $\mu_B = 0$ \cite{Marty:2013ita, Haas:2013hpa, Christiansen:2014ypa, Rougemont:2017tlu, Moreau:2019vhw, Soloveva:2020hpr, Grefa:2022sav}. Therefore, it would be beneficial for the hydrodynamic and transport simulations of the strongly interacting matter for the moderate and high $T$ and $\mu_B$ to have predictions for transport coefficients from lQCD calculations in this region of the phase diagram.

The transport coefficients of the QGP medium have been computed for a wide range of baryon chemical potential for two models with a similar phase structure: the extended $N_f=3$ Polyakov Nambu-Jona-Lasinio (PNJL) model and Dynamical QuasiParticle Model with a CEP (DQPM-CP), where the hypothetical CEP located at $\mu_B = 0.96$ GeV. 
\\
The specific shear viscosity for the QGP phase is shown in Fig. \ref{fig:transport_viscosities} as a function of scaled temperature $T/T_c$ at $\mu_B=0$ (left) and at finite $\mu_B$ (right). At $\mu_B=0$ we show results from the DQPM \cite{Soloveva:2019xph} (solid red line), in comparison with the lQCD results for pure SU(3) gauge theory \cite{Astrakhantsev:2017nrs, Nakamura:2004sy, Meyer:2007ic}, model-averaged results from a Bayesian analysis of the experimental heavy-ion data \cite{JETSCAPE:2020shq} (grey area) and $\eta/s$ employed in hydrodynamic simulations in \cite{Shen:2020jwv} (dashed blue line). For finite $\mu_B \ge 0$ we show the results from the PNJL model and DQPM-CP models obtained by the RTA approach with the interaction rate. 
The estimations from both models show an increase of specific shear viscosities $\eta/s$ and electric conductivities $\sigma_{QQ}/T$ with $\mu_B$. While the specific shear viscosities are in agreement for moderate $\mu_B$ in the vicinity of the phase transition, there is a clear difference in the electric conductivity essentially due to the different description of partonic degrees of freedom \cite{Soloveva:2021quj}.

Furthermore, it has been found that for fixed $\mu_B$, where the phase transition is a rapid crossover, transport coefficients show a smooth temperature dependence while approaching the (pseudo)critical temperature from the high temperature region.  The presence of a first order phase transition changes the temperature dependence of the transport coefficients drastically. 

In order to take into account a proper non-equilibrium description of the entire dynamics through possibly different phases up to the final asymptotic hadronic states, a microscopic treatment is needed. The Parton-Hadron-String Dynamics (PHSD) transport approach~\cite{Cassing:2008sv,Cassing:2009vt,Bratkovskaya:2011wp, Moreau:2019vhw} is an off-shell transport approach based on the Kadanoff-Baym equations in first-order gradient expansion which allows for simulations of both the hadronic and the partonic phases. The microscopic properties of quarks and gluons are described by the DQPM with a crossover phase transition, where the microscopic characteristics of partonic quasiparticles and their differential cross sections depend not only on temperature $T$ but also on the chemical potential $\mu_B$ explicitly. We find that HICs results from the extended PHSD transport approach, where in QGP phase we found that transport coefficients have noticeable  $T$ and $\mu_B$ dependence, have been in agreement with the BES STAR data in case of bulk observables and elliptic flow of charged particles \cite{Soloveva:2020ozg}, and reasonably agrees with the results from hybrid approach \cite{Shen:2020jwv}. It is important to note that, $\eta/s$ used for hydrodynamic evolution is close to the DQPM estimations as shown in Fig. \ref{fig:transport_viscosities} (left). However, results from the PHSD transport approach have shown rather small influence of the $\mu_B$-dependence of the QGP interactions on the elliptic flow than hybrid simulations \cite{Moreau:2019vhw, Soloveva:2020ozg}. This small sensitivity of final observables to the influence of baryon density on the QGP dynamics can be explained by the fact that at high energies, where the matter is dominated by the QGP phase, one probes the QGP at a very small baryon chemical potential $\mu_B$, whereas at lower energies, where $\mu_B$ becomes larger, the fraction of the QGP drops rapidly. Therefore, the final observables for lower energies at order of $1-10$ GeV are in total dominated by the hadrons which participated in hadronic rescattering and thus the information about their QGP origin is washed out or lost.

\subsection{Experimental Efforts\footnote{Prepared by Tetyana Galatyuk}}
\begin{figure}
\centering
\includegraphics[width=0.8\textwidth]{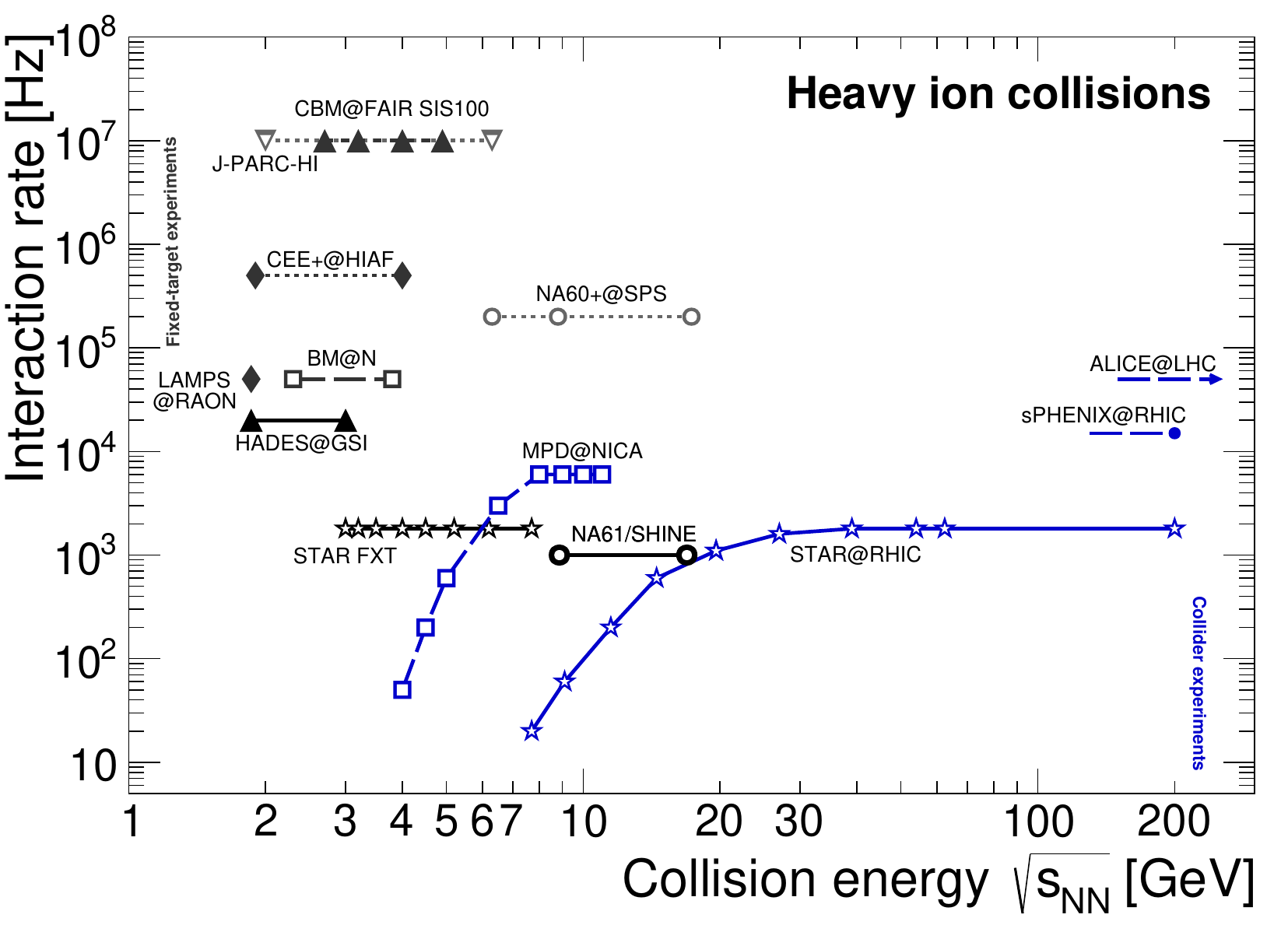}
\caption{Overview over future and past experiments taking data from heavy ion collisions~\cite{high_mub:IRplot,Galatyuk:2019lcf,CBM:2016kpk}.}
\label{fig:exp}
\end{figure}

There is a huge experimental effort  to study the QGP specifically with dileptons \citeTalk{Galatyuk_talk}. Similar as on the theory side, the search for a first order transition and a possible QCD critical endpoint are important research points as well as the general properties of QCD matter around a deconfinement and/or chiral transition. In the near future many experiments are expected to take high statistic data (see figure~\ref{fig:exp} from Ref.~\citeTalk{Galatyuk_talk}) which will allow new inside from statistic hungry probes like dileptons and photons.
Dileptons for example allow answering the fundamental questions related to the mechanism of chiral symmetry restoration in QCD matter and the transition from hadronic to partonic degrees of freedom, the total lifetime of the interacting medium and its average temperature, the evolution of collectivity and the nature of the electromagnetic emission, as well as the transport properties of the medium (i.e., the electrical conductivity). 

\subsection{The Road Ahead\footnote{Prepared by Joerg Aichelin and Elena Bratkovskaya}}
We close the Section with a summary of the lattice issues  which the phenomenological/experimental community considers most urgent: 

\begin{enumerate}
	
	\item The study of fluctuations to identify phase transitions and a possible QCD critical point should be further pursued. As it clearly appeared from the previous discussion, this requires vigorous collaboration between experiments and theoretical work. An important contribution is expected from lattice investigations, but these are hampered by the so-called sign problem, which actually dominates the region of interest in the phase diagram. The solution (or at least an effective mitigation) of the sign problem is thus crucial: this point is addressed in Section \ref{sec:methodological_challenges_spectral_functions_and_sign_problem}. It has to be noted that even existing methods may be stretched to reach the a region candidate for the critical endpoint. 

 Moreover, functional approaches (FA) to QCD offer direct computational results at larger densities. In particular, they can be understood as $mu_B$-exptrapolations of lattice results at lower densities with the maximal dynamical information of QCD in comparison to other extrapolations. This opens a promising route toward a combined LFT-FA analysis of the high density regime of QCD. 
	
	\item The equation of state should be provided for a broad range in temperature and baryon chemical potential. Also, the influence of other parameters like a strangeness chemical potential or a magnetic field should be explored. On the one hand, this could allow a closer comparison to heavy ion collisions where a magnetic field is present especially in off-central collisions as well as strangeness fluctuations if an overall equilibrium is not reached. On the other hand, these parameters offer more theoretically interesting regimes. There is for example hints a critical end point at high magnetic fields (Ref.~\cite{DElia:2021yvk}).
	
	\item  Essential for all phenomenological approaches are the temperature dependence of the pole masses of pseudo scalar and vector bosons
	with zero or finite momentum. This has been partially accomplished however it requires a solid understanding of spectral functions for the identification of
	pole masses. 
	
	\item There are measurements of transport coefficients of heavy quarks in the medium like $D_s$ but the results from different lattice groups
	do not agree (maybe because quenched and not quenched approaches give different results). In addition, in the transport approaches
	we need these coefficients at finite momentum of the heavy quark (with respect to the medium). An improvement in this situation would
	be welcomed. These issues call also for methodological improvements in the computation of spectral functions which will be reviewed in Section \ref{sec:methodological_challenges_spectral_functions_and_sign_problem}.
	
	\item  Another quantity we should urgently know is the pole mass and stability of protons as a function of the temperature. Work in this direction has been done in Ref.
	\cite{Aarts:2017rrl}, however, limited to the pole mass. Information on the stability would be important as well.  To settle this with a physical pion mass
	would be of great help. 
	
	\item  A precise determination of the density (baryon, strangeness) as a function of the temperature - namely of the first derivative of the partition function - would be important.  This would allow us to establish ( what also Nambu-Jona-Lasinio  models predict) whether the hadronization temperature of strange quarks differs from
	that of light quarks.
	
	\item  Any information about the underlying degrees of freedom of a QGP would be of great help. Recent work on unusual symmetries (Ref.~\cite{Rohrhofer:2019qwq}), already mentioned 
 in Section \ref{sec:thermal_phase_transitions_and_critical_points} should be further explored.	

\end{enumerate}

\section{Methodological Challenges: Spectral Functions and Sign Problem\footnote{Editors: Chris Allton and  Christian Schmidt}}
\label{sec:methodological_challenges_spectral_functions_and_sign_problem}

While lattice QCD has been quite successful at Euclidean space-time geometry and zero chemical potential over the past decades, it suffers from severe limitations when it comes to the calculation of expectation values at non-zero baryon number density or quantities related to real time. 
This is due to the fact that lattice QCD calculations crucially rely on the interpretation of the Boltzmann factor as a probability density for the numerical sampling of the path integral. 
Once the Boltzmann factor is no longer strictly positive, or even becomes genuinely complex, this interpretation is lost and standard Monte Carlo methods for the calculation of the path integral cease working. 
This is called the QCD sign problem. 
To deal with or to circumvent the sign problem and to reach out to the expected QCD critical point bears huge methodological challenges. 
Similarly, this is true for the calculation of spectral functions, which provide a way to extract, e.g., transport coefficients, but are also of interest for many other reasons (for example, also at zero temperature several observable quantities are related to spectral densities).
In the following, we will discuss some of those challenges in more detail.

\subsection{Spectral Functions as an  Inverse Problem}

Before entering the details of the inverse problem, we would like to mention an important aspect of functional Renormalization Group studies: fRG can be formulated in real time, 
via a combination of the fRG approach and the formalism of Schwinger-Keldysh path integral, see e.g.
\cite{Fu:2022gou} for a recent review.
The spectral functions thus obtained may be contrasted with lattice results. In this case, 
the sign problem will be solved from scratch, completely bypassing the difficult inversion procedure,
which will be the focus of the remaining part of the discussion. 

The computation of spectral functions begins with lattice correlators \cite{Asakawa:2000tr,Kaczmarek:2022ffn}. It is an ill-posed or at least ill-conditioned problem as 
the task is to reconstruct salient features of the spectral functions (peaks, typically) from a smooth function that is only known in a limited amount of points, with limited accuracy.

Bottomonium has been used as an important case study \cite{Rothkopf:2019ipj,Aarts:2010ek}: first, it is of great physical interest due to the rich production at the LHC. Secondly, the inversion required to compute spectral functions is a ``simple'' inverse Laplace transform, for which a wealth of methods has been designed.  Lattice studies predict the sequential suppression of bottomonium in the QGP, which has
been observed in experiments \cite{Strickland:2021boy}. Despite qualitative coherence among the results, a quantitative agreement has not been reached yet. A comparison of the different methods may be found in Ref.\cite{Spriggs:2021dsb}.

The numerical inversion of the Laplace transform on the real axis is an inverse and ill-posed problem. Usually, methods for the inversion problem require the evaluation of the Laplace function F on some knots; this could be an issue if a closed form of F is not available. In lattice QCD applications, the Laplace transform is known only on pre-assigned samples or measures (and with errors) and an accepted strategy is to design fitting models able to represent this function \citeTalk{Allton_talk}. 

Before entering the details of the inverse problem, we would like to mention an important aspect of functional Renormalization Group studies: fRG can be formulated in real time, 
via a combination of the fRG approach and the formalism of Schwinger-Keldysh path integral, see e.g.
\cite{Fu:2022gou} for a recent review.
The spectral functions thus obtained may be contrasted with lattice results. In this case, 
the sign problem will be solved from scratch, completely bypassing the difficult inversion procedure,
which will be the focus of the remaining part of the discussion. 

Ref.~\citeTalk{Cuomo_talk} is a mathematical introduction to inverse Laplace transforms aimed at physicists. Besides a comprehensive discussion
of different methods, many of them not yet tried in this context, it presents the main numerical issues about the Laplace inversion formulas in the discrete data framework and discusses how to estimate the main sources of errors. Very important in this context is the interpolation of a discrete data set. This latter point is discussed in Ref.~\citeTalk{Conti_talk}, another mathematical review prepared for a physics audience.  Spline models have been widely used in many areas of science and engineering, such as signal and image processing, computer graphics, deep learning, neural networks, or data representation, as important tools to model and predict data trends.
Ref.~\citeTalk{Conti_talk}  aims at providing an introduction to basic spline models-smoothing, regression, and penalized splines-based on polynomial splines but also on exponential-polynomial splines. The latter are particularly suitable for data showing exponential trends as in the framework of the
Laplace transform inversion. In particular,  Ref.~\citeTalk{Conti_talk}  discusses HP-splines, a recently defined penalized regression model, generalization of P-spline, in which polynomial B-splines are replaced by hyperbolic-polynomial bell-shaped basis functions, and a suitably tailored penalization term replaces the classical second-order forward difference operator.

\subsection{Spectral Functions and Effective Field Theories\footnote{Prepared by Nora Brambilla}}

Most important for the control of the results,
and to monitor the approach to the continuum limit, is the interface between lattice 
and effective field theories. Nonperturbative correlators emerge in the nonrelativistic effective field theory (NR EFT) factorization \cite{Brambilla:2004jw}  that should be calculated on the lattice. Ref.~\citeTalk{Brambilla_talk} presents lattice calculation of some of these.

In particular, the EFT called potential nonrelativistic QCD (pNRQCD)
at finite temperature \cite{Brambilla:2008cx}
gives a framework to define the potential, calculate it and systematically calculate energy levels and widths \cite{Brambilla:2010vq}.  Calculations have been made in (resummed) perturbation theory
and then used to compare and check lattice results, for example in the case of the Polyakov loop and the 
Polyakov correlator \cite{Brambilla:2010xn,Bazavov:2016uvm}
establishing the region in which the screening regime is active.

Moreover, combining pNRQCD and an open quantum system \cite{Brambilla:2017zei}, 
it is possible to describe the nonequilibrium evolution of small quarkonia systems (bottomonium) inside the strongly coupled Quark Gluon Plasma with an evolution equation for the singlet and octet density matrix of the Lindblad type
on the basis of two transport coefficients defined as appropriate correlators of electric fields at finite temperature \cite{Brambilla:2020qwo,Brambilla:2021wkt}.
In this way, the EFT works as an intermediate layer that allows to 
use lattice QCD equilibrium input to study the nonequilibrium evolution of bottomonium inside the QGP.
One can also relate these transport coefficients to the thermal modification of the energy levels and to the thermal widths of quarkonium, which allows us to use unquenched lattice calculations of the thermal 
modification of the mass and the width of quarkonium \cite{Brambilla:2019tpt}
as input.  Gradient flow is particularly suitable for the direct lattice calculation of 
these transport coefficients \cite{Brambilla:2022xbd,Brambilla:2020siz}.
Besides the methodological importance, these studies also provide an important input to phenomenology as already mentioned. The same interface between NR EFTs and lattice 
may be used to study a number of problems  ranging from the study of the exotics X Y Z 
\cite{Brambilla:2021mpo,Brambilla:2022fqa} to quarkonium production \cite{Brambilla:2022rjd}. 
This novel alliance of EFTs and lattice, with lattice correlators defined inside the 
EFT  appears to be a novel and promising avenue.

\subsection{QCD at Non-Zero Density: From Taylor Expansions to Lee--Yang Zeros}

The Taylor expansion method \cite{Allton:2002zi} is one of many approaches to circumvent the QCD sign problem and has been very successful in the past. 
Although limited to small baryon chemical potentials $\hat \mu_B\equiv \frac{\mu_B}{T}\lesssim 2$, some results close to the continuum limit have been presented on the QCD equation of state \cite{Bazavov:2017dus, Borsanyi:2021sxv}, the curvature of the transition line \cite{HotQCD:2018pds, Borsanyi:2020fev} and fluctuations of conserved charges \cite{Bazavov:2020bjn, Bollweg:2021vqf}. 
The main idea is the expansion of the dimensionless pressure $p/T^4$ in terms of the three chemical potentials for baryon number, strangeness and electric charge, $\hat\mu_B,\hat\mu_Q\hat\mu_S$,
\begin{equation}
	\frac{p}{T^4}=\sum_{i,j,k}\frac{1}{i!j!k!}\chi^{B,Q,S}_{i,j,k}\hat\mu_B^i\hat\mu_Q^j\hat\mu_S^k\;,
	\label{eq:pressure_expansion}
\end{equation}
where the expansion coefficients are defined as in Eq.~(\ref{eq:fluctuations}). 
The series is even, i.e., the summation runs over all $\{i,j,k\}$ with $(i+j+k)\;\text{mod}\;2=0$. 
It is very tempting to estimate the radius of convergence of the expansion above since, by definition, the radius would be limited by the elusive critical point in the QCD phase diagram. 
However, the limiting singularity can also be located in the complex $\hat\mu_B$ plane. 
A famous example are the Lee-Yang edge singularities \cite{Lee:1952ig}, in the context of lattice QCD and the QCD phase diagram first discussed by \cite{Ejiri:2005ts, Stephanov:2006dn}. 
Estimating the radius of convergence from the lattice results of the Taylor coefficients $\chi^{B,Q,S}_{i,j,k}$ is very challenging, due to the limited number of coefficients, usually $(i+j+k)\le8$, and the increasing statistical error.
A simple rational estimator has been used frequently in the past \cite{Gavai:2004sd, Bazavov:2017dus}, even though it is known to converge slowly \cite{Giordano:2019slo}.

\begin{figure}
	\centering
	\includegraphics[width=0.52\textwidth]{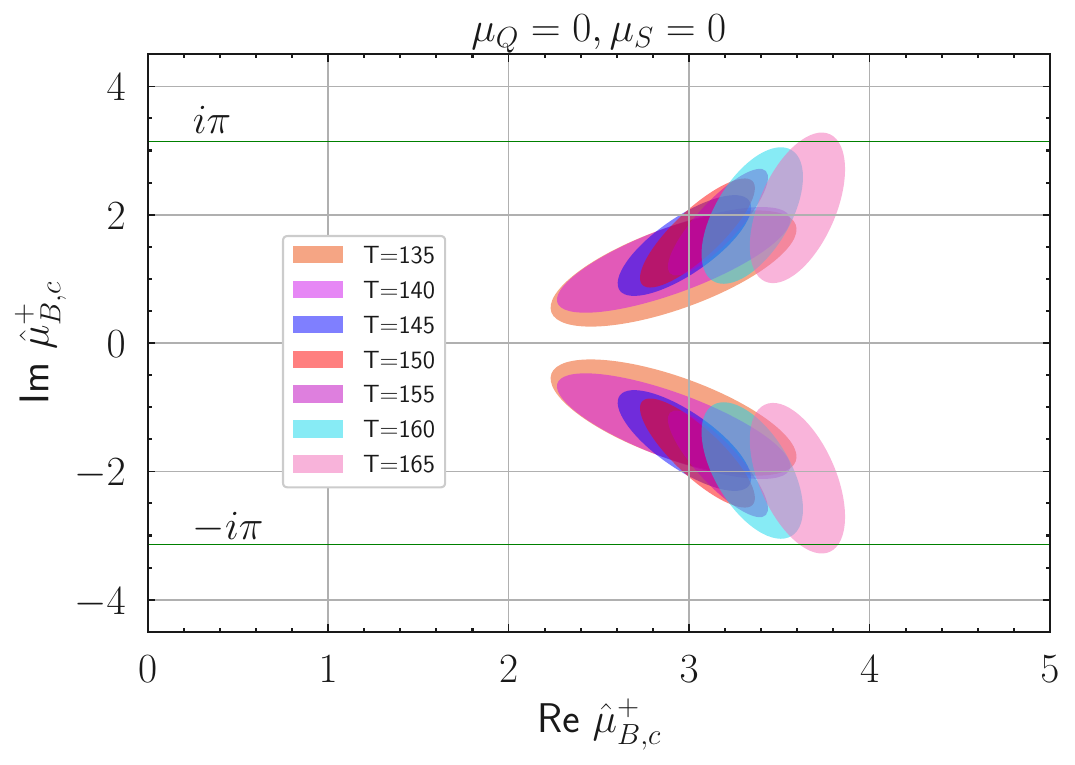}
	\includegraphics[width=0.465\textwidth]{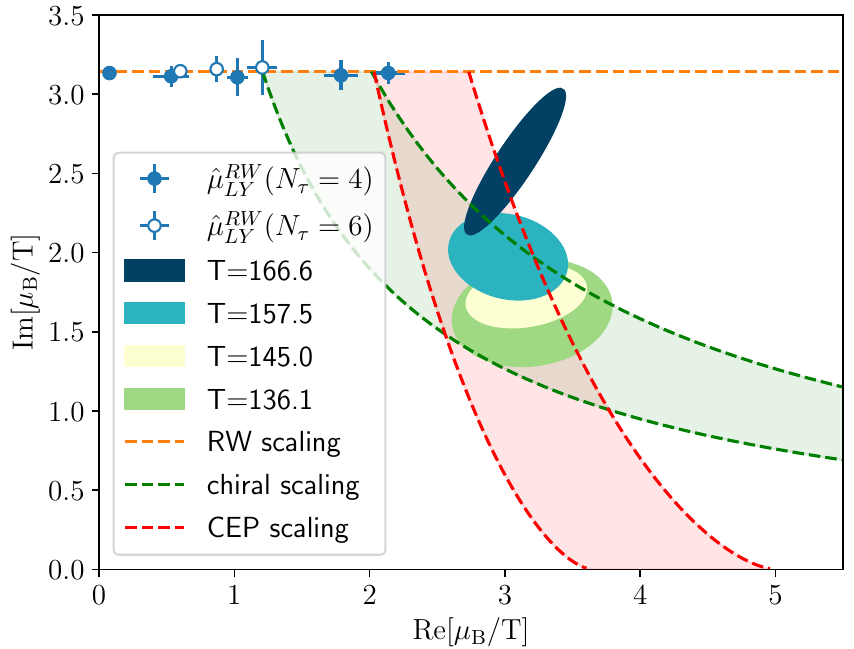}
	\caption{Poles in the complex $\hat\mu_B$ plane from the $[4,4]$-Pad\'e re-summation of the Taylor series about $\hat\mu_B=0$ (left) and from the multi-point Pad\'e approach applied to lattice QCD data at imaginary $\mu_B$ (right). Also shown in the right panel is the expected scaling behavior of the Lee-Yang edge singularities for different critical points, indicated by dashed lines/bands.}
	\label{fig:poles_pade}
\end{figure}

A discussion of  Taylor expansions in (2+1)-flavor QCD for the pressure, net baryon number and the variance of the distribution on
net-baryon number fluctuations is given in \cite{Bollweg:2022rps}, \citeTalk{Karsch_talk}.
The authors obtain series expansions from an eighth-order expansion of the pressure, Eq.~(\ref{eq:pressure_expansion}), which is re-summed by a $[2,2]$ and $[4,4]$ diagonal Pad\'e. 
The poles of those Pad\'es correspond to the Mercer-Roberts estimator \cite{Mercer:1990} of the radius of convergence.
The poles are indeed located in the complex $\hat\mu_B$ plane as shown in Fig.~\ref{fig:poles_pade} (left) and show an apparent approach to the real $\mu_B$ axis with decreasing temperature. 
Corresponding results for a re-organized expansion with zero net strangeness ($n_S=0$) are also discussed. 

Due to the limited number of Taylor coefficients one has at hand for the series about $\mu_B=0$, one needs strategies to compute the Lee-Yang zeros from multi-point Pad\'e approximants obtained from simulations at imaginary $\hat\mu_B$ \cite{Dimopoulos:2021vrk, Schmidt:2022ogw}, \citeTalk{Schmidt_talk}.
This may be achieved by combining continuation from imaginary chemical potential via Pad\'e approximants\cite{Lombardo:2005ks, Cea:2009ba} with Taylor expansion. Analytic continuation in combination with Taylor expansion was proposed in Refs.~\cite{Falcone:2010az,Laermann:2013lma}. For further interesting resummation schemes see \cite{Mondal:2021jxk}.  
The results are shown in Fig.~\ref{fig:poles_pade} (right). Also shown is the expected scaling behavior of the Lee-Yang edge singularities associated with the Roberge-Weiss, the chiral and the QCD critical point. Interestingly, at temperatures close to, but below the Roberge-Weiss transition temperature ($T\lesssim T_{RW}$) the poles follow the expected Roberge-Weiss scaling. 
At temperatures $T\lesssim$ 170 MeV, a qualitative change in the behavior of the singularities is found: they start to approach the real axis. 
If it can be established that the scaling behavior follows the one expected for the QCD critical point, the location of the QCD critical point can be determined by a scaling analysis.  

This method has been successfully applied in the Gross-Neveu model \cite{Basar:2021hdf}, see also \cite{Mukherjee:2021tyg} for further investigations in low energy EFTs with and without fluctuations see \cite{Connelly:2020gwa, Mukherjee:2021tyg, Rennecke:2022ohx, Johnson:2022cqv, Ihssen:2022xjv}.  In particular, the scaling behavior of the location of the edge singularity has been established in \cite{Connelly:2020gwa, Rennecke:2022ohx, Johnson:2022cqv}

It is thus important to understand that these studies not only highlight different numerical strategies to calculate observables at nonvanishing chemical potential via a re-summation of the Taylor series and thus might enhance the results presented in the Section on fluctuations. They also provide, and that is what we have focused on here, a mechanism to locate the elusive QCD critical point (which has been mentioned at the beginning, and will be further discussed in the Section devoted to conformal theories.)

An important input to methodological developments comes from the results on models without the sign problem, as already mentioned.
The same models  can also be used as a test bed. A typical case study is two-color QCD, see Refs.~\cite{Begun:2021nbf}~\citeTalk{Rogalyov_talk}.

\subsection{QCD at Non-Zero Density: Combining Lattice and Functional Approaches\footnote{Prepared by Jan M. Pawlowski}}

Lattice formulations of QCD are based on the formulation of Euclidean QCD on a discrete space-time lattice. The task of solving the infinite-dimensional path integral is converted into controlling both, the thermodynamic and continuum limits of Monte-Carlo simulations of finite but high-dimensional numerical integrals. Typically,  these limits have a polynomial scaling of the numerical costs with the lattice size. However, simulations for real-time QCD, or finite chemical potential require the importance sampling of measures with complex actions, causing sign problems with potentially exponential scaling of the numerical costs that are hard to overcome. This has led to the common strategy for indirect access to QCD at larger density: One simply extrapolates lattice results for a class of correlation functions, mostly the equation of state and higher order fluctuations of conserved charges, at vanishing, small and imaginary chemical potential to larger values by either Taylor expansions, Padé resummations or similar resummation schemes by also taking into account the universality class of the potential CEP. This is a standard inverse problem, and as those encountered for the reconstruction of spectral functions or real-time correlation functions it is ill-conditioned.  

Diagrammatic functional approaches to QCD convert the task of solving the path integral into controlling the infinite hierarchy limit of the solution of a finite hierarchy of closed coupled integral (DSE) or integral-differential equations (fRG) of correlation functions. The numerical costs of this limit are related to the rapidly increasing number of diagrams at higher orders of the hierarchy as well as the linear rise of the interpolation dimension of momenta of higher-order correlation functions. While apparent convergence and quantitative agreement with respective lattice results have been seen for many correlation functions in the vacuum and finite temperature, systematic error control remains an intricate issue that is hard to control. 

In turn, at finite density and for real-time QCD, functional methods allow for direct computations as they are not obstructed by the sign problem. Specifically, finite density or chemical potential correlation functions computed from self-consistent approximations to the hierarchy of correlation functions in functional approaches \textit{define} analytic functions of the chemical potential that carry all required analytic properties of QCD as well as QCD dynamics at larger density. In short, for sufficiently advanced approximations, the results for correlation functions from functional approaches such as the EoS and fluctuations of conserved charges match those obtained from lattice simulations in the validity regime of the latter. 

This suggests a very promising combined approach towards QCD at finite density as well as for real-time computations: one uses the results of functional approaches that meet lattice benchmarks, taking into account their systematic error estimates, for estimates and later predictions of QCD at large chemical potentials, and in particular the existence and location of the potential critical endpoint, see \citeTalk{Pawlowski_talk} and references therein. 
The talk includes a discussion of baryonic effects.  In these studies, the baryon is to some extent approximated
as quark-diquark system. A very detailed study cited in the talk \cite{Eichmann:2015kfa}
found the baryonic effects to be small in the putative region for the existence of the critical endpoint.

This combined approach allows for systematic improvements and hence a reduction of the systematic error. Its results at large density can be readily used as input for transport models, hydrodynamics, and the critical dynamics close to the potential critical endpoint, hence playing an important role in the experimental/theoretical understanding of QCD at large densities.

\subsection{The Road Ahead}

The motivation for going beyond simple importance sampling is very strong and comes from collider experiments and astrophysics.  New methods have been developed and are currently vigorously pursued, and old methods are continuously improved.  We feel that continual interactions with colleagues pursuing analytic approaches on one side, and mathematicians developing advanced methods on the other are beneficial and should be further pursued. 
We have to face strong {\em technical} problems, but this is a road we have to go through. While the material in this Section is the most technical one, much progress actually depends on effective handling of the open problems we addressed (see {e.g.} the conclusions of last Section \ref{sec:nature_and_phenomenology_of_the_quark-gluon_plasma}).

The Density of States may be a promising approach to the solution of the sign problem. In this approach, the Euclidean path integral (or, similarly, the partition function) of a system is evaluated as the integral over the density of a relevant observable (e.g., the action or the Hamiltonian). Similar manipulations~\cite{Langfeld:2015fua} can be performed to evaluate expectations of observables. 
The interest in this approach stems from the fact that a powerful algorithm has been devised~\cite{Langfeld:2012ah} that enables one to evaluate the density of states with exponential error reduction. The method has the potential to overcome most of the limitations of importance sampling, such as topological freezing~\cite{Cossu:2021bgn} and - crucially - the sign problem~\cite{Langfeld:2014nta,Francesconi:2019nph}\citeTalk{Lucini_talk}. 
In the latter case, further developments are needed before the method can be applied to QCD. 

Last but not least: there is an ebullient activity in the field of quantum computing. Quantum link models \cite{Wiese:2021djl}\citeTalk{Wiese_talk}  may well be a successful line of approach.

\section{Conformal Invariance\footnote{Editor: Marco Panero}}
\label{sec:conformal_phase}

The conformal group is defined as the group of transformations that leave the spacetime metric invariant, up to a local rescaling. In $D>2$ spacetime dimensions, it is an extension of the Lorentz--Poincar\'e group to include special conformal transformations and dilations.
In general, conformally invariant field theories represent the ultraviolet or infrared limits of the renormalization group of quantum field theories. Of special interest are strongly coupled conformally invariant theories, which have many important realizations in condensed matter, but also in fundamental particle physics, such as the examples that we describe in the following subsections.

\subsection{The QCD Critical Endpoint}
\label{subsec:the_qcd_critical_endpoint}

It is believed that the phase diagram of quantum chromodynamics, as a function of the baryon-number chemical potential $\mu$ and the temperature $T$, features a critical endpoint exhibiting conformal symmetry~\cite{Rajagopal:2000wf, Stephanov:2004wx, Fukushima:2010bq}. Note that here we are referring to QCD for physical values of the quark masses; for the case of QCD with massless quarks, which is discussed in detail in Section~\ref{sec:thermal_phase_transitions_and_critical_points}, instead, a \emph{tricritical} endpoint~\cite{Stephanov:1998dy} and other interesting features are expected~\cite{Pisarski:1983ms, Pelissetto:2013hqa}.

For QCD with finite quark masses, the conjecture of the existence of a critical endpoint arises from the fact that, while at low net baryon densities the ground state of the theory, characterized by confinement and chiral-symmetry breaking, turns into a deconfined and chirally symmetric quark-gluon-plasma phase through a smooth crossover as the temperature is increased~\cite{Karsch:2001vs, Aoki:2006we}, at large densities many phenomenological models predict a first-order transition line separating the hadronic phase from the QGP and possibly more exotic phases~\cite{Buballa:2003qv}. This line is expected to bend towards the temperature axis, ending at a critical endpoint $(\mucr,\Tcr)$ where the transition should be a continuous one, exhibiting conformal invariance. Although the existence of a critical endpoint is not an \emph{ab initio} prediction of QCD, if it really exists, it would leave remarkable signatures~\cite{Stephanov:1999zu, Stephanov:2008qz, Mukherjee:2015swa, Luo:2017faz}, and this has triggered intense experimental activity~\cite{Spiller:2006gj, Lacey:2006bc, Sissakian:2009zza, STAR:2010vob, Citron:2018lsq, Bzdak:2019pkr, HADES:2020wpc, Hachiya:2020bjg}, as summarized in Sections~\ref{sec:nature_and_phenomenology_of_the_quark-gluon_plasma} and~\ref{sec:methodological_challenges_spectral_functions_and_sign_problem}.

The QCD critical endpoint is expected to be in the conformal universality class of the Ising model in three dimensions (3D)~\cite{Halasz:1998qr, Berges:1998rc}. Despite the deceptively simple nature of the Ising model (a spin model with nearest-neighbor interactions and global invariance under the cyclic group of order two) and the fact that its solution in two dimensions has been known for many decades~\cite{Onsager:1943jn} and can be considered as the prototype for integrable models~\cite{Yang:1967bm, Baxter:1972hz}, it has proven analytically very hard in three dimensions. Until recently, Monte~Carlo calculations were the tool to derive the most precise predictions for the 3D Ising model, but this has drastically changed with the new developments in the conformal bootstrap approach~\cite{El-Showk:2012cjh,El-Showk:2014dwa,Gliozzi:2014jsa} and in the functional renormalization group approach~\cite{Balog:2019rrg, DePolsi:2020pjk}. The description of the QCD critical endpoint in terms of the conformal universality class of the 3D Ising model is an active line of research~\cite{An:2021wof, Parotto:2018pwx, Nonaka:2004pg, Kampfer:2005nt}. An important goal consists in identifying the ``directions'' (in the QCD phase diagram) that correspond to perturbations by ``thermal'' and ``magnetic'' operators in the Ising model~\cite{Caselle:2019tiv, Caselle:2020tjz, Caristo:2021tbk}; this, in particular, would allow one to derive analytical predictions in a finite neighborhood of the critical endpoint using conformal perturbation theory~\cite{Zamolodchikov:1987ti, Guida:1995kc, Gaberdiel:2008fn, Caselle:2016mww, Amoretti:2017aze}. In principle, the procedure to map the Ising-model variables to the $\mu$ and $T$ variables of the QCD phase diagram is relatively straightforward; a recent example of application can be found in Ref.~\cite{Parotto:2018pwx}, that we follow here. The first step consists in modeling the correct scaling behavior of the three-dimensional Ising model close to its critical point: this can be done by parameterizing the magnetization $M$, the reduced temperature $r$ (defined as the difference between the temperature of the Ising model and its critical value, in units of the latter) and the magnetic field $h$, in terms of two variables, denoted as $R$ and $\theta$~\cite{Nonaka:2004pg,Schofield:1969zz,Guida:1996ep}:
\begin{equation}
\label{M_r_h_parameterization}
M = M_0 R^\beta \theta , \qquad r = R (1- \theta^2) , \qquad h = h_0 R^{\beta \delta} \tilde{h}(\theta),
\end{equation}
where $M_0$ and $h_0$ are normalization constants, $\tilde{h}(\theta) = \theta -0.76201 \theta^3 + 0.00804 \theta^5$, while $\beta$ and $\delta$ are the critical exponents of the three-dimensional Ising model. The parameter $R$ is assumed to be a real, non-negative number, while $\theta$ is a real number whose absolute value cannot exceed $\theta_0 \simeq 1.154$, the first non-trivial zero of the function $\tilde{h}(\theta)$. Next, one constructs a function mapping the Ising variables $(h,r)$ to the QCD parameters $(\mu,T)$: under the assumption that this mapping be a linear one (which is expected to be a reasonable approximation in a sufficiently small neighborhood of the critical point), the mapping can be expressed in terms of six parameters~\cite{Rehr:1973zz}:
\begin{equation}
\label{Rehr_mapping}
\mu = \mucr + \Tcr w \left( - r \rho \cos \alpha_1 - h \cos \alpha_2 \right),  \qquad
T = \Tcr \left[ 1 + w \left( r \rho \sin \alpha_1  + h \sin \alpha_2 \right) \right] ,
\end{equation}
where $\alpha_1$ and $\alpha_2$ are the angles between the $r$ and $h$ axes and the horizontal $\mu$ axis in the QCD phase diagram, while the $w$ and $\rho$ parameters respectively encode a global and a relative rescaling of $r$ and $h$. In particular, note that following a thermal perturbation from the critical point of the Ising model into the disordered, paramagnetic phase ($h>0$) corresponds to moving along the crossover branch of the line separating the confining phase and the deconfined phase in the QCD phase diagram. We remark that the mapping between the $(h,r)$ and the $(\mu,T)$ variables is non-universal: ultimately, this is simply related to the fact that QCD and the three-dimensional Ising model are expected to be characterized by the same behavior only at the critical point, where the details about the interactions become irrelevant, while the properties of the two theories off the critical point do differ. It is also important to observe that the correspondence between the $(h,r)$ and the $(\mu,T)$ variables could be more general than the mapping~(\ref{Rehr_mapping}); in particular, including non-linear terms could yield better modeling of the boundary between the hadronic phase and the quark-gluon-plasma phase, which has a small but non-vanishing curvature. This, however, would require additional parameters, to be fixed either using some further theoretical input (e.g., from lattice calculations) or experimental data.

First-principles lattice studies of the QCD critical endpoint are particularly challenging, due to the notorious sign problem affecting simulations at finite $\mu$~\cite{Philipsen:2005mj, deForcrand:2009zkb,Aarts:2015tyj,Gattringer:2016kco}. Popular techniques to tackle the sign problem include Taylor expansions~\cite{HotQCD:2018pds,Bazavov:2017dus,Allton:2002zi,MILC:2008reg,Karsch:2010hm,Bonati:2018nut}, reweighting~\cite{Fodor:2001au, Csikor:2004ik, Fodor:2004nz} (which can be interpreted as a limiting case of non-equilibrium simulations~\cite{Jarzynski:1996oqb, Jarzynski:1997ef, Neal1998, Caselle:2018kap, Caselle:2022acb}), the complex-Langevin method~\cite{Aarts:2012yal, Sexty:2013ica}, Lefschetz thimbles~\cite{Cristoforetti:2012su} (built on an idea originally used for the computation of the partition function of three-dimensional Chern-Simons theory for complex parameters~\cite{Witten:2010cx}), the density-of-states method~\cite{Langfeld:2012ah,Fodor:2007vv}, analytical continuation from imaginary values of the chemical potential~\cite{Borsanyi:2020fev,Alford:1998sd,DElia:2002tig,DElia:2007bkz}, simulations at finite isospin density~\cite{Son:2000xc,Son:2000by}, and simulations in the canonical ensemble~\cite{Alexandru:2005ix,deForcrand:2006ec,Ejiri:2008xt}, but none of them provides \emph{the} final solution to this problem---perhaps for profound reasons~\cite{Troyer:2004ge}. Nevertheless, recently significant progress has been achieved, for example, in the lattice study of fluctuations of conserved charges at finite $\mu$ values~\cite{DElia:2016jqh,Borsanyi:2011sw,HotQCD:2012fhj}, which lead to critical fluctuations in the hadron multiplicity distributions observed in experiments and thus provide an important probe to search for the QCD critical endpoint~\cite{Stephanov:2008qz,Stephanov:2011pb}. Other theoretical studies of the QCD phase diagram are based on functional approaches to QCD~\cite{Fischer:2014ata, Fu:2019hdw, Gao:2020qsj, Gao:2020fbl,Gunkel:2021oya} or on the gauge/string duality~\cite{Maldacena:1997re,Witten:1998qj,Gubser:1998bc,Aharony:1999ti}: some recent examples can be found in Refs.~\cite{DeWolfe:2010he,Evans:2011eu,Alho:2013hsa,Critelli:2017oub}.

We conclude this subsection with some words of caution. The quest for the QCD critical endpoint, and the unambiguous characterization of its properties, is still an open challenge, both from the theoretical and the experimental point of view: as shown, for example, in Ref.~\cite[Fig.~8]{Czopowicz:2020twk}, theoretical predictions obtained with different methods and experimental hints are still scattered across a very wide region of the QCD phase diagram. Finally, as we mentioned earlier, there remains the possibility that the QCD critical endpoint that we discussed so far may not exist at all: this could happen, for instance, if the entire line separating the confining phase and the deconfined one(s) in the QCD phase diagram turned out to be a crossover, as it is at zero and at low densities~\cite{Karsch:2001vs, Aoki:2006we}. We remark that this possibility is not ruled out by symmetry arguments, and, interestingly, some studies based on the extrapolation of lattice results obtained at imaginary values of the chemical potential (where the sign problem not present) and for relatively coarse lattice spacings hint precisely at this scenario~\cite{deForcrand:2002hgr,deForcrand:2007rq,deForcrand:2008vr}. In this case, the only first-order line in the part of the QCD phase diagram directly accessible to laboratory experiments would be the one separating the region of the confining phase at low temperature and low baryon-number chemical potential, which can be described as a dilute hadron gas, from the other region, also in the confining phase (again at low temperatures, but at larger values of the baryon-number chemical potential), that corresponds to a nuclear-matter ``liquid'' and is characterized by larger densities: this is a first-order line that at $T=0$ occurs for $\mu=923$~MeV (this value is obtained from the difference between the nucleon mass and the average binding energy per nucleon of nuclear matter). This first-order line extends also at finite temperatures, up to a critical temperature which can be studied through multifragmentation experiments~\cite{Chomaz:2003dz} and has been estimated to be $16.60(86)$~MeV in Ref.~\cite{Natowitz:2002nw}. While this line extends entirely within the confining phase, the phase diagram might also feature other lines, separating regions of the deconfined phase characterized by different properties, such as one or more superconducting phases~\cite{Alford:1997zt,Rapp:1997zu}, a color-flavor-locked phase~\cite{Alford:1998mk,Alford:2007xm}, a crystalline color superconducting phase~\cite{Alford:2000ze}, etc. A discussion about the existence and location of such lines is beyond the scope of our present discussion, and will not be pursued further here.

\subsection{Conformal Dynamics in Models for Dynamical Electroweak Symmetry Breaking}
\label{subsec:conformal_dynamics_in_extended_technicolor_models}

Another class of elementary-particle theories in which the existence of a conformal phase plays a prominent r\^ole are the theories that may describe physics beyond the Standard Model. Of particular interest are non-supersymmetric, non-Abelian gauge theories, in which a conformally invariant phase exists for some matter field contents (i.e., for suitable gauge group, number of fermion species, and their representation under the gauge group). For the current status of various aspects of this research area, see Ref.~\citeTalk{Sannino_talk}, summarizing the state-of-the-art of strongly coupled theories for physics beyond the Standard Model, Ref.~\citeTalk{Contino_talk}, which discusses an interesting example of a model for dark matter, and Ref.~\citeTalk{Rago_talk}, reporting a calculation of scattering amplitudes in an $\SU(2)$ gauge theory coupled to matter fields in the fundamental representation of the gauge group.

One of the early motivations to investigate strongly coupled gauge theories for physics beyond the Standard Model stems from the fact that they may provide a dynamical realization of the electroweak symmetry breaking mechanism. Given an asymptotically free, strongly coupled gauge theory with fermionic matter fields whose left-handed components are Standard Model weak doublets and form a condensate at a dynamically generated energy scale $\LambdaTC$, the ensuing dynamical symmetry-breaking of the theory leads to Nambu--Goldstone bosons, which can be interpreted as the longitudinal components of the electroweak gauge bosons of the Standard Model. This is the old idea of ``technicolor''~\cite{Weinberg:1975gm,Susskind:1978ms}; while it does not require the existence of a fundamental scalar, it allows one to interpret the experimentally observed Higgs boson as the lightest scalar state in the spectrum (i.e., as the analogue of the $\sigma$ meson in QCD). In a related class of models, the Higgs boson itself is interpreted as a composite particle and as a pseudo-Nambu--Goldstone boson~\cite{Kaplan:1983fs,Kaplan:1983sm,Banks:1984gj}. To accommodate the existing masses of quarks and leptons, technicolor has to be generalized to an ``extended technicolor'' model~\cite{Dimopoulos:1979es, Eichten:1979ah}, with a larger gauge symmetry that is broken down to the technicolor gauge group at an energy scale $\LambdaETC$ (which, due to phenomenological constraints on flavor-changing neutral currents, is expected to be significantly higher than $\LambdaTC$); the quark and lepton masses then arise in the low-energy effective theory obtained by integrating out the heavy degrees of freedom of the extended technicolor theory, through terms suppressed by some inverse power of the $\LambdaETC$ scale. In order to generate the masses of heavy quarks, the fermion condensate should be enhanced by (approximate) scale invariance of the theory between the $\LambdaTC$ and $\LambdaETC$ scales, with a sufficiently large mass anomalous dimension $\gamma$, which would realize a ``walking technicolor'' scenario~\cite{Holdom:1981rm,Yamawaki:1985zg,Appelquist:1986an}. This is indeed possible in the presence of a number of fermion species that is sufficiently large to drive the $\beta$ function of the theory (close) to an infrared-stable fixed point, without exceeding the value that would cause the loss of asymptotic freedom; this defines the so-called ``conformal window''~\cite{Sannino:2004qp,Dietrich:2006cm,Sannino:2009aw,Mojaza:2012zd}. In fact, in an approximately conformal technicolor model an electroweak-symmetry-breaking condensate also breaks scale invariance, and the associated ``dilaton'' may then have properties (and, in particular, a mass) compatible with the one observed experimentally for the Higgs boson~\cite{Bando:1986bg,Bardeen:1985sm,Dzhikiya:1986kk,Goldberger:2007zk,Ryskin:2009kw,Appelquist:2010gy,Grinstein:2011dq,Campbell:2011iw,Matsuzaki:2012gd,Elander:2012fk}.

The literature on strongly coupled models for electroweak symmetry breaking is vast~\cite{Cacciapaglia:2020kgq,Hill:2002ap,Sannino:2009za}. As reviewed in Refs.~\cite{DelDebbio:2010zz,Giedt:2012it,Kuti:2014epa,DeGrand:2015zxa,Nogradi:2016qek,Svetitsky:2017xqk,Witzel:2019jbe,Rummukainen:2022ekh}, in this field lattice calculations remain an essential tool to investigate the properties of candidate theories at a non-perturbative level and from first principles (although interesting complementary approaches, such as holography~\cite{Alho:2013hsa,Jarvinen:2009fe,Jarvinen:2011qe,Alho:2012mh,Alvares:2012kr,Alho:2013dka,Erdmenger:2020flu,Erdmenger:2020lvq,Elander:2020nyd,Elander:2021bmt,Elander:2021kxk} and functional approaches~\cite{Braun:2010qs, Braun:2009ns, Hopfer:2014zna}, have also been used). 

The key questions that lattice studies can answer include: Is a theory confining or nearly conformal in the infrared? What is the phase structure, as a function of the parameters of the theory? What is the spectrum of physical states? What is the value of the mass anomalous dimension? One of the theories that have been most extensively studied through lattice calculations is the $\SU(2)$ gauge theory with two fermions in the adjoint representation of the gauge group, also known as ``minimal walking technicolor'', see Ref.~\cite{Sannino:2004qp}. In the latter work it was also pointed out that for fermions in higher-dimensional representations, the conformal window would already appear in the presence of a small number of fermions. The phenomenological implications of these models were further elaborated upon in Refs.~\cite{Hong:2004td,Evans:2005pu,Dietrich:2005jn}. The first lattice study of minimal walking technicolor was reported in Ref.~\cite{Catterall:2007yx}, which was soon followed by several other works~\cite{Appelquist:2007hu,Catterall:2008qk,Hietanen:2008mr,DelDebbio:2008zf,Bursa:2009we,DelDebbio:2009fd,Hietanen:2009az,DelDebbio:2010hx,DelDebbio:2010hu,DeGrand:2011qd,Catterall:2011zf,Bursa:2011ru,Giedt:2012rj,Rantaharju:2013gz,DelDebbio:2015byq,Rantaharju:2015yva,Rantaharju:2015cne,Bergner:2016hip}. The study of these models also led one to realize that even an $\SU(2)$ theory with just two flavors of Dirac fermions in the fundamental representation could have interesting applications in the context of model building for dark matter~\cite{Ryttov:2008xe}, due to the pattern of chiral-symmetry breaking observed in lattice simulations~\cite{Lewis:2011zb}.

The $\SU(2)$ gauge theory with different numbers of fundamental fermions has been further investigated in Refs.~\cite{Karavirta:2011zg,Hayakawa:2013yfa,Appelquist:2013pqa,Leino:2017hgm,Rantaharju:2021iro}. Similar studies have been carried out also for the $\SU(3)$ gauge theory with a different number of quark flavors~\cite{Appelquist:2007hu,Iwasaki:2003de,Deuzeman:2008sc,Fodor:2009wk,Deuzeman:2009mh,Hasenfratz:2009ea,Appelquist:2009ty,LSD:2009yru,Hasenfratz:2010fi,Fodor:2011tu,Appelquist:2011dp,Cheng:2011ic,Miura:2011mc,Miura:2012zqa,Lin:2012iw,Aoki:2013xza,Ishikawa:2013tua,Lombardo:2014pda,Brower:2015owo,LatKMI:2016xxi,Fodor:2016zil,Hasenfratz:2017qyr,LatticeStrongDynamics:2018hun,Hasenfratz:2019dpr,Hasenfratz:2020ess,Hasenfratz:2022qan} or with fermions in a larger representation of the gauge group, such as the two-index symmetric (sextet) representation~\cite{Fodor:2009ar,DeGrand:2010na,Sinclair:2010be,Kogut:2011bd,Fodor:2012ty,Kogut:2014kla,Hasenfratz:2015ssa} or the adjoint representation~\cite{DeGrand:2013uha}, and for the $\SU(4)$ theory with fermions in the two-index symmetric (decuplet) representation~\cite{DeGrand:2013uha,DeGrand:2012qa,DeGrand:2015lna} or with fermions in two distinct representations~\cite{Ayyar:2017qdf, Ayyar:2018zuk, Ayyar:2018ppa, Ayyar:2018glg, Cossu:2019hse, DelDebbio:2022qgu}.

The study of a (nearly) conformal phase for strongly interacting gauge theories coupled with elementary fermionic fields remains a non-trivial problem in lattice field theory. However, novel, promising techniques have been recently proposed to tackle the challenges in such computations; these include, for example, discretization techniques with a potential to treat multiple length scales in an efficient way~\cite{Brower:2012vg,Neuberger:2014pya,Neuberger:2017dtb,Polyzou:2020ifj}, or methods to extract the anomalous dimensions associated with different operators~\cite{DelDebbio:2010ze,DelDebbio:2010jy,Patella:2012da,Cheng:2013eu,Carosso:2018bmz}.

Finally, we mention that recently a novel method to derive predictions for conformal theories has been proposed, which is based on a large-charge approach~\cite{Hellerman:2015nra,Alvarez-Gaume:2016vff,Banerjee:2017fcx,Orlando:2019skh,Orlando:2020yii}; potentially interesting applications include the analysis of a topological $\theta$ term and axion physics~\cite{Bersini:2022mqn,Bersini:2022jhs}. 

\subsection{The Road Ahead}

The study of conformal dynamics remains a central issue in elementary particle physics. Its relevance extends from the ``low'' energy domain of quantum chromodynamics, with the search for a critical endpoint in the phase diagram discussed in Ref.~\citeTalk{Ratti_talk}, to the ``high'' energy domain of candidate theories for physics beyond the Standard Model discussed in Ref.~\citeTalk{Sannino_talk}. In both of these research directions, significant progress has been achieved during the past few years. Crucially, this has been possible through the combination of analytical insights with numerical calculations; it is reasonable to expect that this type of ``hybrid'' approach will lead to further progress in the forthcoming years.

\section{Cosmology, Topology and Axions\footnote{Editor: Claudio Bonanno}}\label{sec:cosmology_topology_axions}

\subsection{The Peccei--Quinn Axion and QCD Topology}

One of the most intriguing open problems in particle physics is the so-called \emph{strong $\mathrm{CP}$ problem}. While it is well known experimentally that weak interactions are not invariant under the $\mathrm{CP}$ symmetry (consisting of a parity inversion $\mathrm{P}$ plus a charge conjugation $\mathrm{C}$), so far no evidence of $\mathrm{CP}$ symmetry breaking from the strong sector has been observed experimentally. From the theoretical point of view, however, QCD allows for an explicit breaking of this symmetry because of the existence of the dimensionless $\theta$ parameter, coupling the $\mathrm{CP}$-odd topological charge
\begin{equation}\label{eq:topcharge}
Q = \frac{1}{16 \pi^2} \int \Tr\left\{F_{\mu\nu}(x)\tilde{F}^{\mu\nu}(x)\right\} d^4 x
\end{equation}
to the $\mathrm{CP}$-conserving ordinary QCD action. Experimental measures of the neutron electric dipole moment put the extremely stringent upper bound $\vert \theta \vert \lesssim 10^{-9} - 10^{-10}$ on this parameter~\cite{Crewther:1979pi, Guo:2012vf, Abel:2020pzs, Alexandrou:2020mds}, but there is no theoretical reason within the Standard Model (SM) for this parameter to vanish exactly, or to be so unnaturally small\footnote{The $\theta$ angle would be vanishing in the presence of a zero-mass quark, but this scenario has been ruled out by lattice simulations~\cite{Alexandrou:2020bkd,Aoki:2021kgd} and experiments~\cite{PDG2020}. Other exotic scenarios to explain the vanishing of $\theta$ within QCD have been considered, e.g., in~\cite{Ai:2020ptm}.}.

A particularly interesting solution proposed by Peccei, Quinn, Weinberg and Wilczek to solve this issue is the \emph{axion}~\cite{Peccei:1977hh, Peccei:1977ur, Wilczek:1977pj, Weinberg:1977ma}, a hypothetical pseudo-scalar particle introduced as the pseudo-Nambu--Goldstone Boson (NGB) of the spontaneous breaking of a new global $\mathrm{U}(1)_{\mathrm{PQ}}$ axial symmetry, the Peccei--Quinn (PQ) symmetry, which is anomalous under $\mathrm{SU}(3)_{\mathrm{color}}$. Under these assumptions, the axion field, by anomaly matching, directly couples to the QCD topological charge~\eqref{eq:topcharge} and, by virtue of being a NGB, possesses a shift symmetry which dynamically relaxes $\theta$ to zero, solving exactly the strong $\mathrm{CP}$ problem. This is, in brief, the so-called \emph{$\mathrm{PQ}$ mechanism} and it constitutes a very simple, yet powerful, and natural solution to the strong-$\mathrm{CP}$ problem which requires supplementing the SM with just a few new ingredients.

This is not the only intriguing aspect making the PQ axion a promising and well-motivated SM extension. Soon after its introduction, this hypothetical particle has also been recognized as a possible Dark Matter candidate, its couplings with SM particles being suppressed by the axion scale $f_a$, which is expected to be extremely large from astrophysical and cosmological bounds: $10^8~\text{GeV}~\lesssim f_a \lesssim~10^{12}~\text{GeV}$~\cite{Preskill:1982cy,Abbott:1982af,Dine:1982ah,Raffelt:2006cw}. Therefore, the introduction of the PQ axion, motivated by the strong-$\mathrm{CP}$ problem, would also naturally explain (at least partially) a further fundamental missing piece of the SM.

Another crucial property of the PQ mechanism is that, being rooted on anomaly matching and on general properties of NGBs, it holds at low energy scales $\Lambda \ll f_a$ independently of the underlying Ultra-Violet (UV) fundamental dynamics, i.e., of the particular UV-complete SM extension one is considering. In these respects, a particularly well-motivated class of models which has been widely considered in the phenomenology literature to explain PQ axions assumes the existence of a confining strongly-coupled dark sector that possesses an accidental global PQ symmetry~\cite{Contino:2017rkq,Contino:2020god,Contino:2021ayn,Gaillard:2018xgk}~\citeTalk{Contino_talk}. This way, the global $\mathrm{U}(1)_{\mathrm{PQ}}$ is not imposed \emph{ad hoc} but naturally emerges within the newly-introduced non-abelian gauge sector (similarly to flavor symmetries in QCD). In this framework, the axion is described as a composite particle that is made cosmologically stable by the accidental $\mathrm{U}(1)_{\mathrm{PQ}}$ symmetry~\cite{Contino:2017rkq,Contino:2020god,Contino:2021ayn,Gaillard:2018xgk}~\citeTalk{Contino_talk}.

An intriguing aspect of this class of models is that it is in principle amenable to be probed by future experimental interferometric Gravitational Wave (GW) observations, since the spontaneous breaking of the PQ symmetry will create a GW signature~\cite{Croon:2019iuh}~\citeTalk{Houtz_talk}, as it is expected on general grounds for a first-order phase transition~\citeTalk{Ghiglieri_talk}. More precisely, from the analysis of the GW spectra it is possible to infer several interesting properties about the confining strongly coupled dark sector and of its related axion~\cite{Croon:2019iuh}~\citeTalk{Houtz_talk}, thus making GW observations a fascinating tool, alternative to collider experiments, to possibly detect the existence of such hypothetical particle.

Finally, a fundamental aspect of axion physics is constituted by its relation with the topological properties of QCD at finite temperature. As a matter of fact, since the axion field is directly coupled to the topological charge~\eqref{eq:topcharge}, it is possible to relate the temperature-dependent axion effective mass to the QCD topological susceptibility $\chi(T)$ via the well-known relation:
\begin{equation}\label{eq:axion_mass_chi}
	m_a^2(T) f_a^2 = \chi(T) = \underset{V\to\infty}{\lim} \frac{\langle Q^2 \rangle(T)}{V},
\end{equation}
where $V$ is the $4\mathrm{D}$ space-time volume and $T$ is the temperature. Apart from the unknown constant $f_a$, the value of the axion mass in Eq.~\eqref{eq:axion_mass_chi} is completely fixed by the QCD topological susceptibility $\chi$. Moreover, this quantity accounts for the whole temperature dependence of $m_a(T)$. This implies the possibility, once the temperature dependence of the QCD topological susceptibility is known and some cosmological assumptions are made, to put a more stringent upper bound on the value of $f_a$ (which is \emph{a priori} unknown in this model) through the so-called \emph{misalignement mechanism}~\cite{Preskill:1982cy,Abbott:1982af,Dine:1982ah}.

For this reason, the PQ axion has renewed interest in the study of the temperature dependence of the QCD topological susceptibility, as the knowledge of $\chi(T)$ appears to be an essential input for the computation of interesting axion phenomenological observables, which are of the utmost importance for its current and future experimental searches (see, e.g., Refs.~\cite{Lombardo:2020bvn,DiLuzio:2020wdo} for recent reviews).

Given the non-perturbative nature of the topological properties of gauge theories, results about the behavior of $\chi(T)$ in QCD can be obtained analytically only by adopting suitable approximations.

In Chiral Perturbation Theory (ChPT) at leading order and with 2 light quark flavors, it is possible to obtain the following prediction~\cite{DiVecchia:1980yfw,Leutwyler:1992yt,Mao:2009sy,Guo:2015oxa,GrillidiCortona:2015jxo,Bonati:2015vqz,Luciano:2018pbj}~\citeTalk{Villadoro_talk}:
\begin{eqnarray}
\frac{\chi_{\mathrm{ChPT}}(T)}{\chi_{\mathrm{ChPT}}(T=0)} &=& \left[1 - \frac{3}{2}\frac{T^2}{f^2_\pi}J_1\left(\frac{T^2}{m^2_\pi}\right)\right], \\\nonumber
 \\
 \chi_{\mathrm{ChPT}}(T=0) &=& \frac{m_u}{m_u+m_d} m_\pi^2 f_\pi^2,\\\nonumber
 \\\nonumber
J_1(x) &\equiv& \frac{1}{\pi^2}\frac{\partial}{\partial x} \left\{\int_0^{\infty} dq \, q^2 \log\left(1-e^{-\sqrt{q^2+x}}\right)\right\}.
\end{eqnarray}
In the literature, also the NLO results $\chi^{1/4}_{\mathrm{ChPT}}(T=0)=75.5(5)$~MeV (for physical $u$, $d$ quarks) and $\chi^{1/4}_{\mathrm{ChPT}}(T=0)=77.8(4)$~MeV (for degenerate $u$, $d$ quarks) have been computed~\cite{GrillidiCortona:2015jxo,Bonati:2015vqz} (see also~\cite{Gorghetto:2018ocs} for a discussion about NNLO and QED corrections to $\chi_{\mathrm{ChPT}}(T=0)$). However, while the $T=0$ ChPT result is expected (and confirmed from Monte Carlo simulations) to be reliable, leading in particular to the cold axion mass prediction $m_a =5.70(6)~\mu\text{eV}~(10^{12}~\text{GeV}/f_a)$~\cite{GrillidiCortona:2015jxo}, the finite-temperature ChPT result is expected to be unreliable close and above the crossover temperature $T_c \simeq 155$~MeV, since the chiral condensate and thus ChPT are expected to break down in this regime.

Another possible strategy, which is instead expected to be reliable at asymptotically-high temperatures, is to compute $\chi(T)$ by semiclassical methods via the so-called Dilute Instanton Gas Approximation (DIGA). Assuming that instantons can be treated as identical non-interacting pseudo-particles, it is possible to compute $\chi(T)$ at leading order in perturbation theory by performing a Gaussian integration of the fluctuations around a one-instanton configuration, obtaining~\cite{Gross:1980br,Boccaletti:2020mxu}:
\begin{equation}\label{eq:DIGA_prediction_chi}
	\chi_{\mathrm{DIGA}}(T) \sim T^{-d},
\end{equation}
where $d \approx 8$ for 3 light quark flavors.

Although this simple prediction has been customarily used in several computations due to its simplicity and due to the lack of more reliable results, it is expected (and confirmed by numerical simulations) that the DIGA result is not reliable for temperatures close to the crossover and even up to the few GeV region, where deviations from perturbative calculations are still pronounced and non-perturbative effects are still not completely negligible. Corrections to the DIGA due to multi-instanton contribution can be computed systematically \cite{Rennecke:2020zgb}. However, since the temperature-dependence of $m_a(T)$ in this temperature range is needed to accurately compute axion cosmology~\cite{Wantz:2009it}, it has been pointed out in the literature that an independent and fully non-perturbative computation of $\chi(T)$ from lattice simulations would be needed to obtain full control on $m_a(T)$~\cite{Berkowitz:2015aua,Notari:2022zxo}. For this reason, the numerical calculation of $\chi(T)$ has been the goal of several lattice studies in recent years~\cite{Lombardo:2020bvn,Bonati:2015vqz, Frison:2016vuc,Borsanyi:2016ksw, Petreczky:2016vrs, Bonati:2018blm,Athenodorou:2022aay}~\citeTalk{Bonanno_talk}.

\subsection{The QCD Topological Susceptibility at Finite Temperature from the Lattice: Current Status and Future Challenges}

The lattice numerical computation of the topological susceptibility in full QCD at finite temperature is a challenging task in several respects. In the following we will address some of the most severe problems that have to be faced to this end.

One serious numerical problem is posed by the presence of dynamical fermions. In the continuum theory, the contribution of non-zero topological charge configurations in the path integral is suppressed by the fermion determinant as powers of the light quark mass due to the existence of chiral zero-modes in the spectrum of $\slashed{D}$. On the lattice, typically-employed quark discretizations (such as the Wilson or staggered ones) do not preserve the chiral symmetry, which is partially or fully broken explicitly at finite lattice spacing and is only properly recovered in the continuum limit. The explicit breaking of the chiral symmetry prevents the spectrum of the lattice Dirac operator to have exact zero-modes, meaning that lowest-lying modes are shifted by lattice artifacts. This results in somewhat large corrections to the continuum limit when $\chi$ is computed from a standard gluonic definition because the determinant of the lattice Dirac operator does not provide an efficient suppression as in the continuum. Moreover, it is observed that this problem hits hard already at moderate values of $T/T_c$ due to the strong suppression of $\chi$ above the crossover (cf.~Eq.~\eqref{eq:DIGA_prediction_chi}), making it difficult to obtain reliable continuum extrapolations in the high-temperature regime.

Another infamous problem regards the proper sampling of the topological charge distribution during the Monte Carlo evolution. The standard computation of the susceptibility via $\chi = \langle Q^2 \rangle/V$ requires a meaningful sampling of the different relevant topological sectors, i.e., to observe a reasonable number of topological fluctuations during the Monte Carlo evolution. On typically-employed lattice volumes, however, $\langle Q^2 \rangle = \chi V \ll 1$ due to the strong suppression of $\chi$ above $T_c$, meaning that the observed $Q$ distribution is largely dominated by the $Q=0$ sector, and fluctuations of $Q$ above zero become extremely rare. Thus, unreasonably long Monte Carlo histories are needed to compute $\chi$ with reasonable statistical accuracy.

Finally, a notorious and rather general computational problem affects all standard local updating algorithms customarily employed in lattice simulations: the so-called \emph{topological critical slowing down}. On general grounds, local algorithms are expected to become less and less ergodic as the continuum limit is approached, leading the Monte Carlo evolution of all observables to experience a critical slowing down. While for non-topological observables such slowing down is typically polynomial in the inverse lattice spacing, there is plenty of numerical evidence that it is exponential, and thus much more severe, for topological ones~\cite{Alles:1996vn, DelDebbio:2004xh, Schaefer:2010hu, Bonati:2015sqt, Bonati:2016tvi, Bonati:2017woi, Bonanno:2018xtd, Berni:2020ebn,Eichhorn:2022wxn,Bonanno:2022hmz}. In practice, the Monte Carlo Markov chain tends to remain trapped in a fixed topological sector, meaning that the Monte Carlo evolution of the topological charge suffers from unbearably long auto-correlation times. For this reason, this issue is also known as \emph{topological freezing}. Concerning finite-temperature QCD, going below lattice spacings of the order of $\sim 0.03$~fm is extremely challenging because of the freezing problem. Since in the Monte Carlo approach the temperature $T = 1/(a N_t)$ is fixed by the product between the lattice temporal extent and the lattice spacing, this implies that reaching temperatures of the order of $\sim 700$~MeV $-$ 1~GeV or above is a seriously difficult task on typical lattices with $N_t \sim 12-16$, requiring extremely fine lattice spacings of the order of $0.01$~fm or less.

From this brief summary, it is already clear that adopting suitable strategies to deal with such obstacles is necessary for reliable numerical computation of $\chi(T)$ from the lattice. In recent years, several lattice determinations of $\chi(T)$ in finite temperature QCD have appeared in the literature, differing in the methods employed to tackle the difficulties presented so far.

The authors of Ref.~\cite{Borsanyi:2016ksw}, for example, give up the sampling of higher-topological-charge sectors and reduce to the computation of $\chi$ from just the $Q=0$ and $\vert Q \vert=1$ sectors, which is justified on the basis of the DIGA itself, in order to avoid sampling problems related to the dominance of the $Q=0$ sector and/or to the topological freezing:
\begin{equation}
	\chi \sim \frac{2Z_1}{V Z_0}, \quad Z_n = \int_{Q\,=\,n} [dA] e^{-S_{\mathrm{YM}}[A]} \prod_f\det\{\slashed{D}[A] + m_f \}.
\end{equation}
In this approach, the computation of $\chi$ reduces to the computation of the relative weight $Z_1/Z_0$ as a function of $T$. To this end, the authors of Ref.~\cite{Borsanyi:2016ksw} compute suitable observables for lower values of $T$ that still allow observing jumps from $Q=0$ to $Q=\pm 1$ sectors, and obtain $Z_1/Z_0$ at higher values of $T$ by means of a temperature extrapolation, which is expected to be reliable if the variance of the topological charge distribution $\langle Q^2 \rangle$ is sufficiently small (see also Ref.~\cite{Frison:2016vuc} more technical details on this point).\\
Moreover, a reweighting method, based on the expected continuum zero eigenvalues of the Dirac operator, is employed in~\cite{Borsanyi:2016ksw} to restore \emph{a posteriori} the suppression due to the determinant of the continuum Dirac operator and thus to reduce the magnitude of lattice artifacts affecting the gluonic susceptibility:
\begin{eqnarray}
	\chi = \frac{1}{V} \langle Q^2\rangle \rightarrow \frac{1}{V} \frac{\langle Q^2 w(Q)\rangle}{\langle w(Q) \rangle},
\end{eqnarray}
where the weight reads
\begin{equation}
	w(Q) = \prod_f \prod_{i=1}^{2\vert Q \vert} \left(\frac{m_f^2}{m_f^2+\lambda_i^2}\right)^{n_f/4},
\end{equation}
with $\lambda_i$ the lowest-lying eigenvalues of the staggered operator $D_{\mathrm{stag}}$ and $n_f$ the number of quark species with flavor $f$.

The authors of Ref.~\cite{Petreczky:2016vrs}, instead, adopt a fermionic discretization of the topological susceptibility based on the disconnected chiral susceptibility $\chi^{(\mathrm{disc})}$:
\begin{equation}\label{eq:topsusc_chi_disc}
	\chi = m^2_l \chi^{(\mathrm{disc})} = \frac{m^2_l}{V} \left[ \langle(\overline{\psi}_l\psi_l)^2\rangle - \langle\overline{\psi}_l\psi_l\rangle^2\right].
\end{equation}
A similar strategy has been adopted also in Refs.~\cite{Burger:2018fvb,Lombardo:2020bvn,Kotov:2021ujj}. Such definition is based on the assumption that the $\mathrm{U}(1)_{\mathrm{A}}$ flavor symmetry is effectively restored in the deconfined phase, so that the exact continuum relation~\cite{Kogut:1998rh,HotQCD:2012vvd,Buchoff:2013nra}
\begin{equation}
	\chi = m^2_l \chi_5^{(\mathrm{disc})} = \frac{m^2_l}{V} \left\langle\left(\Tr\{\gamma_5 (\slashed{D}+m_l)^{-1}\}\right)^2 \right\rangle
\end{equation}
can be approximated with Eq.~\eqref{eq:topsusc_chi_disc}, being $\chi^{(\mathrm{disc})} = \chi_5^{(\mathrm{disc})}$ by $\mathrm{U}(1)_\mathrm{A}$ invariance.

In Ref.~\cite{Athenodorou:2022aay}, instead, the authors adopt a different fermionic definition of $\chi$, based on spectral projectors~\cite{Luscher:2004fu, Giusti:2008vb, Luscher:2010ik, Cichy:2015jra, Alexandrou:2017bzk} on the eigenmodes of the staggered Dirac operator~\cite{Bonanno:2019xhg}~\citeTalk{Bonanno_talk}, where the same discretizations for sea and valence quarks are taken. In a few words, the bare topological charge is defined as the sum of the pseudo-chiralities of the lowest-lying eigenmodes of $D_\mathrm{stag}$ up to a certain cut-off $M$, whose value is irrelevant in the continuum limit but allows to control the magnitude of discretization corrections to the continuum limit of $\chi$:
\begin{equation}
	Q_{\mathrm{SP}}^{(\mathrm{bare})} = \frac{1}{4}\sum_{\vert\lambda\vert \le M} u_\lambda^\dagger \gamma_5^{(\mathrm{stag})} u_\lambda, \quad i D_{\mathrm{stag}} u_\lambda = \lambda u_\lambda,
\end{equation}
with $\gamma_5^{(\mathrm{stag})}$ the staggered definition of the Dirac $\gamma_5$ matrix. Introducing the spectral projector $\mathbb{P}_M \equiv \sum_{\vert \lambda \vert \le M} u_\lambda u_\lambda^\dagger$, the discretized susceptibility is:
\begin{eqnarray}
	\chi_{\mathrm{SP}} &=& {Z_Q^{(\mathrm{SP})}}^2 \frac{\langle{Q_{\mathrm{SP}}^{(\mathrm{bare})}}^2\rangle}{V} = \frac{1}{16} {Z_Q^{(\mathrm{SP})}}^2 \frac{\langle \Tr^2\{\mathbb{P}_M \gamma_5^{(\mathrm{stag})}\}\rangle}{V}, \\\nonumber
	\\
	{Z_Q^{(\mathrm{SP})}}^2&=& \frac{\langle\Tr\{ \mathbb{P}_M \}\rangle}{\langle\Tr\{ \mathbb{P}_M\gamma_5^{(\mathrm{stag})} \mathbb{P}_M\gamma_5^{(\mathrm{stag})} \}\rangle}
\end{eqnarray}
In addition, to restore a proper sampling of suppressed topological sectors, the authors of~\cite{Athenodorou:2022aay} employ a multicanonical algorithm, consisting in the inclusion of a topological bias potential in the gluonic action~\cite{Bonati:2018blm,Berg:1992qua,Jahn:2018dke,Bonanno:2022dru}. The bias is chosen so as to enhance the probability of visiting suppressed topological sectors, and expectation values with respect to the original distribution are recovered through a standard reweighting procedure.

All the strategies outlined so far have been pursued in the presence of non-chiral fermions (Wilson or staggered discretizations). In Ref.~\cite{Chen:2022fid}, instead, the temperature-behavior of $\chi(T)$ above the crossover has been determined by adopting Domain Wall fermions in the sea sector. The use of the Domain Wall lattice Dirac operator allows for quantitative control of the amount of breaking of the chiral symmetry due to lattice artifacts, at the price of adding a fifth fictitious dimension $s$ (of length $L_s$). In the limit $L_s \to \infty$, chiral symmetry (at finite lattice spacing) is exactly recovered; at finite value of $L_s$ chiral symmetry is instead broken by lattice artifacts. However, it is possible to quantify such explicit breaking in terms of an additive residual quark mass (which depends on $L_s$ and other bare parameters), that adds up to the bare one to give an effective quark mass: $m_{\mathrm{eff}} = m_{\mathrm{quark}} + m_{\mathrm{res}}$. In practical simulations $L_s$ is finite, and thus $m_{\mathrm{res}}$ is non-zero. However, if $L_s$ is large enough, it is possible to achieve $m_{\mathrm{res}} \ll m_{\mathrm{quark}}$, i.e., the explicit breaking of the chiral symmetry due to lattice artifacts is much smaller than the explicit breaking due to a non-vanishing physical quark mass. The aim of using a fermionic discretization which better preserves the chiral symmetry is to try to reduce the large lattice artifacts affecting gluonic definitions of the topological charge at finite temperature. In this respect, the authors employ the standard clover gluonic definition computed after gradient flow to determine $\chi$. In the lattice spacing range they explored ($a \sim 0.68-0.64$~fm), they find it to suffer for milder lattice artifacts.

\begin{figure}[!htb]
\centering
\includegraphics[scale=1.0]{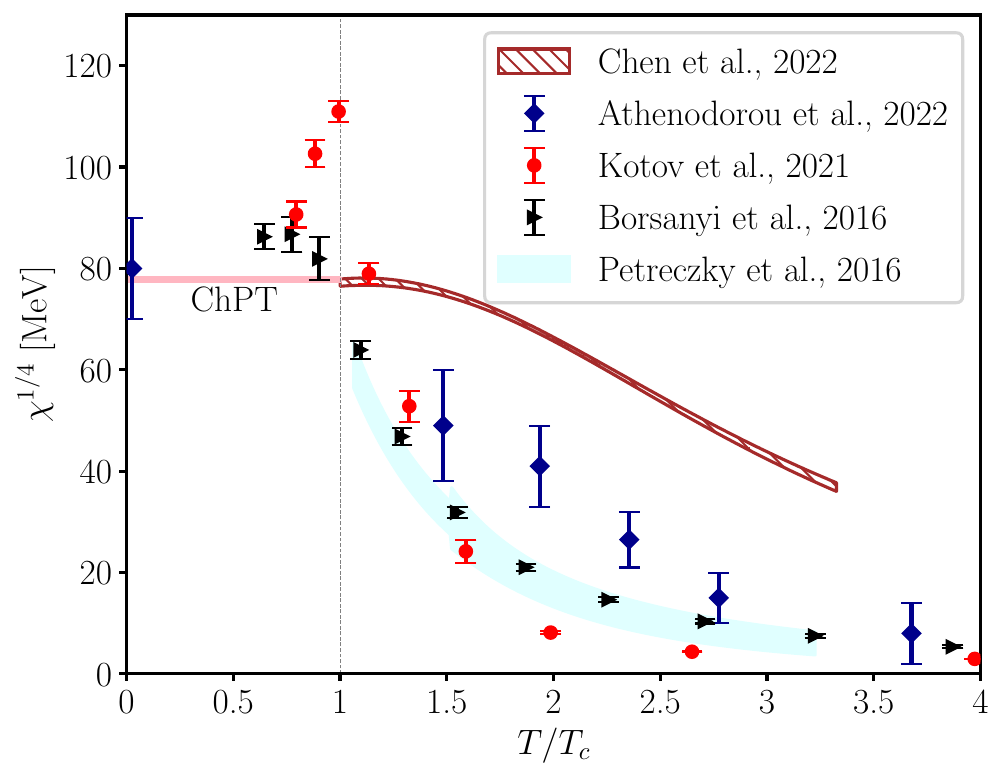}
\caption{Comparison of different determinations of the fourth root of the topological susceptibility $\chi^{1/4}$ in $N_f=2+1$ QCD from lattice simulations. Diamond points are taken from~\cite{Athenodorou:2022aay}, round points from Ref.~\cite{Kotov:2021ujj}, triangle points from Ref.~\cite{Borsanyi:2016ksw} (removing the isospin-breaking factor), the dashed area represents results of Ref.~\cite{Chen:2022fid}, while the shaded area represents results of Ref.~\cite{Petreczky:2016vrs} obtained from the chiral susceptibility for unphysical pion mass and rescaled according to the DIGA prediction $\chi^{1/4} \sim m_\pi$. For the crossover temperature the reference value $T_c=155$~MeV is assumed. For comparison we also report results obtained below the crossover, as well as the NLO two-flavor $T=0$ ChPT prediction for degenerate up-down quarks $\chi^{1/4}_{\mathrm{ChPT}}(T=0)=77.8(4)$~MeV.}
\label{fig:chi_vs_T_comp}
\end{figure}

These recent determinations with $N_f = 2+1$ flavors are displayed and compared in Fig.~\ref{fig:chi_vs_T_comp}. Roughly speaking, there is qualitatively a general common agreement that, for temperatures $T\gtrsim 300$~MeV (i.e., $T/T_c \gtrsim 2$) the behavior of $\chi(T)$ is compatible with a power-law as predicted by the DIGA, with a compatible exponent $\chi^{1/4}(T) \sim (T/T_c)^2$. However, results of Ref.~\cite{Borsanyi:2016ksw} find a very good agreement with the DIGA exponent already soon after the crossover, while Refs.~\cite{Lombardo:2020bvn,Petreczky:2016vrs,Athenodorou:2022aay} point out a change in the effective exponent for $T/T_c \gtrsim 2$.

Since in recent times several works gathered evidence for the existence of a phase of QCD close to the crossover and below $T\approx 300$~MeVs where non-perturbative effects are dominating~\cite{Alexandru:2019gdm,Alexandru:2021pap,Kotov:2021rah,Cardinali:2021mfh}, this aspect surely deserves to be further investigated in the near future, with dedicated studies aiming at addressing the behavior of QCD close to the crossover (see also Ref.~\cite{Lombardo:2020bvn}). These recent findings also match very well with other lattice studies pointing out the abundance of instanton-dyons in this temperature range above the crossover, see, e.g., Refs.~\cite{GarciaPerez:1999hs,Bornyakov:2013iva,Larsen:2018crg}.

Nonetheless, we can still fairly conclude that, while the one-instanton semiclassical computation is likely to be not reliable in the range currently reached by lattice simulations (for example, the authors of Ref.~\cite{Borsanyi:2016ksw} find that the DIGA prediction is about one order of magnitude smaller than their lattice determinations), the assumption of non-interacting instantons becomes reasonably reliable when considering temperatures $T\gtrsim 300$~MeV. This is also in agreement with results of Refs.~\cite{Bonati:2013tt,Borsanyi:2016ksw,Bonati:2015vqz,Bonati:2018blm}, where determinations of the quartic coefficient (related to the quartic axion auto-interaction term)
\begin{equation}
b_2 \equiv -\frac{1}{12} \frac{\langle Q^4\rangle - 3\langle Q^2\rangle^2}{\langle Q^2\rangle}
\end{equation}
for $T/T_c\gtrsim 2$ are in very good agreement with the well-known DIGA prediction $b_2^{(\mathrm{DIGA})}(T) = -1/12$, which only stems from the assumption of dilute instantons alone, and is unrelated to the semi-classical calculation needed in addition to the latter to compute $\chi_{\mathrm{DIGA}}(T)$ (see also Fig.~7 of Ref.~\cite{Lombardo:2020bvn} for a comparison of different high-temperature lattice determinations of $b_2$ in full QCD).

As for a quantitative agreement on the value of the topological susceptibility, no conclusive consensus on its exact behavior as a function of $T$ from different computations performed with different strategies has been reached yet, as it is manifest from Fig.~\ref{fig:chi_vs_T_comp}. Therefore, the determination of $\chi(T)$ in full QCD from the lattice above the crossover still poses a difficult yet stimulating challenge and certainly deserves to be further investigated in the near future.

\begin{figure}[!htb]
\centering
\includegraphics[scale=1.2]{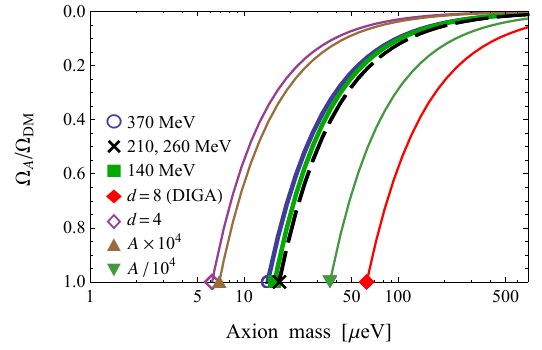}
\caption{Figure taken from Ref.~\cite{Kotov:2021ujj}. Dependence of the ratio $\Omega_{\mathrm{A}}/\Omega_{\mathrm{DM}}$ on the axion mass according to Eq.~\eqref{eq:omega_axion} for physical~\cite{Kotov:2021ujj} and unphysical~\cite{Burger:2018fvb} pion mass ensembles. For the $m_\pi\simeq 140~\text{MeV}$ ensemble, parameters $A$ and $d$ obtained from the best fit of data of Ref.~\cite{Kotov:2021ujj} to Eq.~\eqref{eq:DIGA_prediction_chi} are varied as written in the legend.}
\label{fig:axion_mass}
\end{figure}

However, it is interesting to observe that, although the underlined differences exist, their impact on axion mass windows is not so pronounced. Solving the axion equation of motion in the background of the Friedmann--Lema\^itre--Robertson--Walker metric, and using the simple DIGA parametrization for the $\theta$-dependence of the QCD free energy
\begin{equation}\label{eq:DIGA_theta-dep}
	f_{\mathrm{DIGA}}(\theta) = A \left(\frac{T}{T_c}\right)^{-d} \cos(\theta),
\end{equation}
it is possible to derive:
\begin{equation}\label{eq:omega_axion}
\frac{\Omega_{\mathrm{A}}}{\Omega_{\mathrm{DM}}} \simeq C m_a^{-\frac{3.053-d/2}{2.027-d/2}},
\end{equation}
where $\Omega_{\mathrm{A}}$ is the relic axion energy density, $\Omega_{\mathrm{DM}}$ is the observed Dark Matter energy density, and $C$ is a pre-factor depending weakly on the decay constant $d$ and mainly and on the pre-factor $A$ appearing in Eq.~\eqref{eq:DIGA_theta-dep}, as well as on the details of the axion model. More details on the derivation of Eq.~\eqref{eq:omega_axion} can be found, e.g., in Refs.~\cite{Lombardo:2020bvn,Burger:2018fvb,Kotov:2021ujj,Turner:PRD33889}. Using the parametrization~\eqref{eq:omega_axion}, it is possible to show that even by changing $d$ by a factor of $2$ or $A$ by four orders of magnitude, the axion mass predictions stay essentially in the same ballpark~\cite{Lombardo:2020bvn,Burger:2018fvb,Kotov:2021ujj}, cf.~also Fig.~\ref{fig:axion_mass}.

Nonetheless, further studies to clarify the high-temperature behavior of the QCD topological susceptibility would be welcome, and the current state of the art can be improved in several directions. For instance, it would be interesting to refine existing results closer to the crossover, where non-perturbative effects are more pronounced, and where most of the current tensions take place. It would also be intriguing to probe higher temperatures from lattice simulations, as most of the results obtained so far, with only the exception of Ref.~\cite{Borsanyi:2016ksw}, were limited to the $160~\text{MeV}\lesssim T \lesssim600$~MeV range. In particular, it would be extremely interesting to reach temperatures of the order of $\sim 1$~GeV or above (i.e., $T/T_c\sim 10$), which is also necessary to study axion cosmology.

However, probing such high temperatures is at present an extremely tough challenge from the numerical point of view, because of the topological freezing problem. At present, several promising proposals have appeared in the literature to deal with the topological slowing down in simpler models. As an example, in Ref.~\cite{Kanwar:2020xzo}, machine-learning techniques via the so-called \emph{Equivariant Flows} are employed to mitigate freezing in the $2\mathrm{D}$ $\mathrm{U}(1)$ gauge theory and extensions to more complex gauge theories are expected in the near future. Another proposal can be found in Ref.~\cite{Cossu:2021bgn} (see also~\citeTalk{Lucini_talk}), a strategy based on the adoption of a parallel tempering scheme on the inverse gauge coupling $\beta$ in combination with the \emph{Density of States} approach has been adopted in the pure $\mathrm{SU}(3)$ gauge theory to reduce the large auto-correlation times affecting the Monte Carlo evolution of $Q$. We also note the method adopted in Ref.~\cite{Borsanyi:2021gqg} to avoid topological freezing and the dominance of the $Q=0$ sector. That paper presents simulations at a single beta, using density of states as a function of a 
proxy topological charge. It also uses parallel tempering for the different charge values. Another promising solution is the \emph{parallel tempering on boundary conditions} proposed by M.~Hasenbusch for $2\mathrm{D}$ large-$N$ $\mathrm{CP}^{N-1}$ models~\cite{Hasenbusch:2017unr} and adopted both in the latter case~\cite{Bonanno:2022hmz,Berni:2019bch} and in $4\mathrm{D}$ large-$N$ $\mathrm{SU}(N)$ pure-gauge theories~\cite{Bonanno:2020hht,Bonanno:2022yjr} to mitigate the effects of the topological slowing down by reducing the auto-correlation times of $Q$ by up to several orders of magnitude.

\subsection{The Road Ahead}

Topology in QCD plays a central role in determining the non-perturbative properties of the theory, and has been further revived by the quest for QCD axions. Further discussions on topology will be included in the contributions to a dedicated series of workshops that are planned to be held in Europe in 2023 and following years. For instance, NA6 participants are involved in the new EU COST ``\emph{CosmicWISPers}'', involving both Claudio Bonanno and Maria Paola Lombardo. Moreover, topology will also be the main topic of the dedicated series of workshops ``\emph{Gauge Topology}'', co-led by Massimo D'Elia, that is held in ECT$^\star$ in Trento every two years. Finally, the QCD axion physics will also be covered during the ``\emph{Lattice Gauge Theory Contributions to New Physics Searches}'' workshop in Madrid, which includes Claudio Bonanno in the organizing committee. In the near future, we foresee a more robust limit on the QCD axion mass, as well as more in-depth studies of the axion potential. Such studies will also help in clarifying the role of topology in the Quark Gluon Plasma phase (see also the discussion in Sec.~\ref{sec:nature_and_phenomenology_of_the_quark-gluon_plasma}).

\section{Statistical Field Theory\footnote{Editors: Michele Caselle with  Andrea Pelissetto and Marianna Sorba}
\label{sec:statistical_field_theory}}

In this section, we discuss three topics that show how strong the connection between Quantum Field Theories 
and Condensed Matter/Statistical Mechanics Models is. These advanced topics were covered in \citeTalk{Pelissetto_talk,Sorba_talk,KitchingMorley_talk}.  Discussions on applications of universality classes for continuum transitions and conformal theories 
are in Sections~\ref{sec:thermal_phase_transitions_and_critical_points},
\ref{sec:conformal_phase}.

\subsection{Phase Diagram of Three-Dimensional Abelian-Higgs Models}
Three-dimensional Abelian gauge theories coupled to scalar matter
(Abelian-Higgs models)
have recently drawn significant attention, as they arise as low-energy
effective field theories describing unconventional states of matter 
with fractionalized quantum numbers occurring in two-dimensional 
strongly-correlated quantum systems. They are relevant for 
superconductors, superfluids, and quantum SU($N$)
antiferromagnets~\cite{RS-90, TIM-06, Kaul-12, KS-12, BMK-13,NCSOS-15, WNMXS-17,Sachdev-19}. 
In particular, they are expected to describe the
transition between the N\'eel and the valence-bond-solid state in
two-dimensional antiferromagnetic SU(2) quantum
systems~\cite{Sandvik-07, MK-08, JNCW-08, Sandvik-10, HSOMLWTK-13,CHDKPS-13, PDA-13, SGS-16}, which represents the paradigmatic model
for the so-called deconfined quantum criticality~\cite{SBSVF-04}. 
The behavior in the presence of massless fermionic excitations is also
equally relevant in, e.g., high-$T_c$ superconductors and spin liquids;
see Refs.~\cite{LN-92,KL-99,RW-01,SL-05,GS-06,Lee-08,CS-15,Sachdev-16}
and references therein.

In the lattice scalar 
Abelian-Higgs (AH) model the scalar fields are $N$-component
vectors $\phi_{ x}$ defined on the sites of a lattice. As for the 
gauge fields, two different formulations are possible. In the noncompact
model the gauge field is a real field 
$A_{{ x},\mu}$ defined on the sites of the 
lattice (for definiteness we consider cubic lattices, so that each link
is labelled by a lattice site ${ x}$ and a direction $\mu$) and the 
gauge group is the additive group of the real numbers $\mathbb R$. 
In the compact formulation, the gauge field is a complex phase 
$U_{{ x},\mu}$ and the gauge group is U(1). The action is 
$S = S_\phi + S_g$, where the matter-field part is 
\begin{equation}
S_\phi = - J \sum_{{\bm x}\mu} 
     \phi^*_{\bm x} \cdot \phi_{{\bm x} + \hat{\mu}}\, U_{{\bm x},\mu}^Q +
     \sum_{{\bm x}} V(|\phi_{\bm x}|),
\end{equation}
where $V(x)$ is a generic potential [most of the numerical work considered
fixed-length fields, corresponding to $V(x) = \delta(x-1)$]. 
Here $Q$ is the integer charge of the fields (it is only relevant in the 
compact case) and $U_{{\bm x},\mu} = \exp (i A_{{\bm x},\mu})$ in the
non-compact case. The gauge action $S_g$ is the standard Wilson action 
in the compact case; otherwise, we set
\begin{equation}
S_g = \kappa \sum_{{\bm x},\mu>\nu} 
   (\nabla_\mu A_{{\bm x},\nu} - \nabla_\nu A_{{\bm x},\mu})^2 ,
\end{equation}
where $\nabla_\mu f({\bm x}) = f({\bm x} + \hat{\mu}) - f({\bm x})$ 
is the lattice nearest-neighbor derivative.  The model is invariant 
under U(1)/$\mathbb R$  local and $SU(N)$ global transformations.

AH models have been extensively studied, see Refs.~
\cite{Pelissetto:2019thf,Pelissetto:2019iic,Pelissetto:2020yas,Bonati:2020jlm,Bonati:2020ssr,Bonati:2022oez} 
and references therein. The phase diagram turns out to depend in a 
nontrivial fashion on the compact/noncompact nature of the gauge interactions
and also the charge of the scalar fields (see Ref.~\cite{Fradkin:1978dv} 
for a discussion of the charge dependence of the phase diagram). 
Here we will summarize the behavior along two different transition lines 
where the scalar field condenses and the global $SU(N)$ 
symmetry of the theory is broken. 

For small values of $\kappa$, there is an order-disorder transition at a 
finite value $J_c(\kappa)$ of the scalar coupling $J$. Such a transition is 
always discontinuous, except for $N=2$. For $N=2$ the transition is
continuous, in the O(3) universality class, irrespective of $Q$ and 
of the nature of the gauge fields. These small-$\kappa$ transitions are 
an example of Landau-Ginzburg-Wilson (LGW) transitions. In this case,
an effective description is obtained by  
considering a gauge-invariant scalar field $\Psi$
that represents a coarse-grained version of the microscopic order parameter
that signals the onset of long-range order. 
The effective action is then the most general $\Psi^4$ theory that 
is invariant under the global symmetry group of the AH model.
At LGW transitions the gauge group and the nature of the 
gauge fields do not play any role. Gauge invariance is only relevant in 
defining the set of observables that show a critical behavior. 

For large values of $\kappa$, AH systems 
also undergo an order-disorder transition,
but in this case model details are relevant. For the compact model with 
$Q=1$, the transition has the same nature as for small $\kappa$: it 
is an LGW transition. On the other hand, in the noncompact case or in the 
compact case with charge-$Q$ fields, $Q\ge 2$, a critical transition is 
observed for $N\ge 7(2)$ \cite{Bonati:2020jlm,Bonati:2020ssr,Bonati:2022oez}.
This transition is 
associated with a stable fixed point of 
the renormalization-group flow of the 
continuum AH field theory.  The fixed point is charged---the renormalized
gauge coupling is nonvanishing at the fixed point--- signaling that 
gauge fields play a role in determining the critical behavior.

It is important to extend the present analysis to AH models with 
fermions, which are relevant to understand 
the finite-temperature QCD transition. The analysis pioneered by 
Pisarski and Wilczek \cite{Pisarski:1983ms}
effectively assumes that this transition is a 
LGW one in which only the global symmetry group is relevant. 
However, we cannot exclude {\em a priori} 
the existence of continuous transitions 
with critical gauge excitations as it occurs in the scalar AH model for 
$N\gtrsim 7$. Further work is clearly needed to settle this issue.

\subsection{Interfaces Near Criticality: Results from Field Theory}

In statistical systems exhibiting a phase transition, the coexistence of different phases at criticality naturally leads to the formation of an interface. On the other hand, in particle physics, the confinement of quarks into hadrons is effectively described in terms of a string spanning an interface over time. Since duality relates a lattice gauge theory to a spin model, the two problems turn out to be deeply connected and both are conveniently addressed using the Ising model as a base system.

We thus consider the three-dimensional Ising model in its broken $\mathbb{Z}_2$ symmetry phase below the critical temperature $T_c$, where an interface separating coexisting phases of opposite magnetization is easily induced by a suitable choice of the boundary conditions. The linear size $R$ of the interface must be much larger than the correlation length $\xi$, in order that the two distinct phases to emerge over bulk fluctuations. In \cite{Delfino:2019ohr} the phenomenon is studied for a slab geometry of size $L\times L \times R$ (with $L\to\infty$), in which the magnetization tends to the pure values $\pm M$ as $x\to \pm \infty$ hence creating an interface running between the lines $x=0$, $z=\pm R/2$. Conversely, only the half-volume $x\ge 0$ is considered in \cite{Delfino:2021orq} with the interface pinned along the boundary condition changing lines $z=\pm R/2$ on the impenetrable wall $x=0$. Working in the scaling limit slightly below $T_c$, universal properties emerge and both systems are described by a field theory that admits a particle description. As a consequence, the interface is depicted as the propagation of a string of particle modes and this provides insight on the interfacial tension (i.e. the free energy of the interface per unit area), which is found for both geometries to be related to the particle density along the string and to be fully consistent with an independent Monte Carlo estimation provided in \cite{Caselle:2007yc}. It is then possible to derive analytically the expectation value of any observable with the given boundary conditions $\langle \Phi(x,y,z)\rangle_{\pm}$, using the asymptotic $n$-particle states $|\bf{p}_1,...,\bf{p}_n\rangle$ of the bulk field theory as a basis on which generic excitations can be expanded. More specifically, the configurational averages are expressed in momentum space and the condition $R\gg \xi$ selects the low energy particle modes. The analytic results for the order parameter profile $\langle s(x,y,z)\rangle_{\pm}$ are on one side \cite{Delfino:2019ohr} explicitly confirmed by means of Monte Carlo simulations performed for different values of $R$ and $T\lesssim T_c$, in the total absence of adjustable parameters; on the other side \cite{Delfino:2021orq}, they allow for a simple probabilistic interpretation of the interface as a sharp separation between the two phases. Moreover, in the half-volume system \cite{Delfino:2021orq}, the particle formalism explains the transition from a fluctuating to a binding regime of the interface with respect to the wall, when their interaction becomes sufficiently attractive. Thanks to scattering theory, some key parameters characterizing the binding transition are computed and compared with numerical data coming from the phenomenological wetting theory.

As already anticipated, the study of interfaces in spin models can be useful even when dealing with lattice gauge theories. Concerning the three-dimensional Ising model, we know that it is mapped by duality into the three-dimensional Ising gauge model and, for instance, the interface free energy is analogous to the Wilson loop expectation value. We could hence expect our exact formulation of interfaces close to criticality to be valuable also in characterizing observables close to the critical point in the three-dimensional Ising gauge model. Otherwise, an effective description of the interface behavior can be adopted, resulting in capillary wave theory for the spin model and effective string theory for the corresponding lattice gauge model. In particular, the capillary wave model corresponds (see \cite{Caselle:1994df,Billo:2006zg} for a discussion of this point) to the well known Nambu-Goto effective string theory
\cite{Nambu:1974zg,Goto:1971ce}, which has been shown in the last few years to give a very precise description of  Wilson loops thanks to the so-called low energy universality theorem (see \cite{Aharony:2013ipa,Caselle:2021eir} for a review).

Finally let us mention that, besides the one discussed above, it is also 
possible to give a consistent field theoretic description of interfaces 
with a Landau-Ginzburg type action for the order parameter of the 
model~\cite{K_pf_2008}. In this framework it is possible to obtain 
analytic expressions for the interface profile and width~\cite{K_pf_2008} 
which are fully consistent with those obtained with the capillary wave 
model but do not require the ad hoc cut-offs terms which must be 
introduced in the standard capillary wave model.

\subsection{Infrared Finiteness of Three-Dimensional Super-Renormalisable QFTs}
Three-dimensional super-renormalizable scalar QFTs with fields in the adjoint 
representation of the $\SU(N)$ group attracted lot of interest in the past as 
effective (dimensionally reduced) theories describing the high-temperature limit of four-dimensional Yang-Mills
theories (see, for example, \cite{Appelquist:1981vg,Farakos:1994kx,Kajantie:1995kf,Caselle:1992ic,Billo:1996wv,Billo:1996pu}).
More recently these theories attracted a renewed interest as candidate
holographic models for the very early universe \cite{McFadden:2009fg}.

A relevant open problem in this context is represented by the fact that
massless super-renormalisable
quantum field theories suffer from severe infrared
(IR) divergences in perturbation theory: the same
power counting argument that implies good ultraviolet
(UV) behavior also implies bad IR behavior. 
These IR singularities
were discussed several years ago in \cite{Jackiw:1980kv,Appelquist:1981vg} where it was conjectured that such
theories are nonperturbatively IR finite. The lattice regularization offers a perfect setting to address this issue, which was recently discussed in \cite{Cossu:2020yeg} in the particular case of scalar QFTs with a
$\phi^4$ interaction and fields in the adjoint representation of the $\SU(N)$ group with
$N = 2, 4$.

 When studied in lattice perturbation theory,
these theories exhibit a logarithmic IR divergence for the
critical mass at the two-loop level. However, this divergence was not present in the lattice simulations of~\cite{Cossu:2020yeg}
thus providing  strong evidence for the IR-finiteness of the full
theory. From the lattice results it was also possible to obtain 
a nonperturbative determination of the critical masses which turns out to agree with 2-loop 
perturbation theory, and a determination of the critical exponent which turns out to be close to the leading-order
effective theory prediction~\cite{Cossu:2020yeg}.

These results open the way to a better understanding of the infrared properties of this class of models which could have remarkable implications both for the high $T$ description of Yang--Mills theories and for candidate holographic models of the early Universe.

\subsection{The Road Ahead}
The three examples discussed in this section show how powerful the Statistical Field Theory approach can be when combined with simulations.
This is particularly true for critical systems and more generally for systems in the neighborhood of a phase transition which are the main focus of  this report. Statistical Field Theory allows to propose effective description for the systems of interest (LGW models for the phase diagram of Abelian Higgs models, effective strings for the interfaces, holografic models for the early universe...) which can then be tested and refined with Monte Carlo simulations.  With the improvement of computing power and algorithms (see the next section for a discussion of the remarkable performances of new, machine learning based, algorithms), this virtuous circle between effective theories and simulations will lead to more and more refined models also for QCD related phase transitions and, what is more important, to a more precise description of the relevant degrees of freedom in this context.

\section{Machine Learning\footnote{Editor: Andreas Athenodorou with Gert Aarts, Biagio Lucini and Dimitrios Bachtis}} 
\label{sec:machine_learning}

\subsection{Introduction}
\label{sec:machine_learning_introduction}

Recent advances in the implementation of Machine Learning (ML) techniques for physical systems, especially those which can be formulated on lattices, appear to be suitable for observing the corresponding underlying phase structure of the aforementioned systems~\cite{Carrasquilla2017,vanNieuwenburg2017,PhysRevB.94.195105,Broecker2017,ch2017machine,PhysRevE.96.022140,PhysRevE.95.062122,2017arXiv170700663B,PhysRevB.97.205110,PhysRevLett.120.257204,2018arXiv180402709Z,2018JChPh.149s4109J,PhysRevE.98.022138,2018arXiv180801731Z,Kashiwa:2018jdi,Giannetti:2018vif,Zhou:2018ill,Blucher:2020mjt}. This, was firstly observed in the novel work by J.~Carrasquilla and  R.~G. Melko in 2017~\cite{carrasquilla2017machine} where they used supervised machine learning architectures with fully connected and convolutional neural networks to identify phases and phase transitions in a variety of condensed-matter Hamiltonians. For instance, they can estimate to an adequate precision the critical exponents as well as the critical temperature for the 2D ferromagnetic Ising model. The above advance was followed by a plethora of investigations using Principal Component Analysis (PCA)~\cite{PhysRevB.94.195105,vanNieuwenburg2017,PhysRevE.95.062122,wetzel2017unsupervised,Foreman:2018ktj}, Supervised Machine Learning (ML)~\cite{Broecker2017,2018arXiv180402709Z,morningstar2017deep}, Restricted Boltzmann Machines (RBMs)~\cite{Cossu:2018pxj,Funai:2018esm}, as well as autoencoders~\cite{PhysRevE.95.062122,PhysRevE.96.022140} which appear to successfully identify different phase regions of classical statistical system. Since Quantum Field Theories can be represented in the form of statistical systems it would be reasonable to expect that such methods could apply in Quantum Field Theories. So far only a few investigations dealt with the phase structure of Quantum Field Theories on the lattice. These works will be reviewed throughout this next chapter.

\subsection{Phase Transition Recognition in $\SU(N)$ Gauge Theories and QCD}
\label{sec:machine_learning_suN}

\begin{figure}
	\centering
	\includegraphics[width=0.7\textwidth]{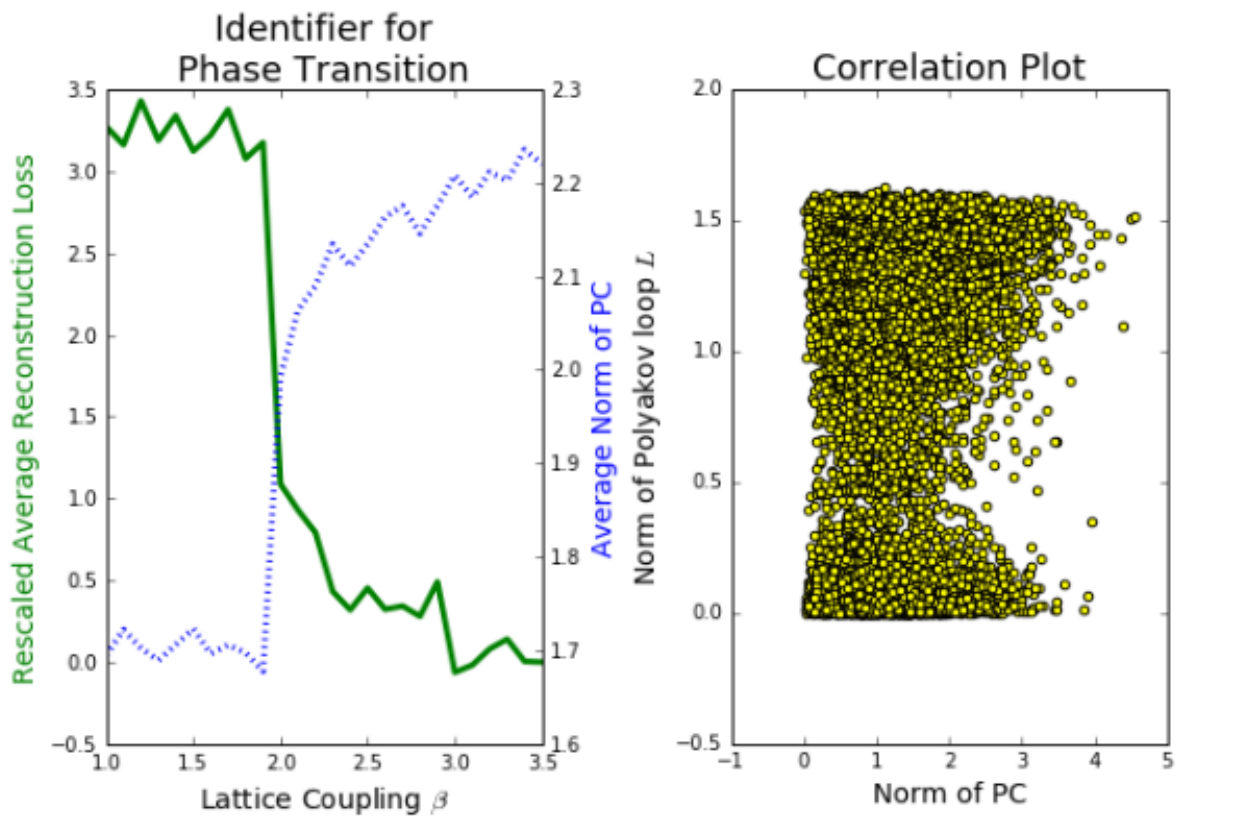}
	\caption{\label{fig:PCAResults}(Ref.~\cite{Wetzel:2017ooo})  Finding a possible phase transition with PCA. {\underline{Left Panel}}: The average mean squared error reconstruction loss as a function of temperature is a universal identifier for a phase transition. This was calculated in 100 independent PCA runs with two principal components (PC), measured in units of $\times 10^{-5}$ and shifted by the value at $\beta =3.5$. The average norm of the PC
also indicates a phase transition. {\underline{ Right Panel}}: A plot indicating that there is no correlation between the principal components and the Polyakov loop.}
\end{figure}

The first investigation which has provided a successful identification of the confining-deconfining transition in $SU(2)$ gauge theory using Machine Learning techniques has been reported in Ref.~\cite{Wetzel:2017ooo}. Namely, the authors using Principal Component Analysis (PCA) on configurations produced for a range of values of $\beta$, demonstrated that even though the $SU(2)$ order parameter Polyakov loop is non-linear, PCA captures indications of a phase transition at the range of $\beta$ $\in$ [1.8, 2.2]. This has been achieved by probing the ``average mean squared error reconstruction loss'' as well as the ``average norm of the PC''. Surprisingly, at the same time they demonstrated that there is no correlation between the Polyakov loop and the principal components. The above is demonstrated in Fig.~\ref{fig:PCAResults}. Bear in mind that $SU(2)$ link matrices can be mapped to four real numbers multiplying the $3$ Pauli matrices and the unity.
Subsequently, the authors turned to the investigation of the phase structure using the Correlation Probing Neural Network. The Correlation Probing Neural Network consists of three types of neural networks stacked on top of each other. The localization network is a fully convolutional neural network which prohibits connections outside of the receptive field of each output neuron. The averaging layer averages over the input from the localization network. The prediction network is a fully connected neural network, which transforms the output of the averaging layer to a prediction probability. The authors trained the correlation probing neural network in a supervised manner on $SU(2)$ Monte Carlo-sampled configurations at lattice couplings $\beta$ $\in$ [1, 1.2] in the deconfining phase and $\beta$ $\in$ [3.3, 3.5] in the confining phase. Then, they tested the neural network for values of lattice coupling $\beta$ $\in$ [1.3, 3.2] and they predicted a phase transition at $\beta = 1.99 \pm 0.10$ for lattice of $T \times L_x \times L_y \times L_z = 2 \times  1 \times 1 \times 1$ and $\beta = 1.97 \pm 0.10$ for a lattice of $2 \times 8 \times 8 \times 8$ while a conventional lattice calculation gives a critical value of $\beta = 1.880 \pm 0.025$. This result has been obtained by probing the average prediction probability.

Finally, the authors move to the more conclusive part of their investigation where they trained a new neural network on the local data samples in order to classify the phases of each local sample. This enables the local neural network to associate a prediction to each patch. The authors performed polynomial regression on the latent prediction of the local neural network. By extracting the weights of the regression they demonstrated that {\it the parameter which quantifies the phase transition is nothing else but the Polyakov loop} on a single spatial lattice site! Subsequently, by acting on the full lattice the decision function takes the form of the Polyakov loop as the argument on a Sigmoid function on the full lattice. This is clear evidence that supervised machine learning can predict the correct order parameter of the theory. 

The work of Ref.~\cite{Wetzel:2017ooo} was followed by the investigation of the phase structure of $SU(2)$ and $SU(3)$~in Ref.~\cite{Boyda:2020nfh}. The authors have developed a supervised Machine Learning network capable of identifying the order parameter of the theory, namely the Polyakov loop. The architecture of the neural network they used can be summarised in the next couple of lines. $SU(2)$ is parametrized by four real numbers while for $SU(3)$ the authors used the full set of 9 complex numbers.  Consider a gauge field with $\left[ N_t, N_s, N_s, N_s, {\rm Dim}, {\vec v}_{\rm SU(N)}  \right]$, where $N_t$, $N_s=N_x=N_y=N_z$ the temporal and spatial lattice extents, Dim the direction $\mu$ of the matrix $U_{\mu}(x)$ at every lattice site $[N_t,
N_s, N_s, N_s]$ and ${\vec v}_{\rm SU(N)}$ the vector of the representation with size 4 and 9 for $SU(2)$ and $SU(3)$ respectively.
The architecture of the neural network for the prediction of the Polyakov loop in the $\SU(N)$ gauge theory is expressed as first for $N_t=2$: Inputlayer: $(N_t, N_s \times N_s, N_s, {\rm Dim} × U)$  $\to$ Convolutional3D $(N_t, N_s \times N_s, N_s, {\rm Dim} × U)$  $\to$ AveragePooling3D  $(N_t, N_s \times N_s, N_s, 16)$  $\to$ Flatten $(1,1,1,16)$ $\to$ Dense $(16)$ $\to$ 1 while for $N_t=4$: Inputlayer: $(N_t, N_s \times N_s, N_s, {\rm Dim} × U)$  $\to$ Convolutional3D $(N_t, N_s \times N_s, N_s, {\rm Dim} × U)$ $\to$  Convolutional3D $(2, N_s \times N_s, N_s, 256)$  $\to$ AveragePooling3D  $(1, 1, 1, 32)$  $\to$ Flatten $(1,1,1,16)$ $\to$ Dense $(32)$ $\to$ 1.

For training purposes, the authors generated 9000 lattice configurations at the one value of $\beta$ of the lattice coupling, $\beta=4$ for $SU(2)$ and $\beta=10$ for $SU(3)$, for lattices with the spatial sizes $N_s = 8, 16, 32$ and the temporal sizes $N_t = 2, 4$. In addition for prediction purposes, they also generated 100 configurations for a number of points at lower values of the coupling $\beta$, which the neural network does not use for training but rather for prediction. 

The network is being trained on the lattice configurations generated in the (volume-induced) deconfinement phase at a point of $\beta$ which is far from the phase transition point. The neural network is trained to predict correctly the value of the Polyakov loop that is already known from the Monte Carlo simulations. The training is done in batches of 10 - 50 configurations. The authors used the mean squared error (MSE) as a loss function.

\begin{figure*}[!htb]
\begin{tabular}{ccc}
\includegraphics[width=0.3\linewidth]{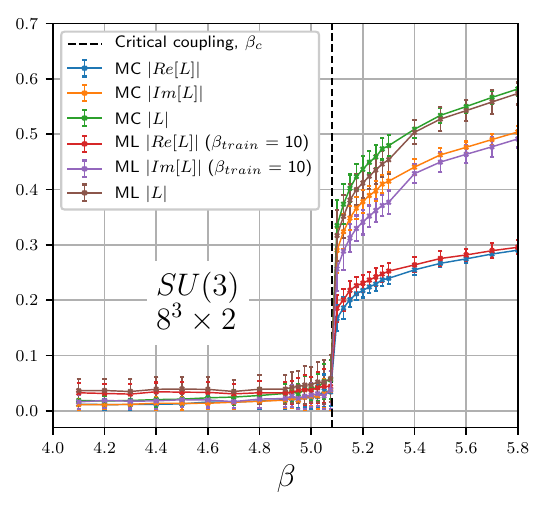} &
\includegraphics[width=0.3\linewidth]{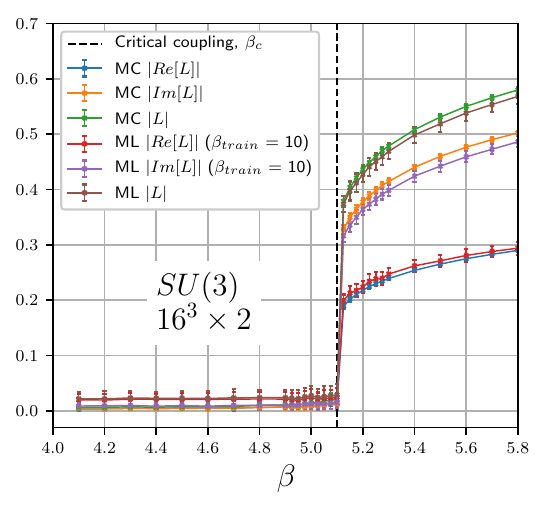} &
\includegraphics[width=0.3\linewidth]{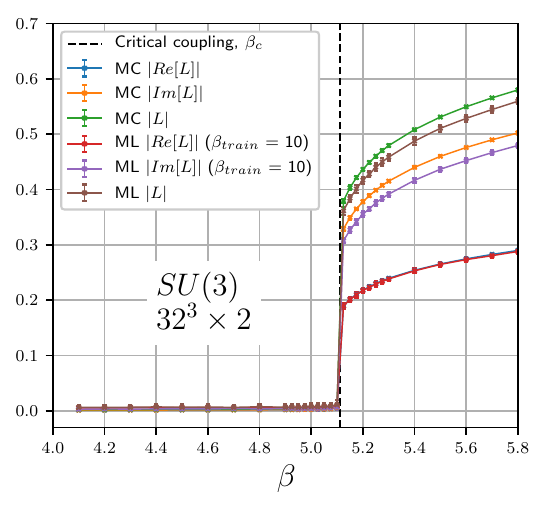} \\[-2mm]
\end{tabular}
\caption{\label{fig:SU3_polyakov}(Ref.~\cite{Boyda:2020nfh}) The results for the Polyakov loop for SU(3) gauge theory at $N_t = 2$ obtained with the Monte Carlo simulations as compared to the neural network prediction. The absolute value, the real and imaginary parts of the loop are shown. The value of the Machine Learning extracted approximation of Polyakov loops $ML\ |L|$ restored from ML predictions of $|Re[L]|$ and $|Im[L]|$ are also presented.}
\end{figure*}

The overall findings resulted out of this investigation demonstrate that the neural network which was trained on a value of $\beta$ located deep in the deconfinement region reproduces the Polyakov loop with an adequate agreement with Monte-Carlo data at all other values of the lattice
coupling constant including the region of the true deconfinement transition. This can be viewed in Fig.~\ref{fig:SU3_polyakov} where results for $SU(3)$ at $N_t=2$ are presented. These data illustrate the agreement between the Polyakov loop and its approximation using Machine Learning. Hence, what the authors demonstrated is that the neural network serves as a successful predictor of the confining-deconfining phase transition obtained by reconstructing the gauge-invariant order parameter in the whole physical region of the $\beta$-parameter space after one performs training on lattice configurations at one unphysical point in this space. It would have been useful to extend this work from the second order phase structure of $SU(2)$ and the weakly first order phase structure of $SU(3)$ to $SU(N > 3)$ where the phase structure is strongly first order.

Recently, the authors of the Lattice 2021 Proceedings~\cite{Palermo:2021jrf} published results on the investigation of the critical temperature on pure $SU(3)$ gauge theory as well as in $N_f=2+1+1$ QCD using Machine Learning techniques.

Instead of probing the actual $SU(3)$ configurations which can be reduced down to eight real numbers, per lattice point, and per Euclidean direction, the authors extracted the temporal Polyakov loop for each time slice. The above set-up corresponds effectively to a 3-dimensional system where the basic degrees of freedom are the values of the Polyakov loop at each point of the effective 3D grid.

To classify configurations of Polyakov loops at different temperatures, the authors built a $3D$-convolutional autoencoder using TensorFlow~\cite{tensorflow2015-whitepaper} and Keras~\cite{chollet2015keras}. The autoencoder is trained, as a whole, to reproduce as output its own input. When this is achieved, the encoded classifier effectively encodes the most important feature(s) describing the variety of the input. The authors simplified the process, by performing a semi-supervised training by pinning some of the input configurations at extreme temperatures to predefined values of the encoded classifier. 

For the pure $SU(3)$ gauge theory the authors used configurations of size $T \times N_x \times N_y \times N_z = 4 \times 8 \times 8 \times 8$ produced using the MILC code. For this choice of geometry and action, the pseudocritical coupling is $\beta_C=5.69(2)$ giving a critical temperature of $T_C \sim$ 260 MeV. The authors analyzed 30 configurations for each temperature. The configurations span a wide range of the coupling parameter $\beta$ from strong to very weak coupling. By training the autoencoder as an unsupervised and semi-supervised classification problem the authors obtained an encoded classifier clearly related to the order parameter which in this case is the Polyakov loop. As a matter of fact, two phase sectors are identified by the encoded classifier, one below $T_C$ and one above. Namely, above $T_C$ the unsupervised scheme
highlights the $Z_3$ symmetry breaking with three different values of the encoded classifier being equally probable while below $T_c$ there is only one possibility with the encoded classifier being zero. When it comes to the semi-supervised learning problem, the authors pinned a fraction of $\sim 20 \%$ of the training configurations at the lowest and highest values of $T$ by assigning an encoded classifier of $0$ and $1$ for confining and de-confining phase respectively. The network appears to successfully recognize the phase transition at $T \simeq T_C$. 

Regarding full QCD with $N_f=2+1+1$ fermions, the configurations have been produced with 
Wilson fermions at maximal twist on a lattice of $32^3$ spatial volume with the strange and charm masses having their physical values while the pion mass being $M_{\rm \pi} \sim 370$ MeV. The pseudocritical temperature is $M_{\rm \pi} \sim 200$ MeV. For each
temperature, the authors used 200 Polyakov loops configurations. For the case of QCD, the Polyakov loop is no longer an order parameter and the identification of
a phase transition based on the Polyakov loop is not theoretically justified as before. The authors studied the semi-supervised problem by assigning to the configurations at low temperatures an encoded classifier of 1 and at higher temperatures of 0. The resulting encoded classifier turns out to be a much smoother function compared to the one for the pure gauge theory. This is somehow expected since the phase transition for the QCD configurations is known to be a crossover. From the encoded classifier the authors could identify two classes separated at temperature $T \sim 1.5 T_C$. 

The authors have successfully used the Convolutional neural networks trained as either unsupervised or semi-supervised classifiers to identify different phases of gauge theories in both pure gauge as well as full QCD. Further work needs to be carried out, namely by moving to a finer temperature scan, a finite-size scaling and a continuum limit. This will improve the performance of the autoencoder. Finally, this will hopefully provide further insight into Machine-Learning approaches to the study of phase transitions.

Recently, in Ref.~\cite{Karsch:2022yka} the authors used Normalizing Flows instead of the traditional rewriting in $\beta$ in order to interpolate the chiral condensate obtained from QCD simulations with five degenerate quarks. Namely, the authors performed calculations in five-flavor QCD $(N_f=5)$ using the HISQ action with quark masses in the range $0.001 \leq m_l \leq 0.016$ and gauge couplings $\beta=4.5-5.4$. They used 4-dimensional lattices with volume $N_s^3 N_t$, with temporal extent $N_t=6$ and spatial volumes $N_s^3 = 16^3, 24^3$. Subsequently, they performed the classical reweighing in $\beta$ to provide an interpolation of the chiral condensate in $\beta$.

Lattice QCD calculations typically are done at a few values of the gauge coupling beta and reweighting in beta is a popular method for interpolating lattice results. This method requires a large number of measurements, performed at a large number of beta values since the observable we are interested in is extracted via the 2D histogram of the action and the observable. As explained before the observable under investigation is the chiral condensate and the interpolation is done in the direction of beta. Applying reweighting reveals reasonable results for small masses $(0.002  \le m_l \le 0.005)$, but exhibits over-fitting for larger values of masses $(m_l=0.006, 0.008)$. 

Followingly, authors turned to normalizing flows which are state-of-the-art ML tools for modeling probability distributions in physical systems. They made use of MAF (Masked Autoregressive Flow)~\cite{papamakarios2017masked} model with eight MADE (Masked Autoencoder for Distribution Estimation)~\cite{germain2015made} blocks. Compared to classical reweighting, this method has the advantage of allowing to interpolate in any parameter. As a matter of fact, in this process there is no need for overlapping distributions of the action density and the method is able to process continuous data. The cost to pay in order to visualize the learned probability distribution is the fact that one needs to draw a large number of samples from the model to fill a two-dimensional histogram. In practice, the model learns to transform a 2D-Gaussian distribution to expectations of the chiral condensate and action $({\bar \psi} \psi, S)$ conditioned on the parameters  ($N_s$, $m_l$, $\beta$). The evaluation of the model was performed for All the integer values of $N_s \in [16, 24]$, $\beta \in [4.5, 5.4]$ in steps of $0.001$ and $m_l \in [0.001, 0.006]$ in steps of 0.001 and for the larger masses $m_l \in [0.008, 0.016]$ in steps of 0.002.  This allowed the authors to fit the entire data set with a single function  $p( {\bar \psi} \psi, S | N_s, m_l, \beta)$ in contrast to the $\beta$-reweighting according to which one needs to do independent reweighting for each mass and volume. A comparison to the reweighting method reveals that the normalizing flow results appear to give a better fit since now the data points support each other also in $m_l$ and $N_s$-directions and not just in $\beta$. As a result, the method of normalizing flows removes over-fitting appearing in the traditional $\beta$-reweighting. 

A glimpse at the 2D histogram in the ${\bar \psi} \psi-S$ plane as well as at the 1D histogram in ${\bar \psi} \psi$ reveal two phases in the small quark mass regime while only one at the large mass regime which manifests as two and one peaks respectively. The double peaks signal the occurrence of a first order phase transition. To determine the quark mass dependence of the double peaks the authors turned to the phase diagram $\langle {\bar \psi} \psi \rangle$ in the $m_l$-$\beta$ plane. This enabled them to demonstrate evidence that the first order region ends in a second order endpoint at about $m_l^c \simeq 0.0045$. As the gap between the peaks
at low and high values $\beta$ becomes smaller, larger lattices will
be needed to resolve these two peaks and establish a gap between them. To locate the endpoint the authors used another ML based approach called the EOS-meter.

According to the Equation-of-State (EOS) meter, one can use convolutional neural network model to create density plots~\cite{Pang:2016vdc}. The authors used a recent approach called the transformer model~\cite{vaswani2017attention}, which is solely based on attention mechanisms and has been shown to outperform convolutional neural networks in translation tasks. The density plots revealed, at the smallest masses, a two peak behavior with a clear gap which is characterized as ``first-order'' while in the largest masses one peak behavior has been spotted characterized as ``crossover''. ``Firstorderness'' and ``crossoverness'' were implemented as categories in one-hot-encoding. The resulting EOS-meter shows that the critical masses marking the borders between first order and crossover regions, extracted via logistic fits
to the ``firstorderness'', indicate a critical mass at $m_c \simeq 0.005(1)$.

As a further investigation, one should move to the extraction of the phase diagram of QCD with $N_f$ flavors in the continuum. For this purpose, one should use larger values of $N_t$. One can also extend the set of interpolating parameters to $(N_f, N_s, N_t, m_l, \beta)$, however, this would require a large amount of training data.

\subsection{Phase Transition Recognition in Other Theories}
\label{sec:presentations}

We now turn to additional investigations which are focussing mostly on simpler theories such as the $\phi^4$ scalar field theory as well as the Ising model.

First, we present the work of Refs.~\citeTalk{Aarts_talk} and~\cite{Bachtis:2021xoh}, where the authors discussed the adoption of Euclidean quantum field theories in machine learning algorithms, which makes inference and learning possible using quantum field dynamics. To do so, it was first demonstrated that the $\phi^4$ scalar field theory satisfies the Hammersley--Clifford theorem. As a result, the quantum field theory can be recast as a machine learning algorithm within the mathematically rigorous framework of Markov random fields. Various applications are then possible. For a fixed target distribution, the parameters of the best approximating $\phi^4$ model are obtained by minimizing the Kullbach--Leibler divergence (which is an asymmetric distance) between the two. In practical applications, the effectiveness of the minimization is an indicator of the goodness of the approximation. Through re-weighting, the analysis can be extended to complex-valued actions with longer-range interactions. Moreover, neural network architectures derived from the $\phi^4$ theory can be viewed as generalizations of conventional neural networks. It is noted that the aims of this work are two-fold: the approach can provide a new perspective on machine learning with continuous degrees of freedom using the language of quantum fields, while also providing a new look at quantum fields when employed as building blocks in neural networks.

\begin{figure}[!ht]
\vspace{0cm}
\begin{center}
    \rotatebox{0}{\includegraphics[width=10cm]{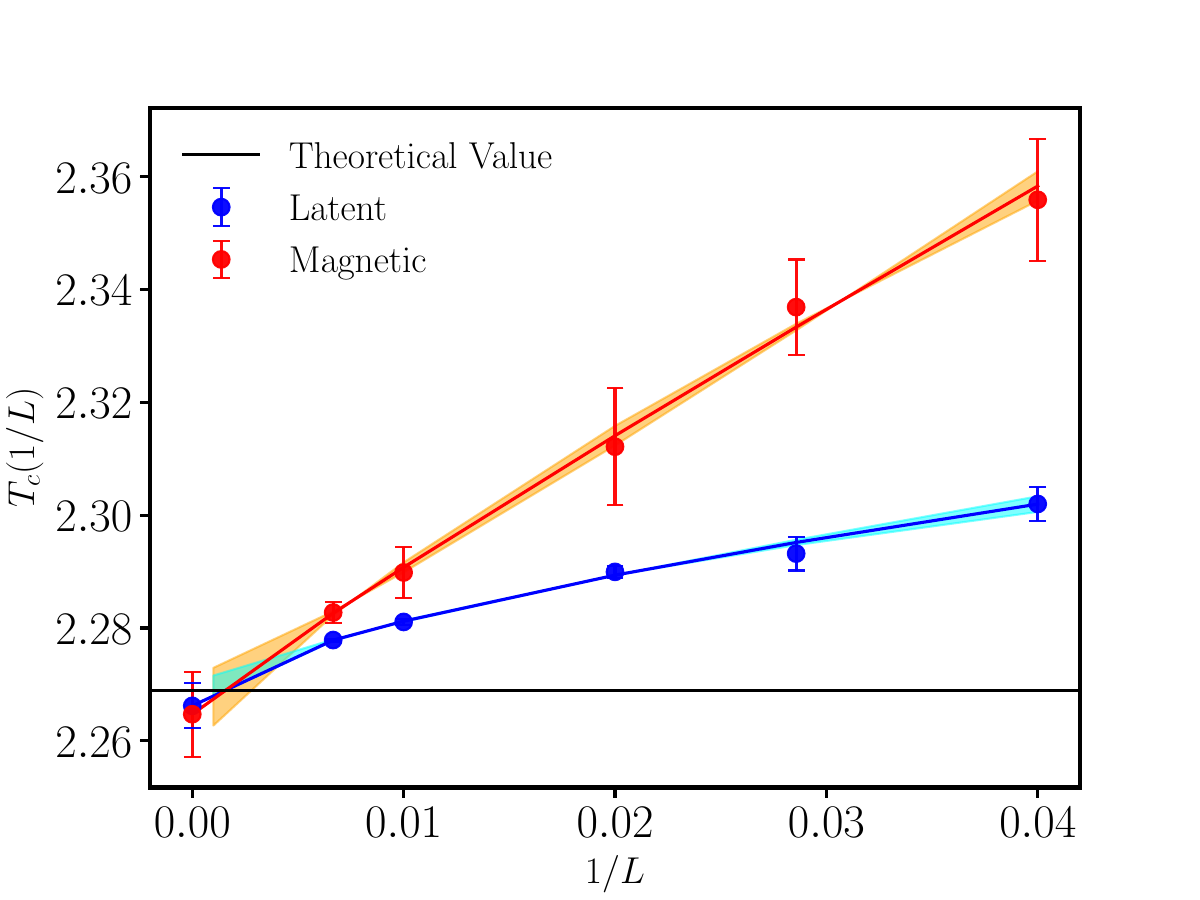}}
\end{center}
\caption{Ref.~\cite{Alexandrou:2019hgt} The critical temperature $T_c(L)$ for the 2D Ferromagnetic Ising model, extracted from fitting  the  magnetic (red) and  the latent (blue) susceptibilities as a function of $1/L$ according to $T_c(L) - T_c(L = \infty) \propto L^{-1/\nu}$. The error bands are estimated using the jackknife fit errors on the fit parameters.}
\label{fig:critical}
\end{figure}

Subsequently, we present the work of Refs.~\citeTalk{Athenodorou_talk} and~\cite{Alexandrou:2019hgt} which discusses deep learning autoencoders for the unsupervised recognition of phase transitions in physical systems formulated on a lattice. Their work elaborates on the applicability and limitations of this deep learning model in terms of extracting the relevant physics. Their results are presented in the context of $2D$, $3D$, and $4D$ Ising models as well as the $XY$ model, and the focus is on the analysis of the critical quantities at $2D$ (anti)ferromagnetic Ising Model. The authors defined it as a quasi-order parameter, the absolute average latent variable, which enabled them to predict the critical temperature to adequate precision. In this way one can define a latent susceptibility from the latent variable and use it to quantify the value of the critical temperature $T_c(1/L)$ at different lattice sizes and that these values suffer from smaller finite scaling effects compared to what one obtains from the magnetic susceptibility. This feature is demonstrated in Fig.~\ref{fig:critical} where the critical temperature extracted from the magnetic as well as the latent susceptibilities are extrapolated to the thermodynamic limit both converging to Onsager's solution. Clearly, $T_c(1/L)$ extracted using latent susceptibility converges much faster to the theoretical prediction as a result of the smaller finite volume effects. Hence, the deep learning autoencoder could potentially provide a tool that can enable the extraction of physical parameters with much better accuracy that the traditional ways.

Finally, we briefly present the work which can be found in~\citeTalk{Bachtis_talk}. This project demonstrates that the combination of renormalization group methods and machine learning algorithms opens up the opportunity to overcome fundamental problems in computational studies of phase transitions, such as the critical slowing-down effect. In this work, the authors discuss applications of machine learning for phase transitions and present a construction of inverse renormalization group transformations that enables the generation of configurations for increasing lattice volumes in the absence of the critical slowing-down effect. Results are presented for the two-dimensional Ising model and the $\phi^{4}$ theory. 

Specifically, the authors demonstrate that the inclusion of a neural network function within the Hamiltonian of the two-dimensional Ising model is able to induce a phase transition by breaking or restoring its symmetry~\cite{Bachtis:2020fly}. Another topic of discussion concerns the implementation of a machine learning approach, based on a set of transposed convolutions, to invert a standard renormalization group transformation in the case of the $\phi^{4}$ theory~\cite{Bachtis:2021eww}. The inverse transformations are then applied consecutively to iteratively increase the volume of the system, without experiencing the critical slowing-down effect. Both methods result in accurate calculations of multiple critical exponents for the aforementioned systems using renormalization group techniques based on matching observables on lattices of different sizes. These methodological advances rely on the observation that machine learning quantities can be interpreted as statistical-mechanical observables. Consequently, the opportunity to apply histogram reweighting to extrapolate them in parameter space is additionally explored~\cite{Bachtis:2020dmf}. Finally, an application of a machine learning technique called {\em transfer learning} is discussed, which indicates similarities in order-disorder phase transitions~\cite{Bachtis:2020ajb}. These similarities extend beyond the notions of symmetry and dimensionality which generally characterize the concept of universality. 

\subsection{The Road Ahead}

In summary, machine learning implementations, when combined with renormalization group approaches, are capable of providing significant computational benefits and novel physical insights into studies of phase transitions. These include the evasion of the critical slowing down effect with the inverse renormalization group, and the inclusion of neural networks within Hamiltonians to induce symmetry-breaking phase transitions in systems. As a result, one envisages the benefits of extending the methods discussed here to more complicated and physically relevant systems, such as lattice gauge theories.

A future workshop {\em Machine Learning approaches in Lattice QCD - An interdisciplinary exchange}~ organized by Nora Brambilla and others at the Institute for Advanced Study of the Technische Universität München will further investigate this topic. The work {\em Density of States approach} conducted by Biagio Lucini in~\citeTalk{Lucini_talk} should also be further discussed. For a recent flow-based density of states application to complex actions see~\cite{Pawlowski:2022rdn}. 

\section{Parting Remarks}\label{sec:conclusions}

We have inserted a few comments at the end of each Section, and we would not reiterate them here. 

We just summarize that this work highlights, and motivates further, interactions with experimentalists and phenomenologists active in
relativistic heavy ion collisions; and with the nuclear astrophysics community, towards the calculation of the equation of state of dense matter and its impact on gravitational waves analysis.  Away from these core hadron physics fields,
relevant directions include physics beyond the standard model, in particular those aspects related to a strongly coupled Higgs Sector, and the broad field of axions and dark matter. The methodological obstacles related to the sign problem would clearly benefit from closer exchanges with mathematicians and computer scientists.

From the point of view of computational techniques, it is worth remarking that new developments in the rapidly expanding field of quantum computing could lead to major scientific breakthroughs in the coming decades. As already envisioned by Richard~P.~Feynman in his 1982 work~\cite{Feynman:1981tf} and remarked in Ref.~\citeTalk{Wiese_talk} (see also Ref.~\cite{Wiese:2021djl}), the use of intrinsically quantum computing devices to simulate the quantum field theories describing the elementary constituents of the physical world could have disruptive scientific potential. In particular, it may open the path to solving some of the most challenging problems, including the study of real-time dynamics of strongly coupled theories, the derivation of the properties of systems at finite fermionic densities, and the strong $\mathrm{CP}$ problem.

More generally, the computational aspects remain of crucial relevance for the research topics covered in this review, at the time of the transition between PRACE and EuroHPC~\cite{EuroHPC}, and the progress towards Exascale computing~\citeTalk{Ryan_talk}. These issues are also relevant for Open Science policies in the LFT community~\cite{Athenodorou:2022ixd},

The lattice community may provide crucial input to this discussion, and in return greatly benefit from the new developments.

\section*{Acknowledgements}

\begin{figure}[!htb]
\centering
\includegraphics[width=6.5cm]{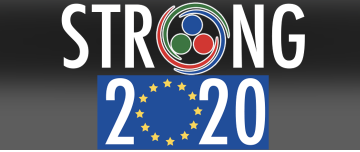}
\\
\includegraphics[width=9cm]{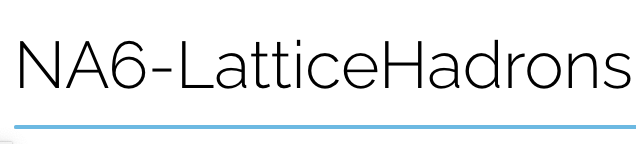}
\end{figure}

This review has been prepared within the framework of
the LatticeHadrons Network of \href{http://www.strong-2020.eu/}{STRONG-2020} ``The Strong Interaction at the Frontier of Knowledge: Fundamental Research and Applications'' as deliverable D17.2. STRONG-2020 has received funding from the European Union's Horizon 2020 research and innovation program under grant agreement No.~824093.  

The layout and general content of the manuscript were discussed during  the meeting
\href{https://www.ggi.infn.it/showevent.pl?id=430}{{\em Phase Transitions in Particle Physics}} held in GGI, Firenze, 28$^{\text{th}}$ March~--~1$^{\text{st}}$ April 2022.

The manuscript has been prepared by some of the organisers and the speakers, and we warmly thank all the other participants\footnote{Workshop participants: Gert Aarts, Joerg Aichelin, Chris Allton, Andreas Athenodorou, Dimitrios Bachtis, Claudio Bonanno, Vitaly Bornyakov, Nora Brambilla, Elena Bratkovskaya, Costanza Conti, Roberto Contino, Salvatore Cuomo, Francesca Cuteri, Tetyana Galatyuk, Jacopo Ghiglieri, Jana N.~Guenther, Tim Harris, Rachel Houtz, Frithjof Karsch,  Benjamin Kitching-Morley, Andrey Yu.~Kotov, Ilya Kudrov, Anirban Lahiri, Biagio Lucini, Lorenzo Maio, Jan Pawlowski, Michael J.~Peardon, Andrea Pelissetto, Owe Philipsen, Antonio Rago, Claudia Ratti, Michele Redi, Roman Rogalyov, Sinéad Ryan, Francesco Sannino, Chihiro Sasaki, Philipp Schicho, Christian Schmidt-Sonntag, Sipaz Sharma, Olga Soloveva, Marianna Sorba, Giovanni Villadoro, Uwe-Jens Wiese} 
and organisers\footnote{Workshop organisers: Claudio Bonati, Mattia Bruno, Michele Caselle, Leonardo Cosmai, Massimo D'Elia,
Petros Dimopoulos, Francesco Di Renzo, Leonardo Giusti, Maria Paola Lombardo, Marco Panero, Mauro Papinutto, Michele Pepe} for the talks and discussions which have 
been most useful for this work. 
In particular we are grateful to Vitaly Bornyakov, Roman Rogaliov and Ilya Kudrov. 

It is a pleasure to thank the other members of the LatticeHadrons Network core group 
Mike Peardon, Gunnar Bali,  Gregorio Herdo\'{\i}za, and Hartmut Wittig for their help and support. 

The GGI hospitality and perfect organization of the workshop are gratefully acknowledged. 

Gert Aarts and Chris Allton are supported by the UKRI Science and Technology Facilities Council (STFC) Consolidated Grant No. ST/T000813/1.

Andreas Athenodorou has been financially supported by the European Union’s Horizon 2020 research and innovation programme ``Tips in SCQFT'' under the Marie Skłodowska-Curie grant agreement No. 791122 as well as by the NI4OS-Europe funded by the European Commission under the Horizon 2020
European research infrastructures grant agreement no. 857645.

Claudio Bonanno acknowledges the support of the Italian Ministry of Education, University and Research under the project PRIN 2017E44HRF, ``Low dimensional quantum systems: theory, experiments and simulations''. The work of Claudio Bonanno is also supported by the Spanish Research Agency (Agencia Estatal de Investigación) through the grant IFT Centro de Excelencia Severo Ochoa CEX2020-001007-S and, partially, by grant PID2021-127526NB-I00, both funded by MCIN/AEI/10.13039/501100011033. Claudio Bonanno also acknowledges support from the project H2020-MSCAITN-2018-813942 (EuroPLEx) and the EU Horizon 2020 research and innovation programme, STRONG-2020 project, under grant agreement No 824093.

Nora Brambilla acknowledges support from the  Deutsche Forschungsgemeinschaft (DFG)  cluster of excellence ``ORIGINS'' under Germany’s Excellence Strategy - EXC-2094 - 390783311 and from the DFG Project-ID 196253076 - TRR 110.

Elena Bratkovskaya, Frithjof Karsch, Christian Schmidt, Olga Soloveva and Sipaz Sharma acknowledge support by the Deutsche Forschungsgemeinschaft (DFG) through the grant CRC-TR 211 ``Strong-interaction matter under extreme conditions'' - project number 315477589 - TRR 211.

The research of Mattia Bruno is funded through the MUR program for young 
researchers ``Rita Levi Montalcini''.

The work of Francesco Di Renzo has received funding from the European Union’s Horizon 2020 research and innovation programme under the Marie Skłodowska-Curie grant agreement No. 813942 (EuroPLEx).

Tetyana Galatyuk acknowledges support by the DFG CRC-TR 211, HFHF, ELEMENTS:500/10.006, GSI F\&E, EMMI.

Frithjof Karsch and Christian Schmidt acknowledge support from the Deutsche Forschungsgemeinschaft (DFG) through the grant 315477589-TRR 211 
``NFDI 39/1'' for the PUNCH4NFDI consortium and from the  grant EU H2020-MSCA-ITN-2018-813942 (EuroPLEx) of the European Union.

The work of Biagio Lucini was supported by the UKRI Science and Technology Facilities Council (STFC) Consolidated Grant ST/T000813/1, by the Royal Society Wolfson Research Merit Award WM170010, by the Leverhulme Foundation Research Fellowship RF-2020-461{\textbackslash}9 and by the European Research Council (ERC) under the European Union’s Horizon 2020 research and innovation programme under grant agreement No 813942.

Jan M. Pawlowski is funded by the Deutsche Forschungsgemeinschaft (DFG, German Research Foundation) under Germany’s Excellence Strategy EXC 2181/1 - 390900948 (the Heidelberg STRUCTURES Excellence Cluster) and the Collaborative Research Centre SFB 1225 (ISOQUANT).

Claudia Ratti acknowledges support by the US National Science Foundation under grants no. PHY1654219, PHY2208724 and PHY-2116686. Her
work was supported in part by the US National Science Foundation (NSF) within the framework of the MUSES collaboration, under grant number OAC-2103680. This material is based upon work supported in part by the U.S. Department of Energy,
Office of Science, Office of Nuclear Physics, within the framework of the Beam Energy Scan Theory (BEST) Topical Collaboration.

Chihiro Sasaki acknowleges partial support by the Polish National Science Centre (NCN) under OPUS Grant No. 2018/31/B/ST2/01663, and by the World Premier International Research Center Initiative (WPI) through MEXT, Japan.

Philipp Schicho has been supported by the European Research Council, grant no. 725369, and by the Academy of Finland, grant no. 1322507.

The research of Uwe-Jens Wiese is supported by the Schweizerischer Nationalfonds.

{\bf Open Access Statement} -- For the purpose of Open Access the authors have applied a Creative Commons Attribution (CC BY) licence to any Author Accepted Manuscript version arising.

\section*{Author contributions}

\noindent
The manuscript has been elaborated and discussed with the authors. 
The main responsibilities are enlisted below, and specific contributions are indicated as footnotes in the text.\vspace{0.5\baselineskip}

% Report coordinators
\noindent
{\sc Report coordinators}
\vskip 0.1 truecm
\noindent Claudio Bonanno, Michele Caselle, Leonardo Cosmai, Massimo D'Elia, Francesco Di Renzo, \underline{Maria Paola Lombardo}, Marco Panero.
\vskip 0.4 truecm

% Editors
\noindent
{\sc Editors}
\vskip 0.1 truecm
\begin{itemize}
\item Sec.~\ref{sec:introduction}, Maria Paola Lombardo
\item Sec.~\ref{sec:thermal_phase_transitions_and_critical_points}, Sipaz Sharma
\item Sec.~\ref{sec:nature_and_phenomenology_of_the_quark-gluon_plasma},  Jana N.~Guenther
\item Sec.~\ref{sec:methodological_challenges_spectral_functions_and_sign_problem}, Chris Allton, Christian Schmidt
\item Sec.~\ref{sec:conformal_phase}, Marco Panero
\item Sec.~\ref{sec:cosmology_topology_axions}, Claudio Bonanno
\item Sec.~\ref{sec:statistical_field_theory}, Michele Caselle
\item Sec.~\ref{sec:machine_learning}, Andreas Athenodorou
\end{itemize}

% BIBLIOGRAPHY

% PAPERS

% TALKS

\end{document}